\documentclass[prc,preprint,superscriptaddress,showkeys,showpacs,
 preprintnumbers,amsmath,amssymb,aps,floatfix]{revtex4-1}
\usepackage{graphicx}
\usepackage[sort&compress]{natbib}
\usepackage{dcolumn}
\usepackage{multirow}
\usepackage{bigstrut}
\usepackage{type1cm}
\usepackage{placeins}
\usepackage{rotating}

\def\bfsig{\boldsymbol{\sigma}}

\begin{document}
\preprint{\today}

\title{Energy-dependent partial-wave analysis of all \\
         antiproton-proton scattering data below 925 MeV/$c$}

\author{Daren Zhou}
\author{Rob G. E. Timmermans}
\affiliation{KVI, Theory Group, University of Groningen, Zernikelaan 25,
                   NL-9747 AA Groningen, The Netherlands}

\date{\today}
\vspace{3em}

\begin{abstract}
\noindent
We present a new energy-dependent partial-wave analysis of
all antiproton-proton elastic ($\overline{p}p\rightarrow\overline{p}p$)
and charge-exchange ($\overline{p}p\rightarrow\overline{n}n$)
scattering data below 925 MeV/$c$ antiproton laboratory momentum.
The long-range parts of the chiral one- and two-pion exchange
interactions are included exactly. The short-range interactions,
including the coupling to the mesonic annihilation channels, are
parametrized by a complex boundary condition at a radius of
$r=1.2$ fm. The updated database, which includes significantly
more high-quality charge-exchange data, contains 3749 scattering
data. The fit results in $\chi^{2}_{\text{min}}/N_{\text{df}}=1.048$,
where $N_{\text{df}}=3578$ is the number of degrees of freedom.
We discuss the description of the experimental data and we
present the antiproton-proton phase-shift parameters.
\end{abstract}


\maketitle

\section{Introduction}
The antinucleon-nucleon ($\overline{N}\!N$) interaction at low energies is of
fundamental interest, but progress towards understanding it has always been
hindered by the lack of scattering data. Major steps forward were taken at the
Low Energy Antiproton Ring (LEAR) at CERN in the 1980's and the early 1990's.
For the first time, good-quality data became available for the total cross section
and the total annihilation cross section as function of antiproton laboratory
momentum ($p_{\rm lab}$), for the analyzing power in antiproton-proton elastic
scattering ($\overline{p}p\rightarrow\overline{p}p$), and for the differential cross section
and analyzing power in charge-exchange scattering ($\overline{p}p\rightarrow\overline{n}n$),
at  antiproton momenta above about 200 MeV/$c$. Unfortunately, LEAR was
closed in 1996 and $\overline{p}p$ scattering experiments came to a halt.
However, the enormous physics potential of a low-energy antiproton beam
is clear, especially when it can be polarized, and in recent years the interest
to investigate $\overline{p}p$ scattering has been revived, for instance by the
collaboration for Polarized Antiproton eXperiments (PAX) \cite{PAX}.

The dominant feature of antiproton-proton scattering at low energy is the annihilation
into mesons, a complex multiparticle process that is difficult to model. In pre-LEAR
days, some qualitative understanding was obtained by using simplified prescriptions,
such as a simple absorptive boundary condition~\cite{Bal58,Spe67,Dal77} or a
state-independent two- or three-parameter optical
potential~\cite{Phi67,Bry68,Myh77,Dov80,Koh86,Hip89,Hip91,Mul95}. 
These models could describe the integrated total, annihilation, and charge-exchange
cross sections, but not the differential observables. Motivated by the experiments at
LEAR, more sophisticated $\overline{N}\!N$ models were developed in order to attempt
a more quantitative fit to the data. Examples are the Paris optical-potential model
\cite{Cot82,Lac84,Pig91,Pig94,ElB99,ElB09} and the Nijmegen~\cite{Tim84,Tim85}
and Pittsburgh~\cite{Liu90} coupled-channels models.

In Refs.~\cite{Tim91,Tim94,Tim95} an energy-dependent partial-wave analysis (PWA)
of all  $\overline{p}p$ scattering data below $p_{\rm{lab}}=925$ MeV/$c$ was developed,
in order to arrive at a model-independent description of the $\overline{N}\!N$ interaction.
The method of analysis was adapted from the Nijmegen PWAs of the $pp$ and $np$
scattering data~\cite{Ber88,Ber90,Sto93,Ren99,Ren03}. These PWAs exploit
as much as possible our knowledge about the interaction in the description of the
energy dependence of the scattering amplitudes. The long-range interactions, which
are responsible for the rapid energy variations of the amplitudes, are included exactly
in the Schr\"odinger equation, while the slow energy variations due to the essentially
unknown short-range interactions are parametrized phenomenologically by a state- and
energy-dependent boundary condition at some radius $r=b$. In this way, an economic
and model-independent high-quality description of the scattering database is possible.
In the $\overline{N}\!N$ case~\cite{Tim91,Tim94,Tim95}, one assumes that the long-range
potential is given by the charge-conjugated version of a corresponding nucleon-nucleon
($N\!N$) potential, and, by implementing a complex boundary condition, one bypasses
with this strategy as well our lack of knowledge of the short-range annihilation dynamics.

There are two important reasons to update the $\overline{p}p$ PWA of Ref.~\cite{Tim94}.
The first and perhaps main motivation is the renewed experimental interest in $\overline{N}\!N$
scattering. The second reason is theoretical and is motivated by the progress reached in
the last two decades in the understanding of the $N\!N$ interaction within the framework of
chiral effective field theory. In particular, the $pp$ and $np$ PWAs have been updated by
including, next to the electromagnetic and the one-pion exchange (OPE) potential, the
long-range parts of the chiral two-pion exchange (TPE) potential~\cite{Ren99,Ren03},
instead of the heavy-boson exchanges of the Nijmegen potential~\cite{Nag78,Sto94},
thereby improving even more the model independence and the quality of the $N\!N$ PWAs
of Refs.~\cite{Ber88,Ber90,Sto93}. Motivated by that success, we include here as well the
charge-conjugated TPE potential in the long-range $\overline{N}\!N$ interaction, instead
of the charge-conjugated heavy-boson exchanges that were used in Ref.~\cite{Tim94}.

At the same time, we take the opportunity to update the database of $\overline{p}p$
scattering data. The database constructed in Ref.~\cite{Tim94} included all scattering data
published in a regular physics journal up to early 1993. A number of high-quality data sets
from LEAR became available only later, in particular differential cross sections and analyzing
powers for the charge-exchange reaction $\overline{p}p\rightarrow\overline{n}n$.
Also the first measurements of the depolarization and spin-transfer observables for
$\overline{p}p\rightarrow\overline{n}n$ were published only later. These data sets
can be included now and they provide significant new constraints on the PWA solution.

The organization of our paper is as follows: In Sec.~\ref{sec:PWA} the method of PWA
developed in Ref.~\cite{Tim94} is reviewed. We summarize only the main points in order
to make this paper self-contained and we emphasize the differences of our PWA with
Ref.~\cite{Tim94}. In Sec.~\ref{sec:BC} we discuss the boundary condition that parametrizes
the short-range interaction. In Sec.~\ref{sec:ChPT} the long-range $\overline{N}\!N$ potential
is discussed, in particular its chiral TPE component. In Sec.~\ref{sec:Data}, the new database
is discussed and the statistical methods are reviewed. In Sec.~\ref{sec:Results} we present
the results of the PWA and discuss the description of the measured observables. In
Sec.~\ref{sec:Phases} we present the $\overline{N}\!N$ $S$ matrix and phase-shift
parameters. We conclude in Sec.~\ref{sec:Summary}. An Appendix is devoted to a
study of the statistical quality of the database.

\section{The method of analysis} \label{sec:PWA}
For states with total angular momentum $J$,
the radial part of the wave function for the antiproton-proton system, $\Phi^{J}(r)$,
is obtained by solving the coupled-channels radial Schr\"odinger equation
\begin{equation}
  \left[ \frac{d^2}{dr^2} - 
         \frac{L^2}{r^2} +  p^{2}-2mV^{J}
         \right]\Phi^{J}(r) = 0~,
         \label{SEq}
\end{equation}
which is a differential equation in channel space. We include the channels $\overline{p}p$
and $\overline{n}n$. It is important to use this physical basis instead of the isospin basis,
in order to be able to include the long-range electromagnetic interactions and to treat
the threshold for charge-exchange scattering $\overline{p}p\rightarrow\overline{n}n$
at $p_{\rm lab} \simeq 99$ MeV/$c$ (or $T_{\rm lab}\simeq 5.2$ MeV) properly, which
gives a much better description of the low-energy charge-exchange data.
In Eq.~(\ref{SEq}), $p$ is a diagonal matrix with the channel momenta $p_a$ in the
center-of-mass system, $m$ is a diagonal matrix with the reduced mass $m_a$
of the two scattered particles in channel $a$ (so $m_a=M_p/2$ or $M_n/2$), and
$V^{J}$ is the potential with matrix elements $\langle \ell's'a'|V^{J}(r)|\ell\,s\,a\rangle$.
For partial waves with $\ell=J$, $s=0,1$, or $\ell=1, J=0$, the matrices are $2\times2$,
and for partial waves with $\ell=J\pm1$ ($J\ge1$), $s=1$, coupled by the tensor force,
the matrices are $4\times4$. The relation between the total energy $\sqrt{s}$ in the
center-of-mass system and the channel momentum is given by the relativistic expression
$\frac{1}{4}s=p_a^2+4m_a^2$.

We solve Eq.~(\ref{SEq}) numerically, starting with the boundary condition at $r=b$, up
to ``$r=\infty$,'' which in practice is a point outside of the range of the strong interaction.
The asymptotic form of $\Phi^{J}(r)$ for $r=\infty$ can be written as
\begin{equation}
   \Phi^{J}_{\rm as}(r) \stackrel{r\rightarrow\infty}{\sim}
   \sqrt{\frac{m}{p}}
   \left[ H_{1}(pr) S^{J} + H_{2}(pr) \right]~,
   \label{Phias}
\end{equation}
where $S^J$ is the partial-wave $S$ matrix and $H_{1}$ and $H_{2}$ are diagonal
matrices.
For the $\overline{p}p$ channel, where the Coulomb force acts, the entries are given by
\begin{eqnarray}
   &&H^{(1)}_{\ell}(\eta,pr) = F_{\ell}(\eta,pr)-iG_{\ell}(\eta,pr)~, \nonumber\\ 
   &&H^{(2)}_{\ell}(\eta,pr) = F_{\ell}(\eta,pr)+iG_{\ell}(\eta,pr)~,
\end{eqnarray}
where $F_{\ell}$ and $G_{\ell}$ are the standard regular and irregular Coulomb wave
functions; $\eta = \alpha/v_{\rm lab}$ is the relativistic Coulomb parameter, where
$\alpha$ is the fine-structure constant and $v_{\rm{lab}}$ is the velocity of the incoming
antiproton in the laboratory frame.
The asymptotic behavior of $F_{\ell}$ and $G_{\ell}$ is
\begin{eqnarray}
   F_{\ell}(\eta,pr) & \stackrel{r\rightarrow\infty}{\sim} &
   \sin\left[pr-\ell\frac{\pi}{2}+\sigma_{\ell}-\eta\ln(2pr)\right]~, \nonumber\\
   G_{\ell}(\eta,pr) & \stackrel{r\rightarrow\infty}{\sim} &
   \cos\left[pr-\ell\frac{\pi}{2}+\sigma_{\ell}-\eta\ln(2pr)\right]~,
\end{eqnarray}
where the Coulomb phase shift is $\sigma_{\ell} \: = \: \arg\,\Gamma(\ell+1+i\eta)$. 
For the $\overline{n}n$ channel, $\eta = 0$, and
\begin{eqnarray}
   F_{\ell}(0,\rho) = \rho j_{\ell}(\rho)~, \hspace{5em}
   G_{\ell}(0,\rho) = -\rho n_{\ell}(\rho)~,
\end{eqnarray}
where $j_{\ell}(\rho)$ and $n_{\ell}(\rho)$ are the spherical Bessel and
Neumann functions. The $S$ matrix is obtained from the matching condition
\begin{equation}
   W\big(\Phi^J(r_{\infty}),\Phi^J_{\rm as}(r_{\infty})\big)\equiv 0~,
\label{eq:match}
\end{equation}
where $\Phi^J$ is the numerical solution of Eq.~({\ref{SEq}) and $\Phi^J_{\rm as}$ is
given by Eq.~({\ref{Phias}). The Wronskian is defined by
\begin{equation}
   W(\Phi_{1},\Phi_{2})  = 
   \Phi_{1}^{\rm T} \frac{\textstyle 1}{\textstyle m} \Phi'_{2} -
   \Phi_{1}^{\prime\,{\rm T}} \frac{\textstyle 1}{\textstyle m} \Phi_{2}~,
\end{equation}
where the prime denotes differentiation with respect to $r$ and ``$\textrm{T}$''
means transposition. This gives for the partial-wave $S$ matrix
\begin{equation}
   S^{J} = - \left[ (\Phi^J)^{\prime\,{\rm T}}
   \frac{\textstyle 1}{\textstyle \sqrt{mp}} H_{1} - (\Phi^J)^{\rm T}
   \sqrt{\frac{\textstyle p}{\textstyle m}} H'_{1} \right]^{-1}
   \left[ (\Phi^J)^{\prime\,{\rm T}}
   \frac{\textstyle 1}{\textstyle \sqrt{mp}} H_{2} - (\Phi^J)^{\rm T}
   \sqrt{\frac{\textstyle p}{\textstyle m}} H'_{2} \right]~;
   \label{eq:smtrx}
\end{equation}
the prime on the Hankel functions denotes differentiation with respect to the
argument $pr$.

Due to the presence of the long-range electromagnetic interaction, care has to be 
taken to define the $S$ matrix ({\it i.e.} the phase-shift parameters)~\cite{Tim94,Ber88,Sto90}.
We include in the potential the long-range parts of the Coulomb, the magnetic-moment,
and the strong (one- and two-pion exchange) interactions, $V = V_C + V_{M\!M} + V_N$.
We integrate the Schr\"odinger equation up to a point outside the range of the strong
interaction, where we match to Coulomb (for $\overline{p}p$) and Bessel (for
$\overline{n}n$) wave functions. The $S$ matrix is therefore defined with respect
to the Coulomb force that acts in the $\overline{p}p$ channel. Because we need to
include the infinite-range Coulomb interaction and part of the magnetic-moment
interaction in all partial waves, but the finite-range nuclear interaction only up to
some maximum value of $J$, we decompose the $S$ matrix in order to split off
the Coulomb part and the magnetic-moment part as
\begin{eqnarray}
   S_{C+M\!M+N}-1 & = & \left(S_C-1\right) +
   S_C^{1/2}\left(S^C_{C+M\!M}-1\right)S_C^{1/2} +
   \nonumber \\ &  & S_C^{1/2}\left(S^C_{C+M\!M}\right)^{1/2}
   \left(S^{C+M\!M}_{C+M\!M+N}-1\right)
   \left(S^C_{C+M\!M}\right)^{1/2}S_C^{1/2} \:\: , \label{smtrx}
\end{eqnarray}
where $S_C$ is the Coulomb $S$ matrix with matrix elements
$\langle \ell's'|S_C|\ell\,s\,\rangle \: = \:
\delta_{\ell\ell'}\delta_{ss'}\, \exp(2i\sigma_{\ell})$ in the $\overline{p}p$
channel and zero in the $\overline{n}n$ channel. In Eq.~(\ref{smtrx}) we used
matrix notation, because the magnetic-moment interaction contains a tensor
part and the $S$ matrix is not diagonal in orbital angular momentum; its
square root is well-defined, however.

The scattering amplitude is correspondingly decomposed as
\begin{equation}
  M_{C+M\! M+N}(\theta)  =  M_C(\theta) + M^C_{C+M\! M}(\theta)
                          + M^{C+M\! M}_{C+M\! M+N}(\theta)~,
\end{equation}
where $M_C(\theta)$ is the Coulomb scattering amplitude, $M^C_{C+M\! M}(\theta)$
is the magnetic-moment scattering amplitude in the presence of the Coulomb interaction,
and $M^{C+M\! M}_{C+M\!M+N}(\theta)$ is the scattering amplitude for the strong
interaction in the presence of the Coulomb and magnetic-moment interactions.
The matrix elements of $M_C(\theta)$ for $\overline{p}p$ scattering are given by
\begin{eqnarray}
   \langle s'm'|M_C(\theta)| s\,m\rangle & = &
   -\delta_{ss'}\delta_{mm'} \,
   \frac{\eta}{p(1-\cos\theta)}
   e^{-i\eta\ln\frac{1}{2}(1-\cos\theta)+2i\sigma_0}
   \nonumber \\ & = &
   -\delta_{ss'}\delta_{mm'}\, \frac{\eta}{2p}
   \frac{e^{2i\sigma_{0}}}{(\sin^2\frac{1}{2}\theta)^{1+i\eta}}~.
\label{eq:coul}
\end{eqnarray}
The matrix elements $M^C_{C+M\! M}(\theta)$ of the magnetic-moment interaction
are calculated in Coulomb distorted-wave Born approximation~\cite{Tim94,Sto90}.
The partial-wave decomposition of the nuclear
scattering amplitude is given by
\begin{eqnarray}
   &&\langle s'm'a'| M^{C+M\! M}_{C+M\! M+N}(\theta) | s\,m\,a \rangle = 
   \sum_{\ell\,\ell' J} \sqrt{4\pi(2\ell+1)} \: i^{\ell-\ell'} \:
   C^{\ell}_{0}\,^{s}_{m}\,^{J}_{m} \:
   C^{\ell'}_{m-m'}\,^{s'}_{m'}\,^{J}_{m} \:
   Y^{\ell'}_{m-m'}(\theta) \nonumber \\
  && \hspace{5em}\langle\ell's'a'|S_C^{1/2}\left(S^C_{C+M\! M}\right)^{1/2}
   \left(S^{C+M\! M}_{C+M\! M+N}-1\right)
   \left(S^C_{C+M\! M}\right)^{1/2}S_C^{1/2}|\ell\,s\,a\rangle / 2ip_a~,
\label{eq:ampl}
\end{eqnarray}
where $a$ denotes the channel $\overline{p}p$ or $\overline{n}n$.
Because $S^{C+M\! M}_{C+M\! M+N}$ is difficult to calculate it is approximated by
$S_{C+M\! M+N}^{C+M\! M} \simeq S_{C+N}^{C}$, where $S_{C+N}^{C}$ is the
$S$ matrix for the strong interaction in the presence of the Coulomb interaction.
From the scattering amplitude on the spin-singlet, spin-triplet basis, all the observables
can be calculated~\cite{LaF92}.

\section{The boundary condition approach} \label{sec:BC}
The coupled-channels Schr\"odinger equation, Eq.~(\ref{SEq}), is solved with a boundary
condition at a radius $r=b$, for each energy and for each partial wave. The fit to the data
is not very sensitive to the exact value of $b$, but in our case an optimal value $b=1.2$ fm
was found. For the specific form of the partial-wave boundary condition we define the
$P$ matrix~\cite{Jaf79,Bak86} by
\begin{equation}
P^J=b\left[(\Phi^J)^{-1}\left(\frac{d\Phi^J}{dr}\right)\right]_{r=b}~,
\label{p-matrix}
\end{equation}
where $\Phi^J(r)$ is the radial wave function. The $P$ matrix parametrizes the complicated
short-range interaction of the $\overline{p}p$ system. The coupling of the $\overline{p}p$
and $\overline{n}n$ channels to the mesonic annihilation channels is taken into account
by a complex $P$ matrix.

The $P$ matrix is a powerful tool in a PWA, since it provides the separation between
the long-range interaction, which is relatively model independent and taken into
account exactly in the Schr\"odinger equation, and the short-range interaction,
which is essentially unknown and parametrized completely phenomenologically.
The long-range interactions cause the rapid energy dependence of the scattering
amplitudes, while the short-range interactions result in slow energy variations. The
results, for that reason, do not depend on the details of the short-range interactions.
We therefore choose a simple parametrization for the $P$ matrix, which corresponds
to a state-dependent, {\it i.e.} spin- and isospin-dependent, short-range optical potential.
We assume that the interaction in each partial wave can be parametrized by a complex
spherical well, the depth of which is different for elastic and charge-exchange scattering,
{\it i.e.} for $I=0$ and $I=1$. For a single-channel partial wave with orbital angular momentum
$\ell$, isospin $I$, and with the spherical well $V_I+iW_I$, the $P$ matrix is given by
\begin{equation}
   P_\ell = p'b\, J'_\ell(p'b) / J_\ell(p'b) \ , \label{pl}
\end{equation}
where $J_\ell(\rho)=\rho j_\ell(\rho)$ and $p'^2=p^2-\overline{M}(V_I+iW_I)$, 
where $\overline{M}=(M_p+M_n)/2$.

The $P$ matrix is calculated on the isospin basis and then transformed to the physical
particle basis with the channels $\overline{p}p$, $\overline{n}n$. For the uncoupled partial
waves with $\ell=J$, $s=0,1$, or $\ell=1$, $J=0$, it is therefore a 2$\times$2 matrix. For the partial
waves with $\ell=J\pm1$ ($J\ge1$), $s=1$, coupled by the tensor force, we introduce for each
value of the isospin $I$ an additional mixing angle $\theta_{I\!J}$ between the partial waves with
$\ell=J-1$ and $\ell=J+1$. We write
\begin{equation}
P^J = \begin{pmatrix}
\cos\theta_{I\!J}  & \sin\theta_{I\!J}  \\
-\sin\theta_{I\!J}  & \cos\theta_{I\!J} 	
\end{pmatrix}
\begin{pmatrix}
P_{J-1} & 0 \\
0 & P_{J+1}	
\end{pmatrix}
\begin{pmatrix}
\cos\theta_{I\!J} & -\sin\theta_{I\!J}  \\
\sin\theta_{I\!J}  & \cos\theta_{I\!J} 	
\end{pmatrix}~,
\label{p-matrix_mix}
\end{equation} 
where $P_{J-1}$ and $P_{J+1}$ are the single-channel $P$ matrices of Eq.~(\ref{pl}) for
$\ell=J-1$ and $\ell=J+1$, respectively. On the particle basis, the $P$ matrix for these coupled
states is 4$\times$4.

\begin{table}
\caption{$P$-matrix parameters for the different partial waves.
         $V_0$ and $V_1$ are the real parts and $W_0$ and $W_1$
         are the imaginary parts of the short-range spherical-well potential,
         for isospin $I=0$ and $I=1$, respectively. The values of the mixing
         angles $\theta_{I\!J}$ that parametrize the off-diagonal $P$ matrix
         for the partial waves coupled by the tensor force are:
         $\theta_{01}$ = $7.6^{\circ}\pm0.4^{\circ}$ and  
         $\theta_{11}$ = $-10.7^{\circ}\pm0.8^{\circ}$
         for the $^3S_1$-$^3D_1$ waves;
         $\theta_{02}$ = $0.0^{\circ}$ and
         $\theta_{12}$ = $-8.8^{\circ}\pm1.6^{\circ}$
         for the $^3P_2$-$^3F_2$ waves; 
         $\theta_{03}$ = $-7.4^{\circ}\pm0.4^{\circ}$ and
         $\theta_{13}$ = $-6.9^{\circ}\pm1.4^{\circ}$
         for the $^3D_3$-$^3G_3$ waves.
         The quoted errors are defined as the change in each
         parameter that gives a rise in $\chi^2_{\rm min}$
         of 1 when the remaining parameters are refitted.}
\tabcolsep=1.3em
\renewcommand{\arraystretch}{1.0}
\begin{tabular}{c|cccc}
\hline
\hline
Partial wave  & $V_0$ (MeV) & $W_0$ (MeV) & $V_1$ (MeV) & $W_1$ (MeV) \\
 \hline
 $^1S_0$     &   $0$            &   $-161.7(25.2)$   &   $-516.1(19.4)$   &  $-132.8(19.9)$              \\
 $^3S_1$     &   $-135.6(9.5)$   &   $-166.9(8.3)$   &   $33.6(5.7)$    &  $-166.3(8.0)$               \\
 $^1P_1$     &   $0$            &   $-374.5(29.6)$   &   $0$          &  $-413.8(40.7)$              \\
 $^3P_0$     &   $-114.9(10.1)$   &   $-142.8(9.3)$   &   $-164.1(4.5)$   &  $-71.9(6.9)$               \\ 
 $^3P_1$     &   $-78.0(4.2)$   &   $-62.2(3.7)$   &   $0$          &  $-382.2(27.6)$               \\
 $^3P_2$     &   $-114.6(5.7)$   &   $-201.4(5.1)$   &   $-41.4(3.0)$   &  $-135.6(5.4)$               \\
 $^1D_2$     &   $-277.8(16.2)$   &   $-330.8(27.0)$   &   $-319.6(30.4)$   &  $-482.8(45.8)$               \\
 $^3D_1$     &   $0$          &   $-96.6(15.5)$   &   $0$          &  $-129.4(19.7)$                 \\
 $^3D_2$     &   $-120.7(17.6)$   &   $-95.5(16.8)$   &   $0$          &  $-338.6(27.3)$               \\
 $^3D_3$     &   $-235.7(7.7)$   &   $-181.1(8.4)$   &   $-102.0(9.1)$   &  $-66.6(7.8)$               \\
 $^1F_3$     &   $-510.0(22.9)$   &   $-312.4(35.6)$   &   $0$          &  $-335.3(82.0)$               \\ 
 $^3F_2$     &   $0$            &   $-356.0(56.6)$   &   $-554.0(26.5)$   &  $-317.1(27.0)$      \\   
 $^3F_4$     &   $-498.4(61.0)$   &   $-423.2(46.6)$   &   $0$   &  $0$      \\
\hline
\hline
\end{tabular}
\label{tab:pars}
\end{table}

In Ref.~\cite{Tim94}, the imaginary parts of the square wells were assumed to be
equal for $I=0$ and $I=1$ in each partial wave. We take these to be different here,
because this choice gives a better fit to the more recent high-quality charge-exchange
data. The fitted values of the $P$-matrix parameters are given in Table \ref{tab:pars}.
The fit to 3749 scattering data requires a total of 46 $P$-matrix parameters.
Almost all the short-range square-well potentials are attractive.
The quoted errors reflect the sensitivity of the fit to variations in the corresponding
parameters. These errors are defined as the change in each parameter that gives
a rise in $\chi^2_{\rm min}$ of 1 when the remaining parameters are refitted.
The lower partial waves all require parameters to obtain a good fit. To decide which
parameters to keep in the fit, a three-sigma criterion is used: When the error turns out
to be more than one third of the parameter value, it implies that $\chi^2_{\rm min}$ rises
by less than 9 when the remaining parameters are refitted. In that case the parameter
is set to zero, {\it i.e.} it is left out. Because of the centrifugal barrier, the fit becomes
progressively less sensitive to short-range parameters for the higher-$\ell$ partial
waves. We assume the parameters in these partial waves to be equal to the ones in
similar lower partial waves. For example, the parameters for  the $^3F_3$ and $^3G_4$
waves are taken to be the same as the ones for $^3D_2$; the ones for $^1G_4$ and
$^1H_5$ are the same as the ones for $^1F_3$; and the ones for $^3G_5$ and $^3H_6$
are the same as the ones for $^3F_4$. We include the partial waves as high as $J=12$,
which is for instance needed to describe the forward ``spike'' in the charge-exchange
differential cross section.

\section{The long-range antinucleon-nucleon potential} \label{sec:ChPT}
The potential tail for $r>b$ includes the electromagnetic and the strong (nuclear) interaction
$V_N$, where the electromagnetic interaction is the one-photon exchange potential, 
{\it i.e.} the Coulomb potential and the magnetic-moment interaction~\cite{Sto90},
\begin{equation}
   V = V_C + V_{M\!M} + V_N~.
\end{equation}
In contrast to the $N\!N$ PWAs, we do not include the vacuum-polarization potential, because
its effects are negligible, except for very low energies~\cite{Ber88}, where there are no
$\overline{p}p$ scattering data available. Two-photon exchange effects~\cite{Aus83} are
not taken into account either.

The Coulomb potential acts only in the $\overline{p}p$ channel and is
given by the expression
\begin{equation}
  V_C(r) = -\frac{\alpha'}{r}~,
\end{equation}
where $\alpha'$ takes care of the main relativistic corrections to the Coulomb potential.
It is defined by the relativistic Coulomb factor $\eta=\alpha'M_p/(2p)$.
The magnetic-moment potential in the $\overline{p}p$ channel is given by
\begin{equation}
   V_{M\!M}(r)  = \frac{\mu^{2}_{p}}{4M^{2}_{p}}\:\:
   \frac{\alpha}{ r^{3}}\:S_{12}
   \: + \: \frac{8\mu_{p}-2}{ 4M^{2}_{p}}
   \frac{\alpha}{ r^{3}}\:\boldsymbol{L}\cdot\boldsymbol{S}~,
   \label{Vmm}
\end{equation}
where $\mu_{p} = 1+\kappa_{p} = 2.793$, with $\kappa_{p}$ the anomalous magnetic moment of
the proton; the tensor operator $S_{12}= 3\,\boldsymbol{\sigma}_{1}\cdot \widehat{\boldsymbol{r}}\,
\boldsymbol{\sigma}_{2}\cdot\widehat{\boldsymbol{r}}-\boldsymbol{\sigma}_{1}\cdot\boldsymbol{\sigma}_{2}$, 
with $\boldsymbol{\sigma}_{1}$ and $\boldsymbol{\sigma}_{2}$ the spin operators of the two nucleons,
$\boldsymbol{L}$ is the angular momentum vector, and $\boldsymbol{S}=(\boldsymbol{\sigma}_{1}+
\boldsymbol{\sigma}_{2})/2$ the total spin. The spin-orbit potential is due to the interaction of the
magnetic moment of one particle with the charge of the other particle and includes a relativistic
correction from the Thomas precession. The tensor force is due to the interaction between the
magnetic moments of the two particles. The magnetic-moment interaction in the $\overline{n}n$
channel contains only the tensor-force part of Eq.~(\ref{Vmm}) with $\mu_n=\kappa_n=-1.913$ and $M_n$.

The nuclear potential $V_N$ contains the OPE and TPE potentials
for $\overline{N}\!N$ scattering. Since the strong interaction is invariant under charge conjugation
$C$, the $\overline{N}\!N$ potential can be obtained from the $N\!N$ potential by using the operator
$C$. If one assumes that isospin symmetry $SU(2,I)$ is exact, one can also use the $G$-parity operator,
which is defined as $G=C\exp(i\pi I_2)$, and thus contains charge conjugation and a rotation in isospin
space. The OPE potential is isospin dependent, while the TPE potential contains both isospin-independent
and isospin-dependent parts. When we define the nuclear potential in isospin space for the $N\!N$
system by
\begin{equation}
   V_N(N\!N) = W_\pi\,\vec{\tau}_1\!\cdot\!\vec{\tau}_2 + V_{2\pi} + 
                          W_{2\pi}\,\vec{\tau}_1\!\cdot\!\vec{\tau}_2 \ ,
   \label{opetpeNN}
\end{equation}
the potential for the $\overline{N}\!N$ system is given by
\begin{equation}
   V_N(\overline{N}\!N) = -W_\pi\,\vec{\tau}_1\!\cdot\!\vec{\tau}_2 + V_{2\pi} +
                                              W_{2\pi}\,\vec{\tau}_1\!\cdot\!\vec{\tau}_2 \ ,
\end{equation}
which implies for elastic and charge-exchange scattering, respectively,
\begin{eqnarray}
	V_{N}(\overline{p}p\rightarrow\overline{p}p) &=& W_\pi + V_{2\pi} - W_{2\pi}~, \nonumber \\
	V_{N}(\overline{p}p\rightarrow\overline{n}n) &=& 2\,(W_\pi-W_{2\pi})~,
\label{opetpennbar}	
\end{eqnarray}
where the factor 2 is due to isospin symmetry.

The TPE potential for $N\!N$ scattering has been derived from the effective nonlinear chiral
Lagrangian density, which implements the spontaneously broken $SU(2,L)\otimes SU(2,R)$
chiral symmetry of QCD~\cite{Ord92,Kai97,Ren99}.
The leading order of this effective Lagrangian density is the nonlinear Weinberg model,
\begin{equation}
   {\mathcal L}^{(0)}
   = -\overline{N}\left[\gamma_\mu{\mathcal D}^\mu + M
   + ig_A\gamma_5\gamma_\mu\,\vec{\tau}\cdot\vec{D}^\mu\right]N~,
\label{eq:PVWT}
\end{equation}
with the chiral-covariant derivative
\begin{eqnarray}
  {\mathcal D}^\mu N & = & \Big( \partial^\mu + \frac{i}{F_\pi}
  c_0\,\vec{\tau}\cdot\vec{\pi}\!\times\!\vec{D}^\mu \Big) N~,
\label{eq:covder}
\end{eqnarray}
where $\vec{D}^\mu=D^{-1}\partial^\mu\vec{\pi}/F_\pi$ and $D=1+\vec{\pi}^2/F_\pi^2$;
$M$ is the mass of the nucleon, $g_A=1.269$ is the Gamow-Teller coupling constant in
neutron $\beta$ decay, and $F_\pi=185$ MeV is the pion decay constant.
The subleading-order chiral Lagrangian density is
\begin{eqnarray}
  {\mathcal L}^{(1)}
   =  -\overline{N}\big[
  8 c_1 D^{-1}m_\pi^2\,\vec{\pi}^2/F_\pi^2 +
  4 c_3\,\vec{D}_\mu\!\cdot\!\vec{D}^\mu 
  + 2 c_4\,\sigma_{\mu\nu}\,\vec{\tau}\cdot
  \vec{D}^\mu\!\times\!\vec{D}^\nu \big]N~.
\label{eq:c134}
\end{eqnarray}
The constant $c_{0}=1$ multiplying the Weinberg-Tomozawa $N\!N\pi\pi$ interaction is
fixed by chiral symmetry. However, the coupling constants $c_j$ ($j=1,3,4$) are low-energy
constants that have to be determined from experimental data. These constants are of order
${\mathcal O}(1/M)$ and their values contain contributions from the ``integrated-out'' heavy
hadrons, in particular the $N$- and $\Delta$-isobars, and the two-pion resonances
$\varepsilon$(760) and $\varrho$(770). (The constant $c_2$ does not contribute to
$N\!N$ scattering at this order.)

\begin{figure}[htbp]
   \centering
   \includegraphics[width=0.1\textwidth]{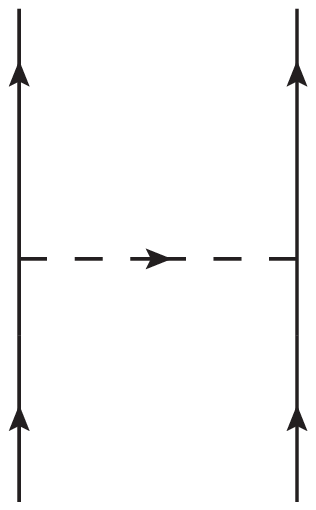}\hspace{3em}
   \includegraphics[width=0.1\textwidth]{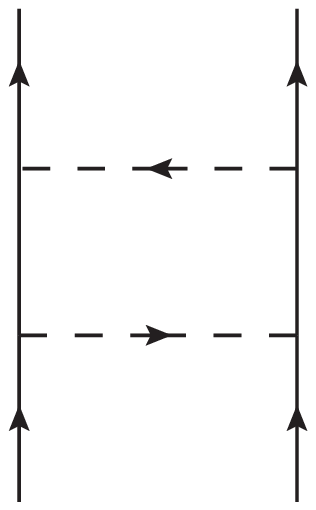}\hspace{3em}
   \includegraphics[width=0.1\textwidth]{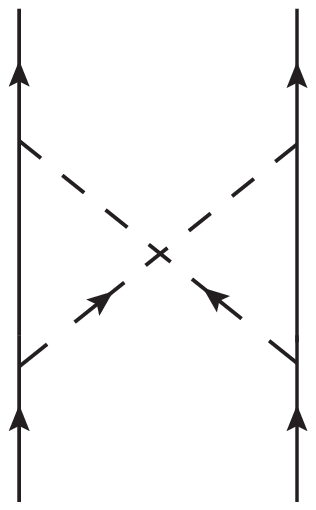}\hspace{3em}
   \includegraphics[width=0.1\textwidth]{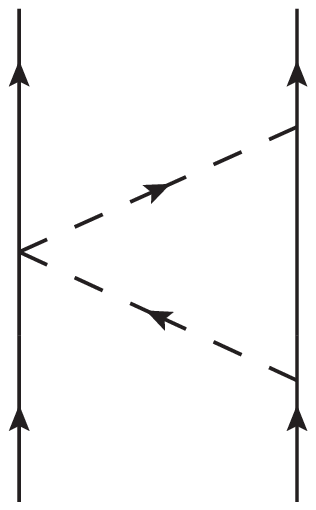}\hspace{3em}
   \includegraphics[width=0.1\textwidth]{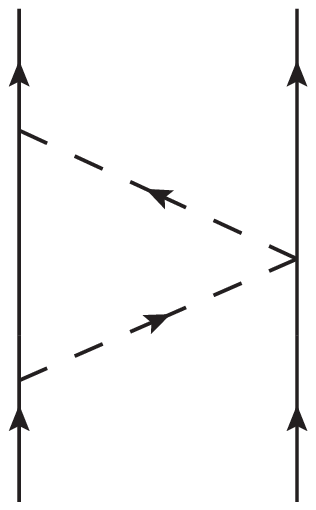}\hspace{3em}
   \includegraphics[width=0.1\textwidth]{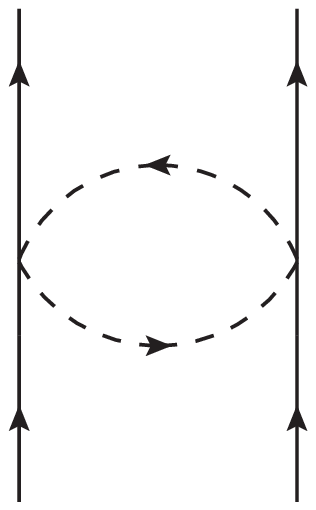}
\caption{\label{pion-exchange}The Feynman diagrams for one- and two-pion exchange.}   
\end{figure}

The Feynman diagrams for the OPE and TPE processes are shown in Fig. \ref{pion-exchange},
where the OPE diagram and the planar- and crossed-box TPE diagrams come from Eq.~(\ref{eq:PVWT}),
the ``triangle'' and ``football'' TPE diagrams containing the Weinberg-Tomozawa $N\!N\pi\pi$ interaction
also come from Eq.~(\ref{eq:PVWT}), while the other ``triangle'' TPE diagrams containing the $c_j$
($j=1,3,4$) $N\!N\pi\pi$ interactions come from Eq.~(\ref{eq:c134}). The pion-exchange potentials of
Eq.~(\ref{opetpeNN}) contain isospin-independent and isospin-dependent central, spin-spin, tensor,
and spin-orbit terms,
\begin{eqnarray}
  V_N &=& V_C + V_S\,\bfsig_1\cdot\bfsig_2 + V_T\,S_{12}
                   + V_{SO}\,\boldsymbol{L}\cdot\boldsymbol{S} \nonumber \\
        && + \left( W_C + W_S\,\bfsig_1\cdot\bfsig_2 + W_T\,S_{12}
                   + W_{SO}\,\boldsymbol{L}\cdot\boldsymbol{S}\right) 
                   \,\vec{\tau}_1\!\cdot\!\vec{\tau}_2~,
\label{VV}
\end{eqnarray}
where for OPE only the coefficients $W_S$ and $W_T$ are nonzero, and TPE contains
in leading order only the terms $V_S$, $V_T$, and $W_C$, whereas in subleading order
all the terms are nonzero. The coefficients in Eq. (\ref{VV}) are
written in terms of dimensionless functions as
\begin{equation}
  V_i(r) +W_i(r) \,\vec{\tau}_1\!\cdot\!\vec{\tau}_2 =
       f^{2n}\,\xi^{2n} \big[v_i(x) + w_i(x)\vec{\tau}_1\!\cdot\!\vec{\tau}_2 \big] m_\pi~,
\label{eq:V}
\end{equation}
with $n=1$ for OPE and $n=2$ for TPE, $i=C,\,S,\,T,\,SO$, and $x=m_\pi r$.
We use the conventional rationalized ``pseudovector'' $N\!N\pi$ coupling constant $f$,
normalized such that $f^2\simeq 0.075$~\cite{Tim93,Swa97}. This means that we
introduced the scaling mass $m_s$, chosen to be numerically equal to the
charged-pion mass $m_s=m_{\pi^{+}}$, and we defined $\xi=m_\pi/m_s$.
If the Goldberger-Treiman relation were exact, one would have that
$g_A/F_\pi = \sqrt{4\pi}f/m_s$.

The OPE potential contains isospin-dependent spin-spin and tensor parts, with
\begin{eqnarray}
   w_S(x) & = & e^{-x}/3x~, \nonumber \\
   w_T(x) & = & \left(1+x+x^2/3\right) e^{-x}/x^3~.
\end{eqnarray}
For the leading- and subleading-order TPE potential, the isospin-independent and the
isospin-dependent parts can be written as
\begin{eqnarray} 
   v_i(x) &=& \frac{2}{\pi}\,v_{i,1}(x) + \frac{m_\pi}{M_p}\, v_{i,2}(x)~, \nonumber \\
   w_i(x) &=& \frac{2}{\pi}\,w_{i,1}(x) + \frac{m_\pi}{M_p}\, w_{i,2}(x)~,
\end{eqnarray}
where the subscript $1$ indicates leading-order and the subscript $2$ subleading order. 
The leading-order, static TPE potential contains isospin-independent spin-spin and  tensor
terms and an isospin-dependent central term, with
\begin{eqnarray}
  v_{S,1}(x) & = & 12 K_0(2x)/x^3 +
             (12+8x^2)K_1(2x)/x^4~, \nonumber \\
  v_{T,1}(x) & = & -12 K_0(2x)/x^3 -
             (15+4x^2)K_1(2x)/x^4~,  \nonumber\\
  w_{C,1}(x) & = & \left( \tilde{c}_0^2+10\tilde{c}_0-23
             - 4x^2 \right) K_0(2x)/x^3  \nonumber \\
             &   & +\left[ \tilde{c}_0^2+10\tilde{c}_0-23 +
             (4\tilde{c}_0-12)x^2 \right] K_1(2x)/x^4~,
\label{eq:TMO}            
\end{eqnarray}
where $\tilde{c}_0=c_0/\tilde{g}_A^2$ with $\tilde{g}_A=F_\pi\sqrt{4\pi}f/m_s$ and
$K_n(2x)$ ($n=0,1$) are the modified Bessel functions (the hyperbolic Bessel functions)
of the second kind, which have asymptotic behavior $K_n(2x)\sim\sqrt{\pi/4x}\,e^{-2x}$ for
$x\rightarrow\infty$. The subleading-order potential contains nonstatic terms from
Eq.~(\ref{eq:PVWT}) and the leading-order terms from Eq.~(\ref{eq:c134}), which can
be written as
\begin{equation}
   v_{i,2}(x) = \textstyle{\sum_{k=1}^6}\,a_k\,e^{-2x}/x^k~,
 \label{eq:coef}
\end{equation}
and similarly for the $w_{i,2}(x)$ terms. The coefficients $a_k$ are listed in
Table~\ref{tab:TPE}, where we defined $\tilde{c}_{j}=c_{j}M_p/\tilde{g}_A^2$ ($j=1,3,4$)
and $\tilde{c}_{04}=\tilde{c}_0+4\tilde{c}_4$.

\begin{table}
\caption{The coefficients of the subleading-order
         TPE potential of Eq.~(\ref{eq:coef}) for the central, spin-spin, tensor, and spin-orbit terms~\cite{Ren99};
         we define $\tilde{c}_0=c_0/\tilde{g}_A^2$; $\tilde{c}_{j}=c_{j}M_p/\tilde{g}_A^2$ for $j=1,3,4$,
         and $\tilde{c}_{04}=\tilde{c}_0+4\tilde{c}_4$.}
\tabcolsep=0.4em        
\begin{tabular}{r|cccccc}
\hline
\hline
  & $a_{1}$ & $a_{2}$ & $a_{3}$ & $a_{4}$ & $a_{5}$ & $a_{6}$\\
\hline
$v_{C,2}$ & $3/4$
             & $9+48\tilde{c}_1+24\tilde{c}_3$
             & $27+96\tilde{c}_1+96\tilde{c}_3$
             & $99/2+48\tilde{c}_1+240\tilde{c}_3$
             & $54+288\tilde{c}_3$
             & $27+144\tilde{c}_3$ \\
$v_{S,2}$  & & $-3$  & $-9$   & $-33/2$ & $-18$ & $-9$ \\
$v_{T,2}$  & & $3/2$ & $27/4$ & $15$    & $18$  & $9$ \\
$v_{SO,2}$ & &       & $-12$  & $-36$   & $-48$ & $-24$ \\

$w_{C,2}$ & $3/2$
             & $4-2\tilde{c}_0$
             & $14-8\tilde{c}_0$
             & $31-20\tilde{c}_0$
             & $36-24\tilde{c}_0$
             & $18-12\tilde{c}_0$ \\
$w_{S,2}$ & & $-2/3$
             &   $-14/3+8\tilde{c}_{04}/3$
             &   $-31/3+20\tilde{c}_{04}/3$
             &   $-12+8\tilde{c}_{04}$
             &   $-6+4\tilde{c}_{04}$ \\
$w_{T,2}$ & & $1/3$
             &   $17/6-4\tilde{c}_{04}/3$
             &   $26/3-16\tilde{c}_{04}/3$
             &   $12-8\tilde{c}_{04}$
             &   $6-4\tilde{c}_{04}$ \\
$w_{SO,2}$ & & &
              &   $8-8\tilde{c}_0$
              &   $16-16\tilde{c}_0$
              &   $8-8\tilde{c}_0$ \\
\hline
\hline
\end{tabular}
\label{tab:TPE}
\end{table}

The OPE and TPE potentials for $\overline{p}p\rightarrow\overline{p}p$ and for
$\overline{p}p\rightarrow\overline{n}n$ are now given by Eq.~(\ref{opetpennbar}).
In the OPE potential, we take $m_\pi$ for $\overline{p}p$ and $\overline{n}n$ elastic scattering
to be the neutral-pion mass $m_{\pi^0}$ and for charge-exchange scattering the charged-pion
mass $m_{\pi^+}$. In the PWAs of Refs.~\cite{Tim91,Tim94}, the pion-nucleon coupling
constant $f_c^2=f_{pn\pi^+}f_{np\pi^-}/2$ was determined from the charge-exchange
data. In Ref.~\cite{Tim91} $f_c^2=0.0751(17)$ was found, and in Ref.~\cite{Tim94}
$f_c^2=0.0732(11)$. The values were consistent with the values for $f^2_{pp\pi^0}$
and $f_c^2$ found in the $pp$ and $np$ PWAs~\cite{Tim93}, resulting in the
recommended value $f^2=f^2_{N\!N\pi}=0.0750(9)$ for the pion-nucleon coupling
constant, with no significant evidence for isospin breaking~\cite{Swa97}. We have
taken here the values $f^2_{pp\pi^0}=0.075$ and $f_c^2=0.075$ for the OPE potential
for elastic and charge-exchange scattering, respectively. In the TPE potential we use
for $m_\pi$ the average pion mass $(2m_{\pi^+}+m_{\pi^0})/3=138.04$ MeV and the
charge-independent coupling constant $f^{2}=f^{2}_{N\!N\pi}=0.075$.
The strong potentials for $\overline{n}n\rightarrow\overline{n}n$ and
$\overline{n}n\rightarrow\overline{p}p$ are equal to the ones for
$\overline{p}p\rightarrow\overline{p}p$ and $\overline{p}p\rightarrow\overline{n}n$,
respectively.

The values of $c_j$ ($j=1,3,4$) were determined in the $pp$ and $np$
PWAs~\cite{Ren99,Ren03}. The $c_1$ term in Eq.~(\ref{eq:c134}) breaks chiral symmetry
explicitly, since it is proportional to $m_\pi$. The value of $c_1$ cannot be determined
accurately from the $N\!N$ data. It was fixed theoretically at $c_{1}=-0.76$/GeV by assuming
a value for the pion-nucleon sigma term~\cite{Ren99}. We take the same value here.
It is interesting, however, to probe the sensitivity of our results to variations in $c_3$
and $c_4$. It is difficult to determine $c_3$ and $c_4$ and their statistical
errors by a fit to the database. Since they are parameters in the long-range interaction
for $r>b$, this would require that for each small step in varying $c_3$ or $c_4$, the
Schr\"odinger equation would have to be solved for all the energies.
However, we found that very good results were obtained for the values
$c_{3}=-5.8$/GeV and $c_{4}=4.0$/GeV, where we estimate the uncertainties to be of
the order of 0.5. This means that the values we found are remarkably consistent with
the values determined in the $pp$ PWA to 350 MeV: $c_3=-5.08(28)$/GeV and
$c_4=4.70(70)$/GeV \cite{Ren99}. In the $pp$ and $np$ PWA to 500 MeV the values
$c_3=-4.78(10)$/GeV and $c_4=3.96(22)$/GeV were found \cite{Ren03}. One could
interpret this as a demonstration of charge conjugation invariance of the TPE interaction.
We leave a more careful study of the chiral OPE and TPE potential tail in $\overline{N}\!N$
scattering for the future.

The resulting long-range OPE and TPE potentials should be compared to the ones
of Ref.~\cite{Tim94} where the charge-conjugated version of the high-quality soft-core
Nijmegen one-boson exchange (OBE) potential \cite{Nag78,Sto94} was used as long-range
interaction. In both cases,
OPE is included, so one should compare TPE to the exchange of the heavy bosons, in
particular the two-pion resonances $\varepsilon$(760) and $\varrho$(770).
Since the vector mesons have negative charge parity, the coupling constants
of $\varrho(770)$ and $\omega(782)$ change sign when going from nucleons
to antinucleons. When we write schematically for the $pp$ potential
\begin{equation}
    V(pp\rightarrow pp) = W_\pi + V_\varepsilon + W_\varrho + V_\omega  + \dots \ ,
\end{equation}
we obtain for the OBE potential for elastic $\overline{p}p\rightarrow\overline{p}p$ and
charge-exchange $\overline{p}p\rightarrow\overline{n}n$ scattering 
\begin{eqnarray}
    V(\overline{p}p\rightarrow\overline{p}p) & = &
                 W_\pi + V_\varepsilon - W_\varrho - V_\omega + \dots \ , \nonumber \\
     V(\overline{p}p\rightarrow\overline{n}n) & = & 2\,(W_\pi - W_\varrho + \dots) \ ,
\end{eqnarray}
respectively. This should be compared
to Eq.~(\ref{opetpennbar}). It implies that for the $N\!N$ case the central potential is
relatively weak, because there is a cancellation between the repulsion due to the
vector mesons and the attraction due to the scalar mesons, there is a strong
coherent spin-orbit force from the exchange of the scalar and vector mesons,
and the tensor forces due to OPE and $\varrho(770)$ exchange have opposite sign. 
For the $\overline{N}\!N$ case, a strong coherent central attraction results due
to scalar- and vector-meson exchange and a relatively weak spin-orbit potential.
Moreover, a strong coherent tensor potential acts in $\overline{N}\!N$ due to
OPE and $\varrho(770)$ exchange. This strong tensor force dominates the
charge-exchange $\overline{p}p\rightarrow\overline{n}n$ and strangeness-exchange
$\overline{p}p\rightarrow\overline{\Lambda}\Lambda$ processes, where no
neutral mesons can be exchanged~\cite{Tim88,Swa89,Tim90,Tim92}.

The chiral TPE potential in subleading order has qualitatively a number of similar
features. Because the values of $c_3$ and $c_4$ are large, the corresponding
``triangle'' diagrams with an $N\!N\pi\pi$ ``seagull'' interaction lead to relatively
strong potentials. The $c_3$ term gives rise to a strong central attraction, while
the $c_4$ term gives rise to a strong tensor force with the same sign as the
tensor force due to OPE. This results in a strong attractive central force in the
elastic process $\overline{p}p\rightarrow\overline{p}p$ and a strong coherent tensor
force in the charge-exchange process $\overline{p}p\rightarrow\overline{n}n$. This can
be understood because the $c_3$ and $c_4$ terms contain effects from ``integrated-out''
$\varepsilon$(760) scalar-isoscalar and $\varrho$(770) vector-isovector mesons,
respectively.
In fact, these two mesons are prominent broad two-pion resonances. In the potential
of Refs.~\cite{Nag78,Sto94} their widths are treated in a two-pole approximation; the
lowest-mass poles correspond to mesons of masses of about 550 and 650 MeV,
respectively, resulting in relatively long-range potentials.

\section{Antiproton-proton database and statistics} \label{sec:Data}
The antiproton-proton database was constructed for the first time in Ref.~\cite{Tim94}.
It included all available scattering data below antiproton laboratory momentum 925 MeV/$c$
published up to early 1993 in a regular physics journal, {\it i.e.} total and annihilation cross
sections, differential cross sections and analyzing powers for elastic and charge-exchange
scattering, total cross sections for charge-exchange scattering, and (very few) differential
depolarizations for elastic scattering. At that time, most of the experiments at LEAR were
finished. However, some more data sets were published after the completion of the PWA
of Ref.~\cite{Tim94}. We include these data sets here, along with a few data sets for which
the numerical values were not available back then. The present, new database is summarized
in Table~\ref{database}. The data sets that were not included in Ref.~\cite{Tim94} are marked
with an asterisk in the left column. We always consult the original publications for information
about the data and their statistical and systematic uncertainties.

Statistical tools are an essential part of the data analysis in a PWA. We use exactly the same methods
as in the $N\!N$ PWAs~\cite{Ber88}. We mention here only the main relevant points, more details
can be found in Ref.~\cite{Tim94}. We perform a least-squares fit of the model parameters to
the total database, which contains individual data sets labelled by $A$. One data set contains
$N_A$ individual data points labelled by $i$. The $\chi^{2}$ of the fit is correspondingly defined as
\begin{equation}
\chi^{2}(\boldsymbol{p})=\sum_{A}\chi^{2}_{A}(\boldsymbol{p})=\sum_{A}\text{min}\left[\sum_{i=1}^{N_A}
\left(\frac{M_{A,i}(\boldsymbol{p})-\nu_{A}E_{A,i}}{\epsilon_{A,i}}\right)^2
+\left(\frac{\nu_{A}-1}{\epsilon_{A,0}}\right)^2\right]~,
\label{chi2}
\end{equation}
where $\boldsymbol{p}$ is the parameter vector with $N_{\text{par}}$ entries, $M_{A,i}(\boldsymbol{p})$
is the value predicted by the model for the measured observable $E_{A,i}$ labelled $i$ in set $A$ with
statistical error $\epsilon_{A,i}$ (in several cases, point-to-point systematic errors were added in
quadrature to the statistical errors in the experimental papers). In most cases, the data sets have
an overall normalization uncertainty,
denoted by $\epsilon_{A,0}$, specified by the experimentalists. For each of these sets we introduce
a normalization parameter $\nu_{A}$ that multiplies the measured values $E_{A,i}$ of the entire set.
In the case that the experimental data sets are only relative, or in the case that the normalization
error was underestimated, the error $\epsilon_{A,0}$ is taken to be $\infty$ (in practice very large)
and the corresponding normalization parameter $\nu_{A}$ is ``floated.'' The contributions to $\chi^{2}$
of these normalizations are then zero. In a few cases the normalizations are absolute, {\it i.e.}
$\epsilon_{A,0}=0$, and the contributions to $\chi^{2}$ of these normalizations are again zero.

By using a sophisticated numerical fitting code, the value of $\chi^{2}(\boldsymbol{p})$ is minimized
with respect to the model parameters. By using the definition Eq.~(\ref{chi2}), the normalization
parameters are adjusted implicitly. According to the theory of least-squares fitting, the expectation
value of the minimum is $\langle\chi^{2}_{\rm{min}}\rangle=N_{\rm{df}}\pm\sqrt{2N_{\rm{df}}}$,
where $N_{\rm df}$ is the number of degrees of freedom, provided the data points are distributed
statistically ({\it i.e.} they do not contain systematic errors) and provided they are Gaussian
(which is the case for counting experiments with enough events per bin). The error matrix $E$
of the model parameters is defined by
\begin{equation}
   (E^{-1})_{\alpha\beta}=\frac{\partial^{2}\chi^{2}(\boldsymbol{p})}
     {2\partial p_{\alpha}\partial p_{\beta}}\Big |_{\boldsymbol{p}=\boldsymbol{p}_{\rm{min}}}~,
\label{errmatrix}
\end{equation}
where $\boldsymbol{p}_{\rm{min}}$ are the values of the model parameters in the minimum value of
$\chi^{2}$. The error matrix allows us to determine the error in the model parameter $p_{\alpha}$
as $\sqrt{E_{\alpha\alpha}}$. This error corresponds to the variation in that parameter that gives
a rise in $\chi^2_{\rm min}$ of 1 when the remaining parameters are refitted. As mentioned in
Sec.~\ref{sec:BC}, when
the error is more than one third of the parameter value, it implies that $\chi^2_{\rm min}$ rises
by less than 9 when the remaining parameters are refitted. In that case the parameter is set to
zero, {\it i.e.} it is left out. The error matrix allows us also to provide statistical uncertainties on
our predictions for the observables. In the plots of the differential observables below, the PWA
result is given as a full red line with an area bordered by blue dotted lines that indicate the
one-standard-deviation uncertainty in the prediction.

\begin{table}
\caption{Reference table of antiproton-proton scattering data with $p_{\textrm{\,lab}}\leq923$ MeV/$c$. 
         The asterisks in the leftmost column indicate the data sets that were not included in Ref.~\cite{Tim94},
         because the data are more recent or because the values of the data points were not available.
         The meanings of the superscripts in the heading and the comments in the rightmost column are given at the end
         of the table.}
\tabcolsep=0.4em 
\renewcommand{\arraystretch}{0.94}
\label{database}        
\begin{tabular}{cccccccc}
\hline
\hline
$p_{\rm lab}$& $\rm No.^a$ &
              &  Norm  &  Pred.    &          &           &          \\
(MeV/$c$) &  $\rm type^b$ & $\chi^2_{\rm min}$ &
 $\rm error^{v}$ & $\rm norm^c$ & $\rm Rejected^d$ & Ref. & Comment   \\
\hline
 119.0--923.0 &   50  $\sigma_{\rm ce}$   &    46.5
                          &4\%& 1.058 & $\leq$385.0, \#=8; 468.0
                                                &\cite{Ham80a}& k, m  \\
 176.8--396.1 &    5  $\sigma_{\rm ann}$  &     9.4
                           &4.4\%& 0.949 & 176.8&\cite{Bru90} &       \\
 181.0   &   46 d$\sigma_{\rm el}$   &    .
                           &  5\%& . &  all
                                         &\cite{Bru86a,Bru91a}& j, l, o  \\
 183.0   &   13 d$\sigma_{\rm ce}$   &     13.3
                           &  5\%& 1.002 & 0.940,$-$0.170,$-$0.574 &\cite{Bru86b}&       \\
 194.8   &   19 d$\sigma_{\rm el}$   &          .
                           &  4\%&  .    & all  &\cite{Spe70} &f, i, o \\
 200.0--588.2 &   48  $\sigma_{\rm ann}$  &    52.5
                           &2.2\%& 0.989 &      &\cite{Bru90,Bru87}&  \\
\hline
 \end{tabular}
\end{table}
\addtocounter{table}{-1}
 
\begin{table}
\caption{({\it Continued\/}).}
\tabcolsep=0.3em
\renewcommand{\arraystretch}{0.81}
\begin{tabular}{cccccccc}
\hline
\hline
$p_{\rm lab}$& $\rm No.^a$ &
              &  Norm  &  Pred.    &          &           &          \\
(MeV/$c$) &  $\rm type^b$ & $\chi^2_{\rm min}$ &
 $\rm error^{v}$& $\rm norm^c$ & $\rm Rejected^d$ & Ref. & Comment   \\
\hline
 392.4   &   15 d$\sigma_{\rm ce}$   &   8.6
                           &  5\%(6.4\%)&  1.103    &  0.985, 0.954, 0.244, 0.099, \\
                           &&&&& $-$0.059, $-$0.239, $-$0.616 &\cite{Nak84b}&   w   \\ 
 395.9--737.4* &   28  $\sigma_{\rm tot}$  &  19.4
                           &0.8\%&  1.007    &   737.4   &\cite{Kam80,Nak84a}&  \\
 404.3   &   38 d$\sigma_{\rm el}$   &    36.3
                           &  4\%& 0.974 & 
                                                &\cite{Con68} & i, o  \\
 406.0--922.0 &   30 d$\sigma_{\rm el}$   &    29.9
                           &$\infty$& 0.757 &  norm    &\cite{Als79} & h, n   \\

 406.0   &  119 d$\sigma_{\rm el}$   &    100.5
                           &  5\%& 1.002 & 0.991, 0.750, 0.579
                                                &\cite{Cre83} & j, o  \\
\hline
 \end{tabular}
\end{table}
\addtocounter{table}{-1}
 
\begin{table}
\caption{({\it Continued\/}).}
\tabcolsep=0.7em
\renewcommand{\arraystretch}{1.01}
\begin{tabular}{cccccccc}
\hline
\hline
$p_{\rm lab}$& $\rm No.^a$ &
              &  Norm  &  Pred.    &          &           &          \\
(MeV/$c$) &  $\rm type^b$ & $\chi^2_{\rm min}$ &
 $\rm error^{v}$& $\rm norm^c$ & $\rm Rejected^d$ & Ref. & Comment   \\
\hline
 505.0   &   54 d$\sigma_{\rm el}$   &          .
                           &  5\%&  .    & all
                                         &\cite{Bru86a,Bru91a}&j, l, o\\
 505.0   &   14 d$\sigma_{\rm ce}$   &    17.3
                           &  5\%& 1.021 &  &\cite{Bru86b}&       \\
 508.0   &  119 d$\sigma_{\rm el}$   &   106.2
                           &  5\%& 0.998 & 0.663, 0.530
                                                &\cite{Cre83} & j, o  \\
 508.9   &   39 d$\sigma_{\rm el}$   &    29.6
                           &  5\%& 1.005 &      &\cite{Sak82} &   o   \\
 516.0   &    8 d$\sigma_{\rm el}$   &     5.1
                           &  5\%& 1.006 & 0.996&\cite{Iwa81} & j, o  \\
 523.0   &   15  $A_{y{\rm ,el}}$    &     11.5
                        &$4.5\%$& 1.037 &      &\cite{Kun88,Kun89}& \\
\hline
 \end{tabular}
\end{table}
\addtocounter{table}{-1}
 
\begin{table}
\caption{({\it Continued\/}).}
\tabcolsep=0.7em
\renewcommand{\arraystretch}{1.01}
\begin{tabular}{cccccccc}
\hline
\hline
$p_{\rm lab}$& $\rm No.^a$ &
              &  Norm  &  Pred.    &          &           &          \\
(MeV/$c$) &  $\rm type^b$ & $\chi^2_{\rm min}$ &
 $\rm error^{v}$& $\rm norm^c$ & $\rm Rejected^d$ & Ref. & Comment   \\
\hline
 565.5   &    8 d$\sigma_{\rm el}$   &     5.4
                           &  5\%& 0.994 & 0.996&\cite{Iwa81} & j, o  \\
 568.4   &   37 d$\sigma_{\rm el}$   &    35.4
                           &  5\%& 1.025 & $-$0.675, $-$0.825
                                                &\cite{Sak82} & i, o  \\
 577.2   &   36 d$\sigma_{\rm el}$   &    33.9
                           &  4\%& 0.969 &      &\cite{Con68} & i, o  \\
 578.1   &    9 d$\sigma_{\rm el}$   &     6.0
                           &  5\%& 1.001 & 0.999&\cite{Iwa81} & j, o  \\
 578.3   &  119 d$\sigma_{\rm el}$   &   133.0
                           &  5\%& 1.022 &      &\cite{Cre83} & j, o  \\
 584.0   &   10 d$\sigma_{\rm ce}$   &    6.3
                        &$\infty$& 1.043 &      &\cite{Tsu83} &   g   \\
\hline 
\end{tabular}
\end{table}
\addtocounter{table}{-1}
\begin{table}
\caption{({\it Continued\/}).}
\tabcolsep=0.8em
\renewcommand{\arraystretch}{1.01}
\begin{tabular}{cccccccc}
\hline
\hline
$p_{\rm lab}$& $\rm No.^a$ &
              &  Norm  &  Pred.    &          &           &          \\
(MeV/$c$) &  $\rm type^b$ & $\chi^2_{\rm min}$ &
 $\rm error^{v}$& $\rm norm^c$ & $\rm Rejected^d$ & Ref. & Comment   \\
\hline
 658.6   &    9 d$\sigma_{\rm el}$   &     8.9
                           &  5\%& 0.999 & 0.998&\cite{Iwa81} & j, o  \\
 670.0   &   10 d$\sigma_{\rm ce}$   &     6.2
                        &$\infty$& 1.150 &      &\cite{Tsu83} &   g   \\
 671.5   &    9 d$\sigma_{\rm el}$   &     3.8
                           &  5\%& 0.981 & 0.998&\cite{Iwa81} & j, o  \\
 679.0   &   26 d$\sigma_{\rm el}$   &          .
                        &$\infty$&  .    & all  &\cite{Kun89} &  h, l    \\
 679.0   &   27  $A_{y{\rm ,el}}$    &    25.1
                        &$4.5\%$& 1.005 & 0.540&\cite{Kun88,Kun89}& \\
 679.0   &    1  $D_{yy{\rm ,el}}$   &     3.2
                           & --  &  --    &      &\cite{Kun91} &  q  \\
\hline 
\end{tabular}
\end{table}
\addtocounter{table}{-1}
\begin{table}
\caption{({\it Continued\/}).}
\tabcolsep=0.7em
\renewcommand{\arraystretch}{1.0}
\begin{tabular}{cccccccc}
\hline
\hline
$p_{\rm lab}$& $\rm No.^a$ &
              &  Norm  &  Pred.    &          &           &          \\
(MeV/$c$) &  $\rm type^b$ & $\chi^2_{\rm min}$ &
 $\rm error^{v}$& $\rm norm^c$ & $\rm Rejected^d$ & Ref. & Comment   \\
\hline
 767.0*   &   10  d$\sigma_{\rm ce}$    &    9.1
                           &  15\%&  1.441 &       &\cite{Ahm95} &       \\ 
 767.0*   &   8  d$\sigma_{\rm ce}$    &    9.6
                           &  15\%&  1.231 &       &\cite{Ahm95} &       \\ 
 767.0   &   22  $A_{y{\rm ,ce}}$    &    28.0
                           &  4\%& 1.106 &      &\cite{Bir91} &       \\ 
 780.5   &   39 d$\sigma_{\rm el}$   &          .
                           &  5\%&  .    & all  &\cite{Kag87} &  l    \\
 780.5   &   15 d$\sigma_{\rm ce}$   &    6.7
                           &  5\%(7.1\%)& 0.963 & 0.982, $-$0.143     &\cite{Nak84b}&    w   \\
 783.0   &   30 d$\sigma_{\rm el}$   &          .
                        &$\infty$&  .    & all  &\cite{Kun89} &  h, l    \\
\hline
\hline
\end{tabular}
\end{table}
\FloatBarrier
\begin{itemize}
\item[a] The number includes all published data, except those given
         as 0.0$\pm$0.0 (see Comment i), and those having
         $p_{\mbox{\scriptsize lab}}>923$ MeV/$c$ (see Comment m).
\item[b] The subscripts ``el'' and ``ce'' denote observables
         in the elastic $\overline{p}p \rightarrow \overline{p}p$
         and charge-exchange $\overline{p}p \rightarrow \overline{n}n$
         reactions, respectively. ``d$\sigma$'' denotes
         a differential cross section d$\sigma$/d$\Omega$,
         ``$A_y$'' a polarization-type datum (asymmetry or
         analyzing power), ``$D_{yy}$'' a depolarization type
         datum, and ``$K_{yy}$'' a spin-transfer type datum. ``$\sigma_{\rm tot}$''
         stands for total cross section,
         ``$\sigma_{\rm ann}$'' for total annihilation cross section,
         and ``$\sigma_{\rm ce}$'' for total charge-exchange
         cross section.
\item[c] Normalization, predicted by the analysis, with which the
         experimental values should be multiplied before
         comparison with the theoretical values.
\item[d] Tabulated is $p_{\mbox{\scriptsize lab}}$ in MeV/$c$,
         $\cos\theta$, ``norm'' or ``all.'' The notation ``$\leq$385.0, \#=8''
         e.g. means that the 8 points with $p_{\textrm{lab}}\leq$385.0 MeV/$c$
         are rejected. The ``norm'' means that the given normalization is rejected 
         and a ``floated'' normalization is used instead. 
         The ``all'' means that all of the data points in this set are rejected.
\item[e]  Group rejected due to improbable low $\chi^2_{\rm min}$.
\item[f]  Group rejected due to improbable high $\chi^2_{\rm min}$.
\item[g]  ``Floated'' normalization. Data are relative only.
\item[h]  Normalization ``floated'' by us, since the norm contributes
          much more than 9 to $\chi^2_{\rm min}$.
\item[i]  Data points given as 0.0$\pm$0.0 not included.
\item[j]  Coulomb-nuclear interference measurement. Data points
          in the extreme forward angular region are rejected
          when they contain multiple-scattering effects.
\item[k]  Data points at low momenta rejected.
\item[l]  Problematic differential cross sections. Not included
          in the database. For detailed explanation, see Sec. VIIIB
          and Tables II and III of Ref.~\cite{Tim94}.
\item[m]  Part of a group of data with points having
          $p_{\mbox{\scriptsize lab}}>923$ MeV/$c$.
\item[n]  Elastic differential cross sections as a function of
          momentum taken at backward angle $\cos\theta=-0.994$.
\item[o]  Normalization error assumed by us, since no
          clear number is stated in the reference.
\item[p]  Depolarization data. Not included in the fit, in view
          of the large error bars.
\item[q]  Normalization error taken to be zero, in view
          of the large error bars of these data.
\item[r]  Data points taken at the same angles averaged.
\item[s]  Data taken from the website {\tt http://hepdata.cedar.ac.uk}.
\item[t]  Data not available.
\item[u]  The momentum is the average of 700 MeV/$c$ and 760 MeV/$c$.
\item[v]  Normalization errors as used in the fitting, as deduced from the
              experimental articles; when not explicitly given, a reasonable value was assumed by us.
\item[w]  The $x\%(y\%)$ notation means that $x\%$ is the overall normalization error and
           $y\%$ is the point-to-point systematic error. 
\end{itemize}

The total $\chi^2_{\rm min}$ is only a global measure for the quality of the fit. In the Appendix,
we discuss in more detail the statistical quality of the final database, by examining the final
$\chi^2$ distribution of the data points and how it compares to theoretical expectations.

Data selection is a necessary ingredient of a PWA. In PWAs of large amounts of scattering data,
a significant minority of the data sets turns out to be inconsistent with the rest of the database and
with the PWA solution. In these cases, the data sets usually suffer from large systematic errors, which
cannot be traced and corrected for. Examples in our case are the elastic differential cross sections
measured at LEAR, which are inconsistent among themselves and with earlier measurements,
and which in many cases cannot even be fitted properly with Legendre polynomials, as discussed
at length in Ref.~\cite{Tim94}.
Including these flawed data sets would seriously bias the PWA solution. To decide whether a data
set or an individual data point is acceptable, we use the standard statistical criteria outlined in
Ref.~\cite{Ber88} and already applied in Ref.~\cite{Tim94}. They are generalized three-sigma criteria:
Any single data point with $\chi_{A, i}^{2}>9$ is rejected, as well as any data set with significantly
too high or too low $\chi_{A}^{2}$, according to the limits given in Ref.~\cite{Ber88}. Some data or
data groups are rejected because of other reasons, as mentioned also in the Comments column
of Table~\ref{database}.

\section{Description of the data} \label{sec:Results}

\begin{figure}
   \centering
   \includegraphics[width=0.9\textwidth]{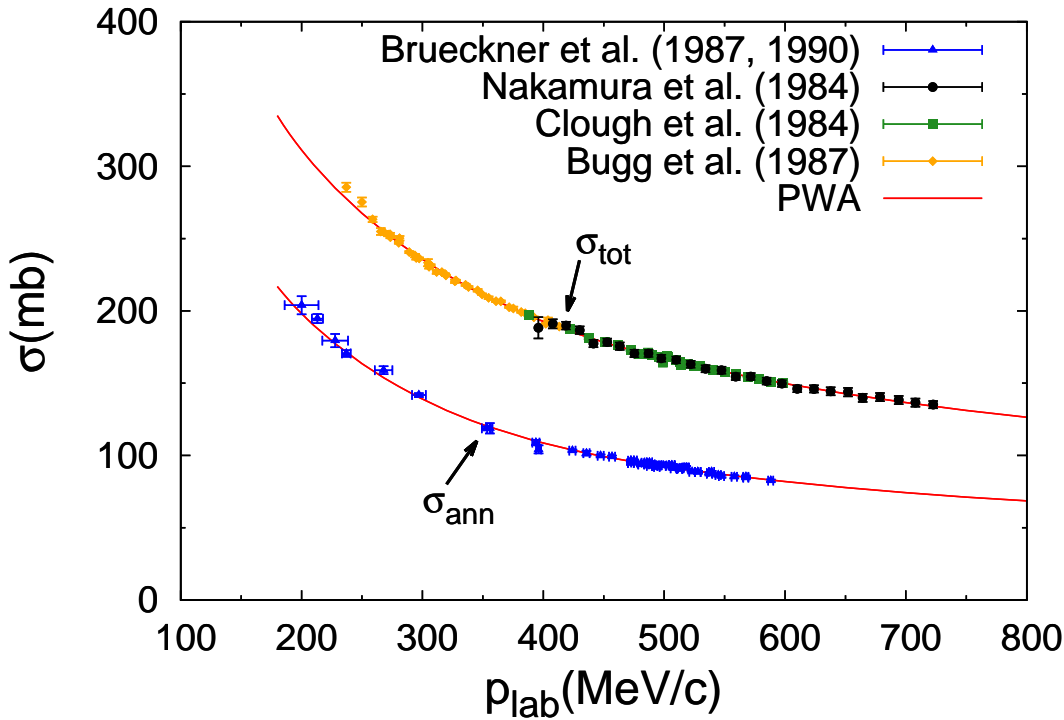} 
\caption{\label{Tot} (Color online)
Total cross sections and total annihilation cross sections as
function of antiproton laboratory momentum. The PWA fit has 
for Br\"uckner \textit{et al}.~\cite{Bru87,Bru90}
$\chi^{2}_{\textrm{min}}=9.4$ for 4 points $\sigma_{\textrm{ann}}$ and
$\chi^{2}_{\textrm{min}}=52.5$ for 48 points $\sigma_{\textrm{ann}}$;
for Nakamura \textit{et al}.~\cite{Nak84a}
$\chi^{2}_{\textrm{min}}=19.4$ for 27 points $\sigma_{\textrm{tot}}$;
for Clough \textit{et al}.~\cite{Clo84}
$\chi^{2}_{\textrm{min}}=35.2$ for 28 points $\sigma_{\textrm{tot}}$;
for Bugg \textit{et al}.~\cite{Bug87}
$\chi^{2}_{\textrm{min}}=55.3$ for 38 points $\sigma_{\textrm{tot}}$.}
\end{figure}

The final $\overline{p}p$ database contains $N_{\text{obs}}=3636$ scattering observables.
The details for each of the data sets can be found in In Table~\ref{database}. We need
$N_{\text{par}}=46$ model ($P$-matrix) parameters for an optimal fit. In the fit we must
determine at the same time $N_{\text{n}}$ normalization parameters, so the total number
of free parameters is $N_{\text{fp}}=N_{\text{par}}+N_{\text{n}}$. Of the $N_{\text{n}}$
normalization parameters $N_{\text{ne}}$ have a finite error, while the rest,
$N_{\text{nf}}=N_{\text{n}}-N_{\text{ne}}$, is the number of ``floated'' normalizations.
In our case, the total number of normalizations is 131, but we fixed the normalizations
for the five depolarization $D_{yy}$ and for the one spin transfer $K_{yy}$
measurements, because these data sets have relatively large error bars. Therefore,
$N_{\text{n}}=125$. Of these, $N_{\text{nf}}=12$ normalizations are ``floated,''
either because the data sets are relative only, or because the normalization errors were
underestimated in the experimental papers. Thus, the number of normalizations with errors
is $N_{\text{ne}}=N_{\text{n}}-N_{\text{nf}}=113$. This implies that the total number of free
parameters is $N_{\text{fp}}=N_{\text{par}}+N_{\text{n}}=171$, the total number of
data is $N_{\text{dat}}=N_{\text{obs}}+N_{\text{ne}}=3749$, and the number of degrees
of freedom is $N_{\text{df}}=N_{\text{dat}}-N_{\text{fp}}=3578$. The fit results in a minimum
$\chi^{2}$ value of $\chi^{2}_{\text{min}}=3750.6$.
Therefore, the minimum $\chi^{2}$ per datum is $\chi_{\text{min}}^{2}/N_{\text{dat}}=1.000$,
and the minimum $\chi^{2}$ per degree of freedom is $\chi_{\text{min}}^{2}/N_{\text{df}}=1.048$.
When the model is perfect and the database is a perfect statistical ensemble, one expects
$\langle\chi_{\text{min}}^{2}/N_{\text{df}}\rangle=1.000\pm0.024$, hence our result
for  $\chi_{\text{min}}^{2}/N_{\text{df}}$ is only two standard deviations too high. The
quality of the fit implies in particular that the charge-conjugated chiral OPE and TPE
potential provides an excellent long-range $\overline{N}\!N$ interaction.

\begin{figure}[t]
   \centering
   \includegraphics[width=0.7\textwidth]{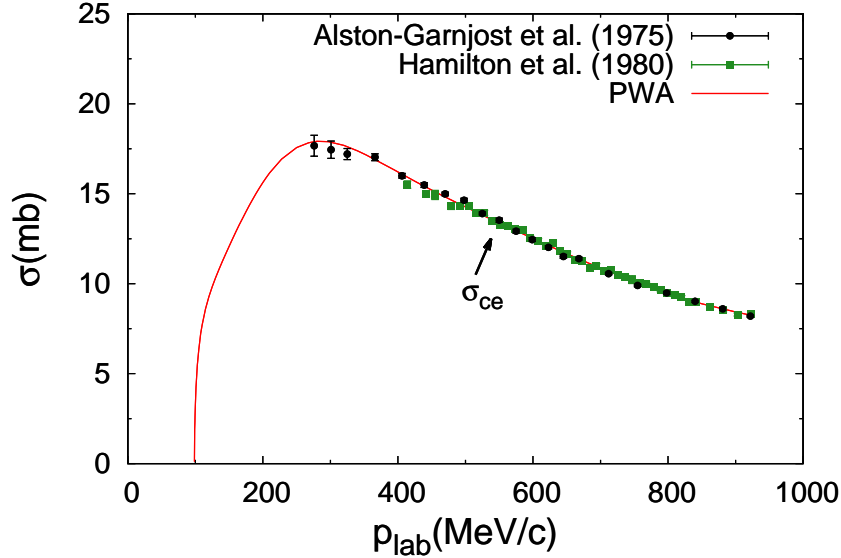}
\caption{\label{Tot_ce} (Color online)
Total charge-exchange cross sections $\sigma_{\textrm{ce}}$ as function of
antiproton laboratory momentum. The PWA fit has
for Alston-Garnjost \textit{et al}.~\cite{Als75} 
$\chi^{2}_{\textrm{min}}=26.2$ for 21 points;
for Hamilton \textit{et al}.~\cite{Ham80a} 
$\chi^{2}_{\textrm{min}}=46.5$ for 41 points.}
\end{figure}

\begin{figure}[ht]
   \centering
   \includegraphics[width=0.7\textwidth]{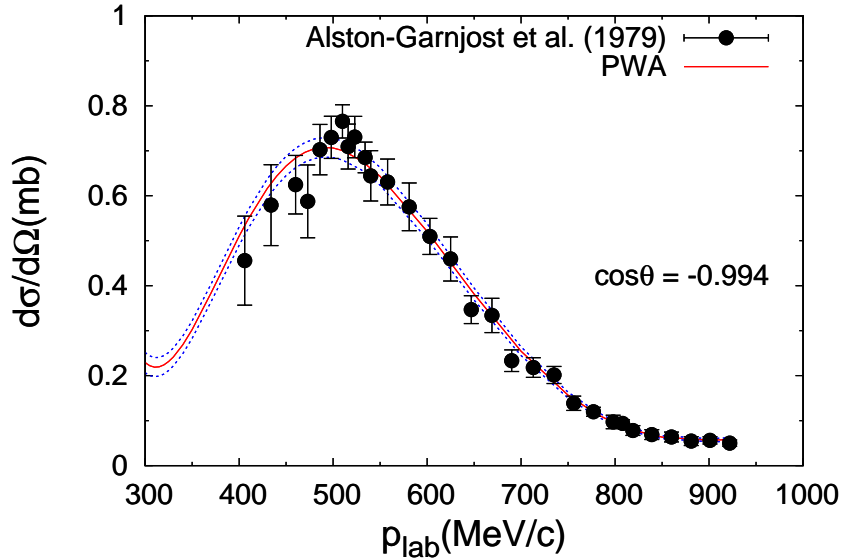}
\caption{\label{back_el} (Color online)
Elastic differential cross sections $d\sigma/d\Omega$
at backward angle, $\cos\theta = -0.994$,
as function of antiproton laboratory momentum.
The PWA result is given by the drawn red line and the dotted blue lines
indicate the one-sigma uncertainty region. The fit has for
Alston-Garnjost \textit{et al}.~\cite{Als79} $\chi^{2}_{\textrm{min}}=29.9$ for 30 points.}
\end{figure}

A detailed discussion of most of the data sets can be found in Ref.~\cite{Tim94}. Here
we will show the results for a number of important data sets, and in particular address
the high-quality data sets that were not available in Ref.~\cite{Tim94}. The data sets
in the figures have been multiplied by the predicted normalization factors given in
Table~\ref{database}. The rejected outliers are not plotted in the figures. In case
point-to-point systematic errors were added in quadrature to the statistical errors,
we plot these total errors.

In Fig.~\ref{Tot} the total cross sections $\sigma_{\rm tot}$
and the total annihilation cross sections $\sigma_{\rm ann}$
are plotted as function of $p_{\textrm{lab}}$, the antiproton momentum in the
laboratory frame. For the annihilation cross sections, we introduced two different
normalization parameters for the data taken with a thin and with a thick target, {\it cf.}
Table \ref{database}. In Fig.~\ref{Tot_ce} the total charge-exchange cross sections
$\sigma_{\rm ce}$ are plotted as function of $p_{\textrm{lab}}$. 
Unfortunately, there are no good data that map out the rise of the cross
section above the $\overline{p}p\rightarrow\overline{n}n$ threshold at
$p_{\textrm{lab}}\simeq 99$ MeV/$c$. In Fig.~\ref{back_el} the elastic differential
cross sections $d\sigma/d\Omega$ at backward angles with $\cos\theta = -0.994$
are plotted as function of the momentum in the laboratory frame. These
data are described well, but the normalization of the data set was ``floated.''
At low energies, the theoretical uncertainty of the PWA is significantly smaller
than the errors of the data points.

\FloatBarrier
\begin{table}[thbp]
\caption{Partial-wave elastic and charge-exchange cross sections, total cross sections, and
total annihilation cross sections, in mb, for $p_{\textrm{lab}}=$ 200, 400, 600, and 800 MeV/$c$.}
\tabcolsep=1em
\renewcommand{\arraystretch}{0.80}
\begin{tabular}{c|rrrr|rrrr}
\hline
\hline
 &\multicolumn{4}{c|}{\hspace{0.9em}$\overline{p}p\rightarrow\overline{p}p$}
 &\multicolumn{4}{c}{\hspace{1.4em}$\overline{p}p\rightarrow\overline{n}n$} \\
 $p_{\mbox{\scriptsize lab}}$ (MeV/$c$)
                               &   200 &   400 &   600 &  800  &   200 &   400 &   600 &  800 \\
 \hline
 $^1S_0$                       &  15.7 &   7.9 &   4.1 &  2.1  &   0.7 &   0.1 &       &      \\
 $^1P_1$                       &   0.9 &   2.5 &   4.5 &  5.6  &   0.8 &   0.1 &       &      \\
 $^1D_2$                       &   0.1 &   0.4 &   1.4 &  3.1  &   0.1 &   0.3 &   0.1 &      \\
 $^1F_3$                       &       &   0.1 &   0.2 &  0.5  &       &   0.1 &   0.1 &  0.1 \\
 $^1G_4$                       &       &       &       &  0.1  &       &       &   0.1 &  0.1 \\
 $^3P_0$                       &   4.9 &   5.4 &   5.0 &  3.5  &   1.5 &   0.8 &   0.1 &      \\
 $^3P_1$                       &   1.8 &   4.9 &   4.0 &  3.5  &   4.9 &   2.9 &   0.2 &  0.1 \\
 $^3D_2$                       &   0.1 &   0.3 &   1.0 &  1.5  &   0.3 &   2.4 &   2.5 &  1.0 \\
 $^3F_3$                       &       &   0.1 &   0.1 &  0.2  &       &   0.4 &   1.1 &  1.4 \\
 $^3G_4$                       &       &       &   0.1 &  0.1  &       &   0.1 &   0.3 &  0.5 \\
 
 $^3S_1$                       &  66.1 &  26.0 &  13.2 &  8.8  &   3.0 &   1.0 &   0.5 &  0.2 \\
 $^3S_1\rightarrow$$^3D_1$     &   0.3 &   0.4 &   0.2 &  0.1  &   0.8 &   1.5 &   1.1 &  0.6 \\
 $^3D_1\rightarrow$$^3S_1$     &   0.3 &   0.4 &   0.2 &  0.1  &   2.0 &   2.0 &   1.2 &  0.7 \\                       
 $^3D_1$                       &   0.1 &   0.5 &   0.8 &  1.0  &   0.1 &   0.5 &   0.6 &  0.4 \\                        
 
 $^3P_2$                       &   7.0 &  17.0 &  13.9 &  9.6  &   0.9 &   1.4 &   0.4 &  0.1 \\
 $^3P_2\rightarrow$$^3F_2$     &   0.1 &   0.1 &   0.1 &       &   0.1 &   0.5 &   0.5 &  0.5 \\
 $^3F_2\rightarrow$$^3P_2$     &   0.1 &   0.1 &   0.1 &       &   0.3 &   0.8 &   0.6 &  0.5 \\                                              
 $^3F_2$                       &       &       &   0.1 &  0.4  &       &       &   0.1 &  0.1 \\
 
 $^3D_3$                       &       &   1.6 &   5.9 &  7.0  &       &   0.5 &   1.3 &  0.6 \\
 $^3D_3\rightarrow$$^3G_3$     &       &   0.1 &   0.1 &       &       &   0.2 &   0.3 &  0.3 \\
 $^3G_3\rightarrow$$^3D_3$     &       &   0.1 &   0.1 &       &       &   0.3 &   0.5 &  0.4 \\                                             
 $^3G_3$                       &       &       &       &       &       &       &       &  0.1 \\
 
 $^3F_4$                       &       &       &   0.3 &  0.8  &       &       &   0.1 &  0.3 \\
 $^3F_4\rightarrow$$^3H_4$     &       &       &       &  0.1  &       &       &   0.2 &  0.2 \\
 $^3H_4\rightarrow$$^3F_4$     &       &       &       &  0.1  &       &   0.1 &   0.2 &  0.3 \\                    
 $^3H_4$                       &       &       &       &       &       &       &       &      \\
 Rest                          &       &       &   0.1 &  0.3  &       &   0.1 &   0.4 &  0.8 \\
\hline
Singlet                        &  16.7 &  10.9 &  10.2 &  11.3 &   1.6 &   0.6 &   0.4 &  0.3 \\
Triplet                        &  80.8 &  56.9 &  45.1 &  37.1 &  14.0 &  15.6 &  12.1 &  9.2 \\
 Total                         &  97.5 &  67.9 &  55.3 &  48.4 &  15.6 &  16.2 &  12.5 &  9.4 \\
 \hline 
 \hline
 &\multicolumn{4}{c|}{\hspace{0.8em}$\overline{p}p\rightarrow$ all}
 &\multicolumn{4}{c}{\hspace{1.3em}$\overline{p}p\rightarrow$ mesons} \\
 $p_{\mbox{\scriptsize lab}}$ (MeV/$c$)
                       &   200  &   400  &   600  &   800  &   200  &   400  &  600  & 800  \\
 \hline
                       &  311.2 &  192.6 &  149.8 &  126.4 &  198.1 &  108.5 &  81.9 & 68.6 \\
\hline
\hline 
\end{tabular}
\label{tab:partxs}
\end{table}

\begin{figure}
   \centering
   \includegraphics[width=0.48\textwidth]{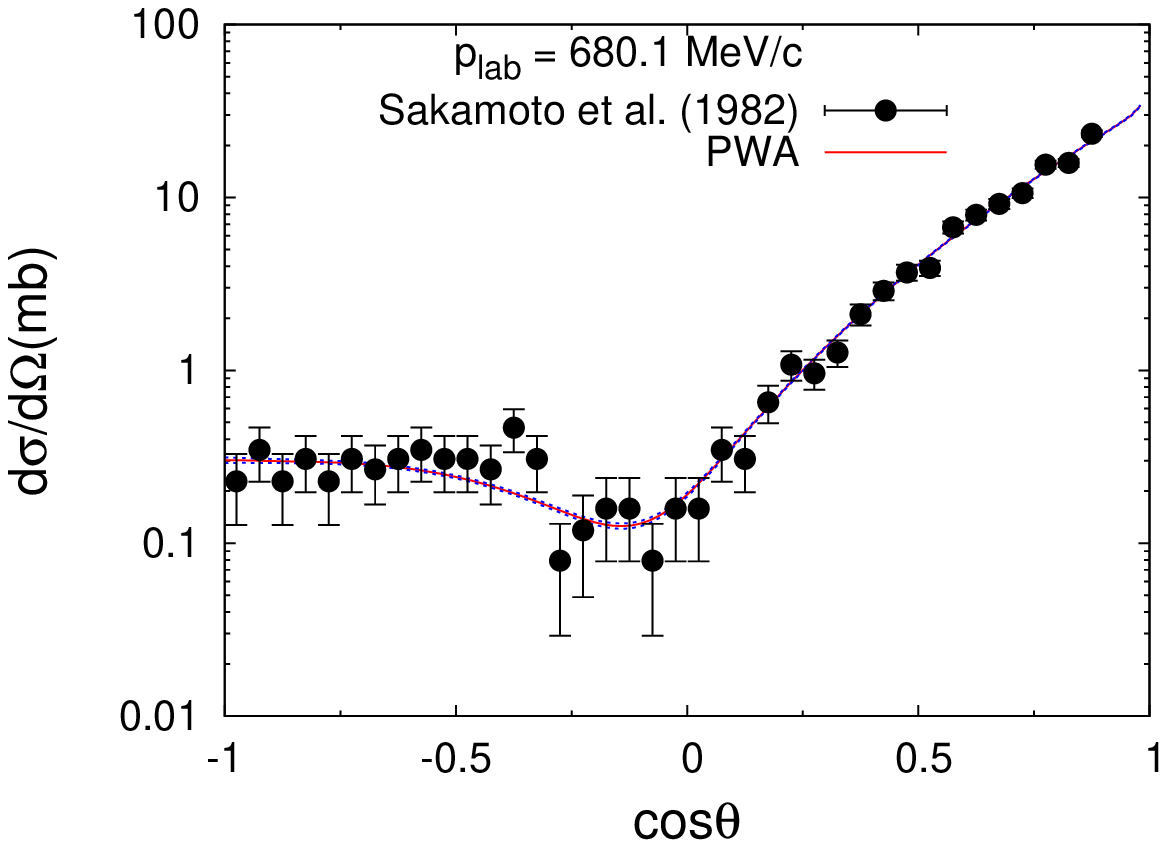}\hspace{1em}
   \includegraphics[width=0.48\textwidth]{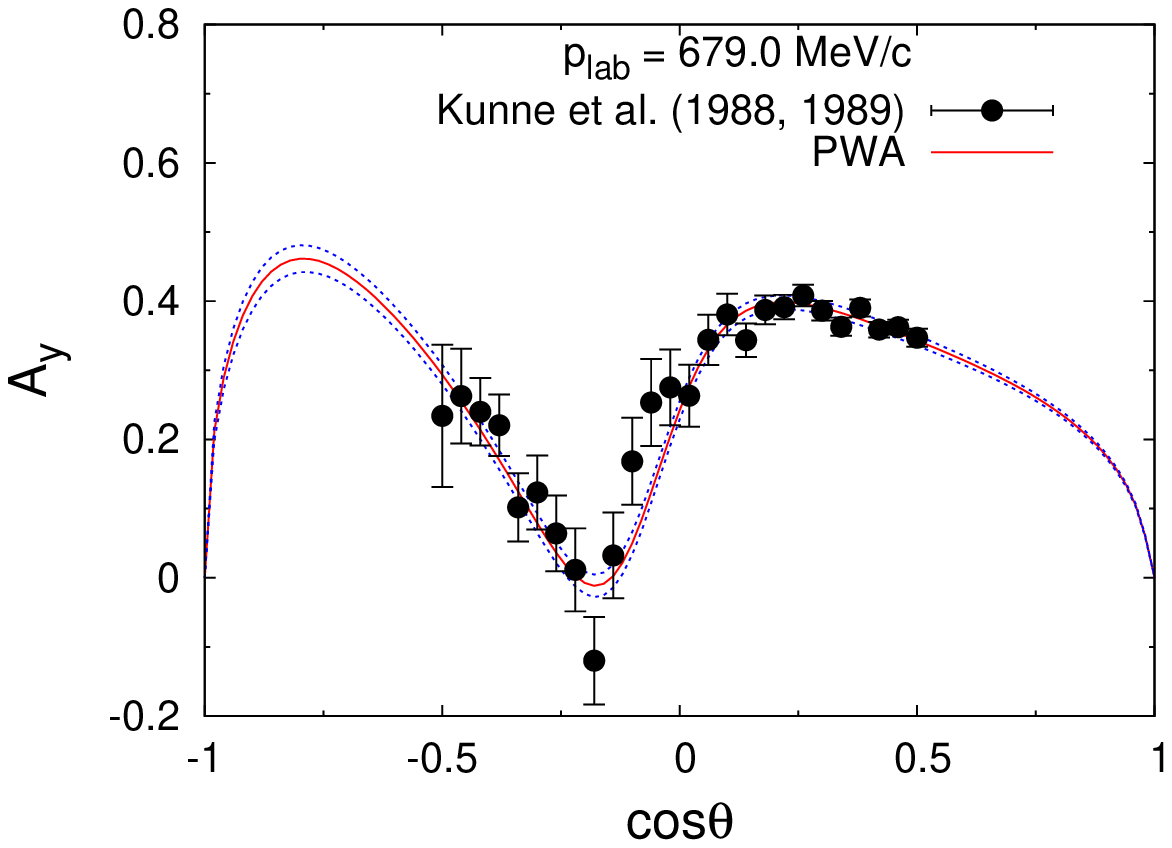}\\ \vspace{0.4em}
   \includegraphics[width=0.48\textwidth]{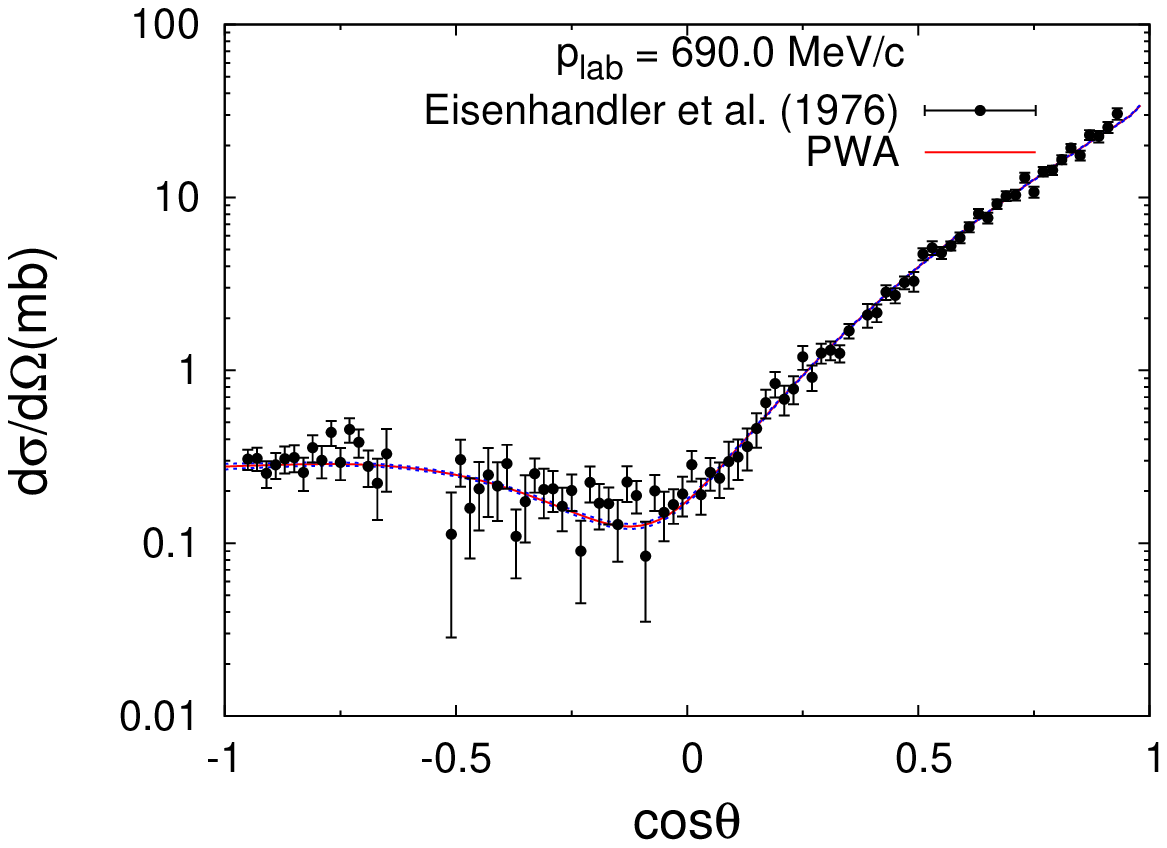}\hspace{1em}
   \includegraphics[width=0.48\textwidth]{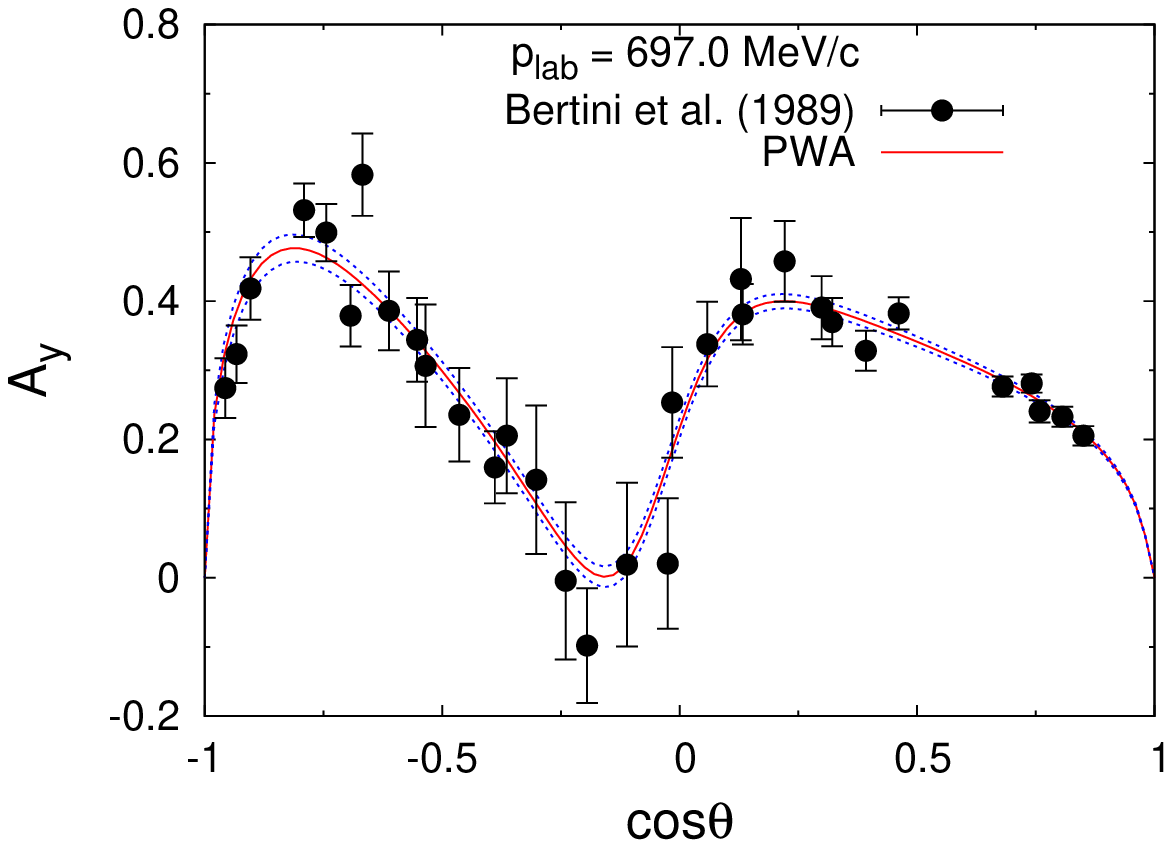}
\caption{\label{679_el} (Color online)
Differential cross sections and analyzing powers
for elastic scattering as function of angle in the center-of-mass system.
The PWA result is given by the drawn red line and the dotted blue lines
indicate the one-sigma uncertainty region. The fit has
for Sakamoto \textit{et al}.~\cite{Sak82}
$\chi^{2}_{\textrm{min}}=39.2$ for 38 points $d\sigma/d\Omega$;
for Kunne \textit{et al}.~\cite{Kun88,Kun89}
$\chi^{2}_{\textrm{min}}=25.1$ for 26 points $A_{y}$;
for Eisenhandler \textit{et al}.~\cite{Eis76}
$\chi^{2}_{\textrm{min}}=94.5$ for 88 points $d\sigma/d\Omega$;
for Bertini \textit{et al}.~\cite{Ber89}
$\chi^{2}_{\textrm{min}}=20.8$ for 32 points $A_{y}$.}
\end{figure}

The partial-wave cross sections for both elastic and charge-exchange scattering
at $p_{\textrm{lab}} =$ 200, 400, 600, and 800 MeV/$c$ are given in Table \ref{tab:partxs}.
It is clear that, in contrast to $N\!N$ scattering, many partial waves contribute to
$\overline{N}\!N$ scattering already at low energies. The reason is that the $\overline{N}\!N$
potentials, in particular the central and tensor components, are very strong.
The dominance of the tensor force is seen in particular in the charge-exchange
$\overline{p}p\rightarrow\overline{n}n$ reaction. For low energies of the final-state
$\overline{n}n$ system the strong tensor force leads to large cross sections for the
transitions $\ell(\overline{n}n)=\ell(\overline{p}p)-2$, in particular $^3D_1\rightarrow\,^3S_1$
and $^3F_2\rightarrow\,^3P_2$. This is similar to the strangeness-exchange reaction
$\overline{p}p\rightarrow\overline{\Lambda}\Lambda$, where these off-diagonal
tensor-force transitions due to $K(494)$ and $K^*(892)$ exchange dominate the cross
section in the $\overline{\Lambda}\Lambda$ threshold region~\cite{Tim90,Tim92}.
For these transitions, there is a large overlap between the wave functions of the
initial $\overline{p}p$ state and the final $\overline{n}n$ or $\overline{\Lambda}\Lambda$
state~\cite{Tim92} at low energy.
The contributions from the spin-triplet states are much larger than the contributions
from the spin-singlet states, especially for $\overline{p}p\rightarrow\overline{n}n$.
The total annihilation cross section is large, and decreases from a fraction of about 2/3
of the total cross section at $p_{\textrm{lab}} = 200$ MeV/$c$  to about 1/2 of the total
cross section at $p_{\textrm{lab}} = 800$ MeV/$c$. 

\begin{figure}
   \centering
   \includegraphics[width=0.48\textwidth]{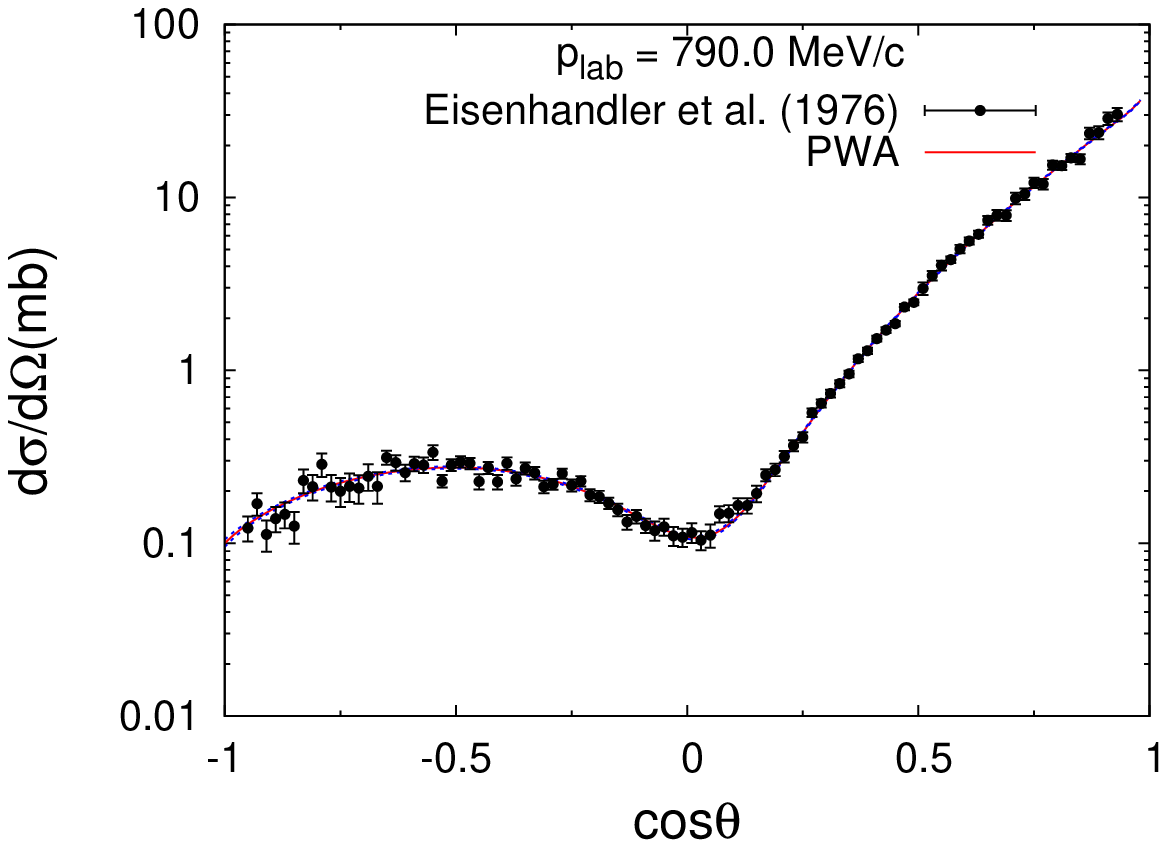}\hspace{1em}
   \includegraphics[width=0.48\textwidth]{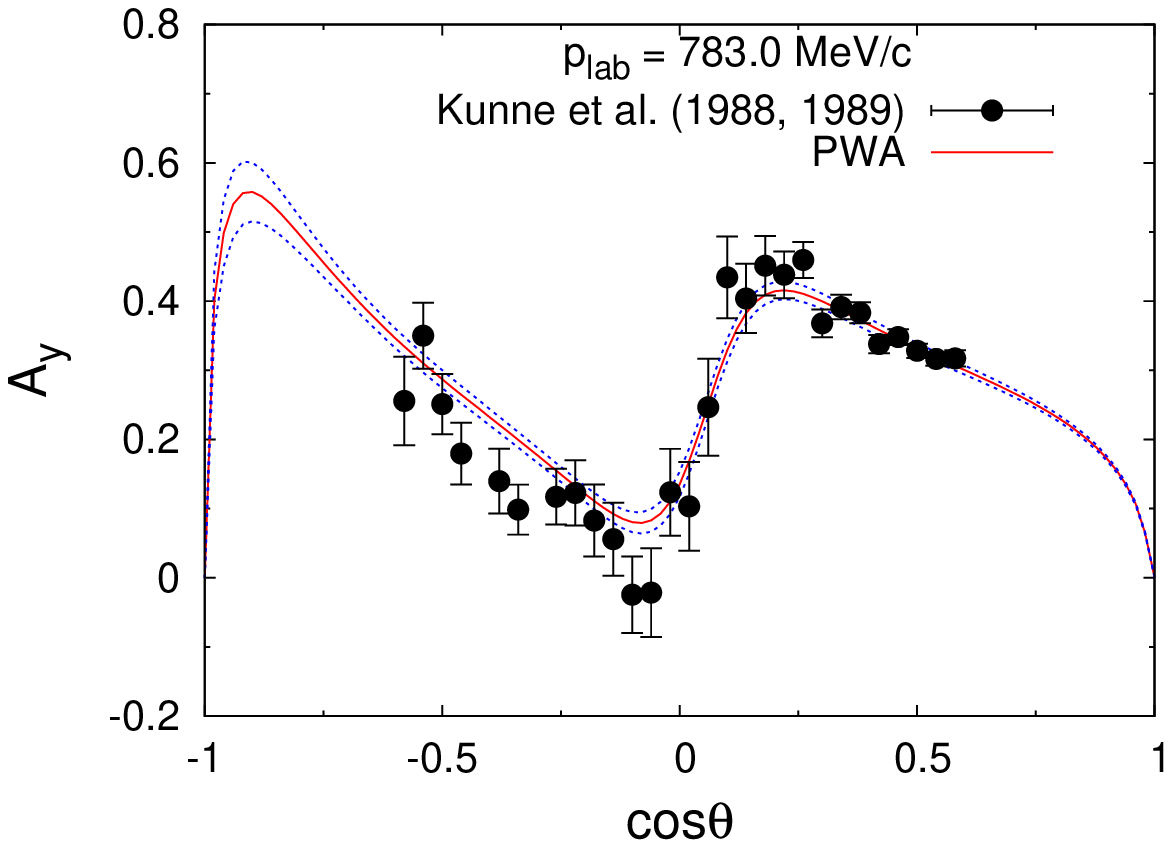}
\caption{\label{790_el} (Color online)
Differential cross sections and analyzing powers for elastic
scattering as function of angle in the center-of-mass system.
The PWA result is given by the drawn red line and the dotted blue lines
indicate the one-sigma uncertainty region. The fit has
for Eisenhandler \textit{et al}.~\cite{Eis76}
$\chi^{2}_{\textrm{min}}=95.3$ for 95 points $d\sigma/d\Omega$;
for Kunne \textit{et al}.~\cite{Kun88,Kun89}
$\chi^{2}_{\textrm{min}}=36.2$ for 28 points $A_{y}$.}
\end{figure}

\begin{figure}
   \centering
   \includegraphics[width=0.48\textwidth]{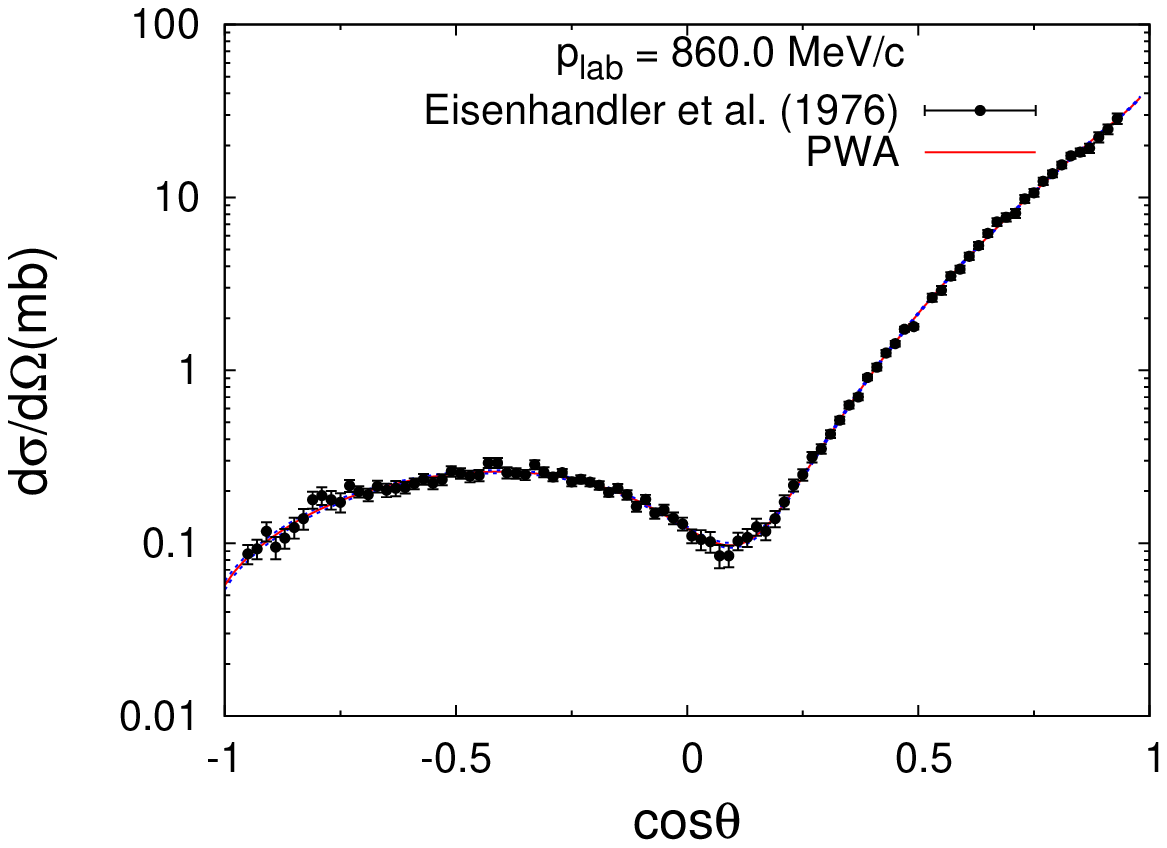}\hspace{1em}
   \includegraphics[width=0.48\textwidth]{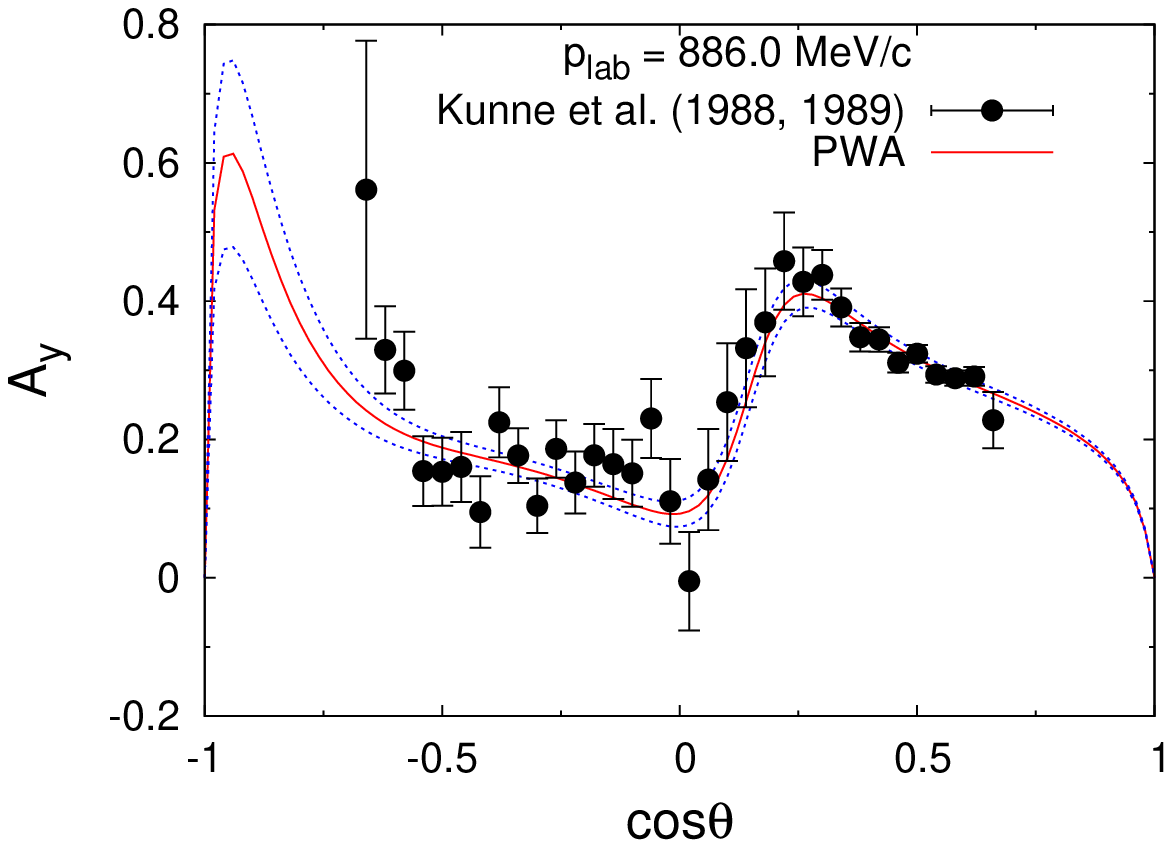}
\caption{\label{860_el} (Color online)
Differential cross sections and analyzing powers for elastic
scattering as function of angle in the center-of-mass system.
The PWA result is given by the drawn red line and the dotted blue lines
indicate the one-sigma uncertainty region. The fit has
for Eisenhandler \textit{et al}.~\cite{Eis76} 
$\chi^{2}_{\textrm{min}}=61.0$ for 94 points $d\sigma/d\Omega$;
for Kunne \textit{et al}.~\cite{Kun88,Kun89} 
$\chi^{2}_{\textrm{min}}=34.1$ for 34 points $A_{y}$.}
\end{figure}

\begin{figure}
   \centering
   \includegraphics[width=0.50\textwidth]{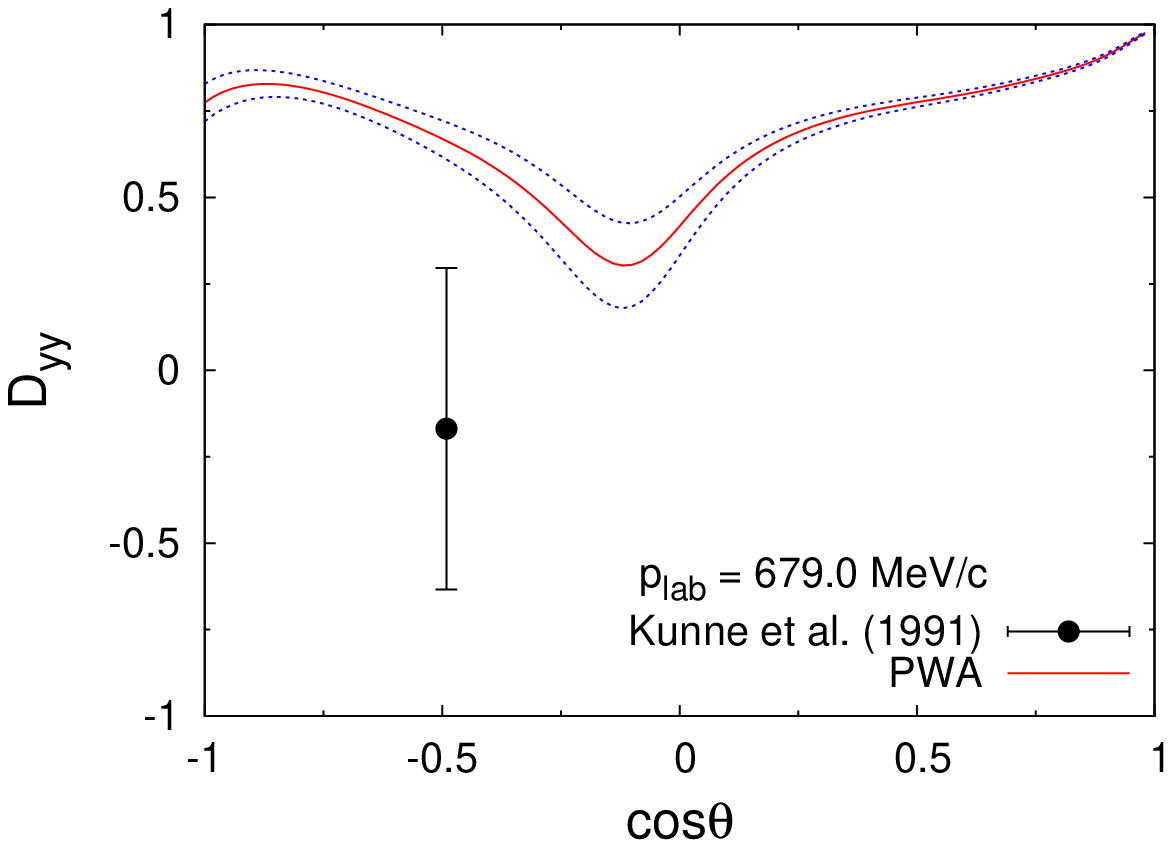}\\ \vspace{0.4em}
   \includegraphics[width=0.50\textwidth]{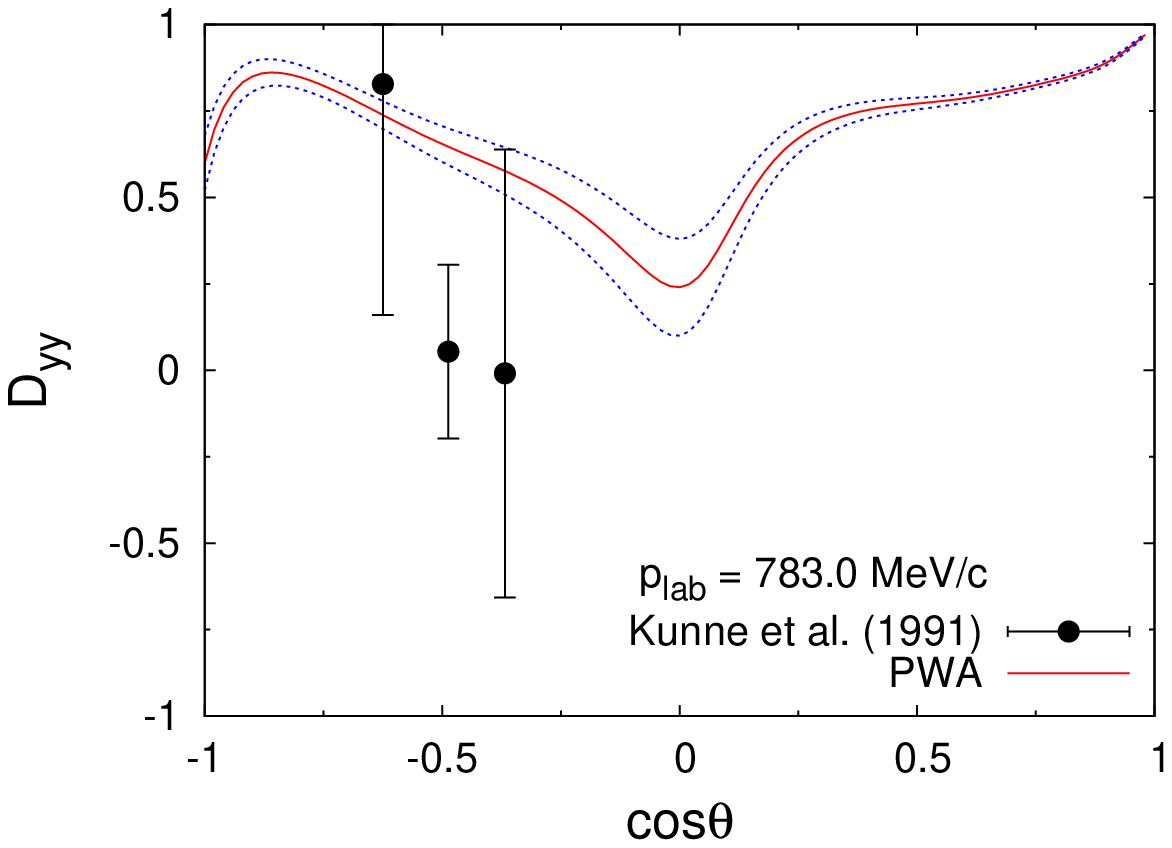}\\ \vspace{0.4em}
   \includegraphics[width=0.50\textwidth]{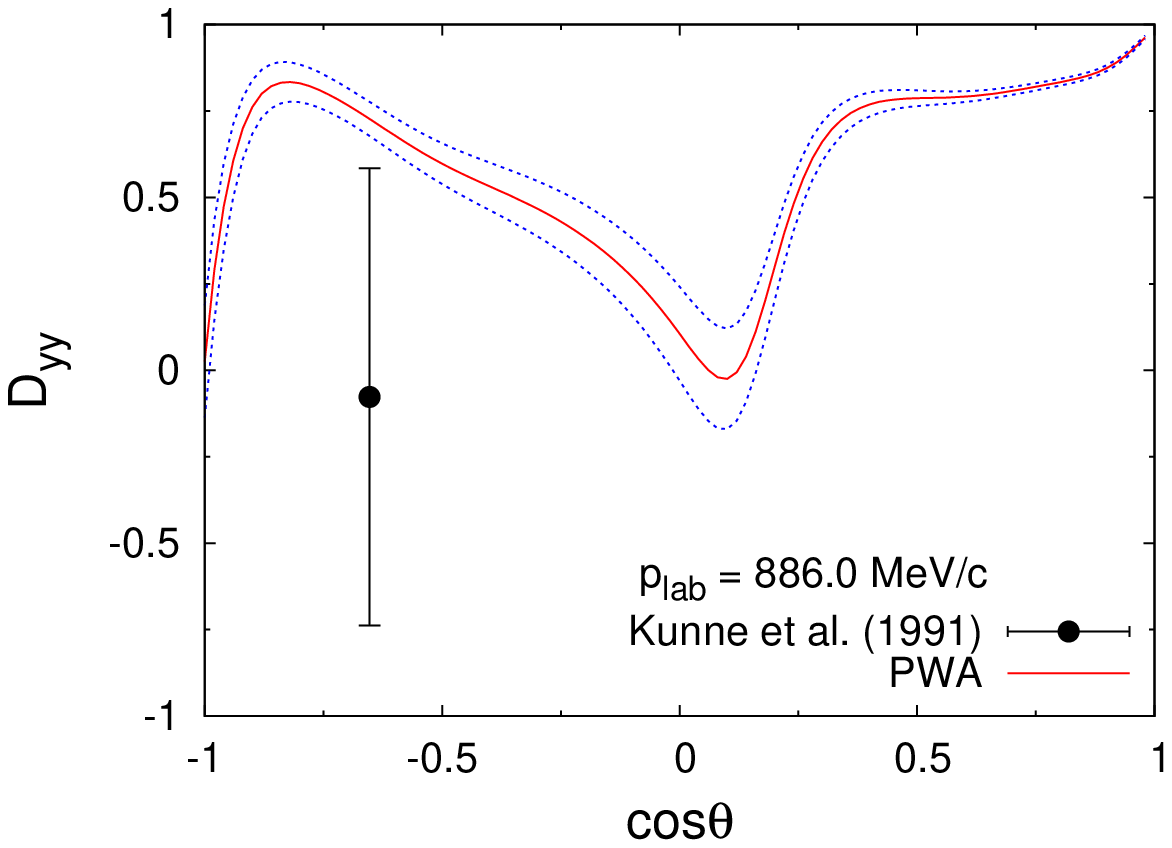}
\caption{\label{Dyy_el} (Color online) Differential
depolarizations $D_{yy}$ for elastic
scattering as function of angle in the center-of-mass system.
The PWA result is given by the drawn red line and the dotted blue lines
indicate the one-sigma uncertainty region. 
The fit has for Kunne {\it et al.}~\cite{Kun91}
at $p_{\textrm{lab}}=679.0$ MeV/$c$
$\chi^{2}_{\textrm{min}}=3.2$ for 1 point,
at $p_{\textrm{lab}}=783.0$ MeV/$c$  
$\chi^{2}_{\textrm{min}}=6.4$ for 3 points,
at $p_{\textrm{lab}}=886.0$ MeV/$c$ 
$\chi^{2}_{\textrm{min}}=1.5$ for 1 point.}
\end{figure}

\begin{figure}
   \centering
   \includegraphics[width=0.48\textwidth]{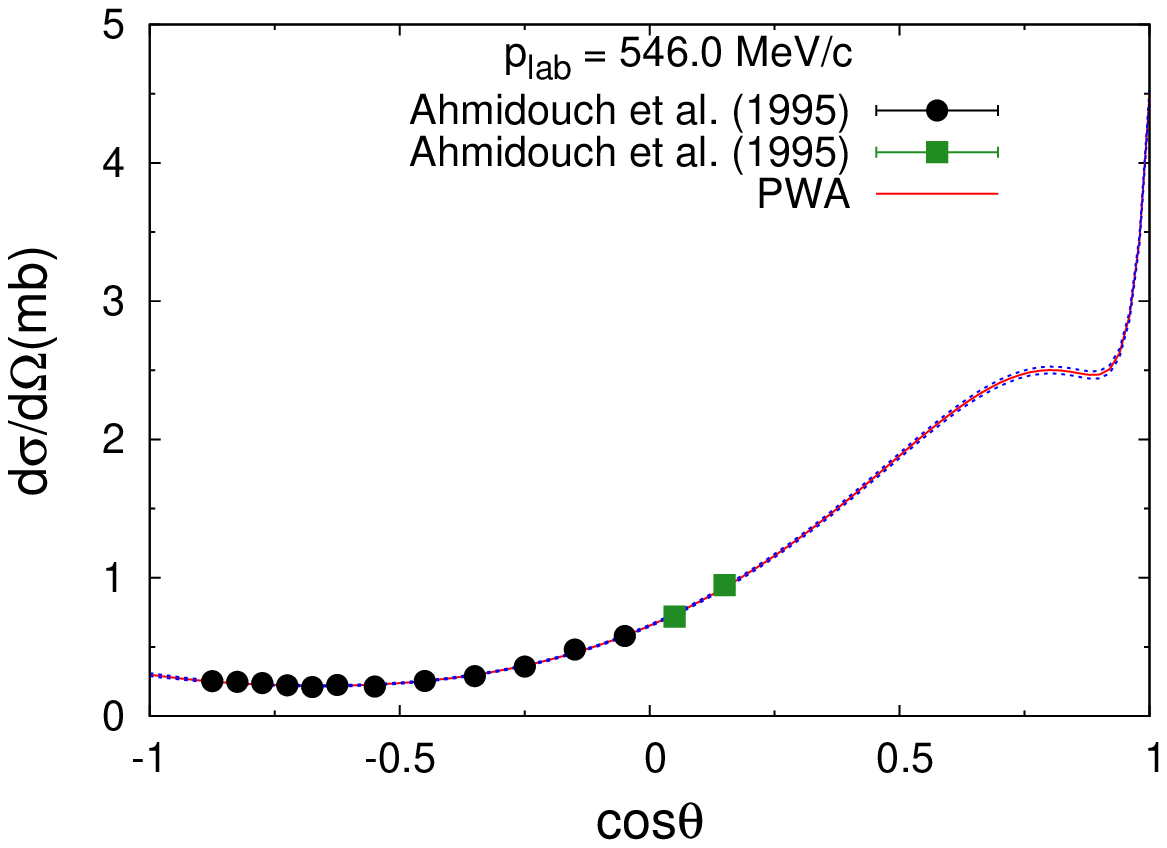}\hspace{1em}
   \includegraphics[width=0.48\textwidth]{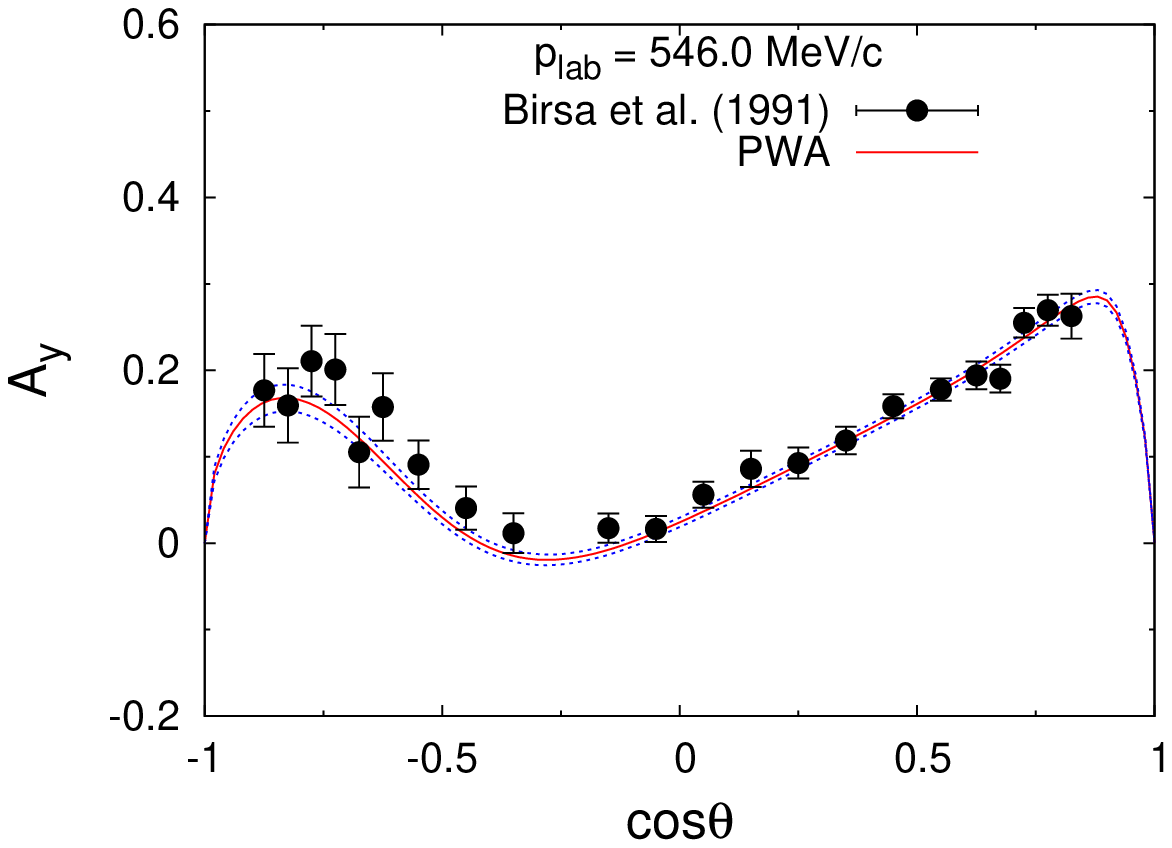}
\caption{\label{546_ce} (Color online)
Differential cross sections and analyzing powers for charge-exchange
scattering as function of angle in the center-of-mass system.
The PWA result is given by the drawn red line and the dotted blue lines
indicate the one-sigma uncertainty region. The fit has
for Ahmidouch \textit{et al}.~\cite{Ahm95}
$\chi^{2}_{\textrm{min}}=12.7$ for 12 points $d\sigma/d\Omega$ at backward angles,
$\chi^{2}_{\textrm{min}}=1.0$ for 2 points $d\sigma/d\Omega$ at forward angles;
for Birsa \textit{et al}.~\cite{Bir91}
$\chi^{2}_{\textrm{min}}=23.3$ for 22 points $A_{y}$.}
\end{figure}

\begin{figure}
   \centering
   \includegraphics[width=0.48\textwidth]{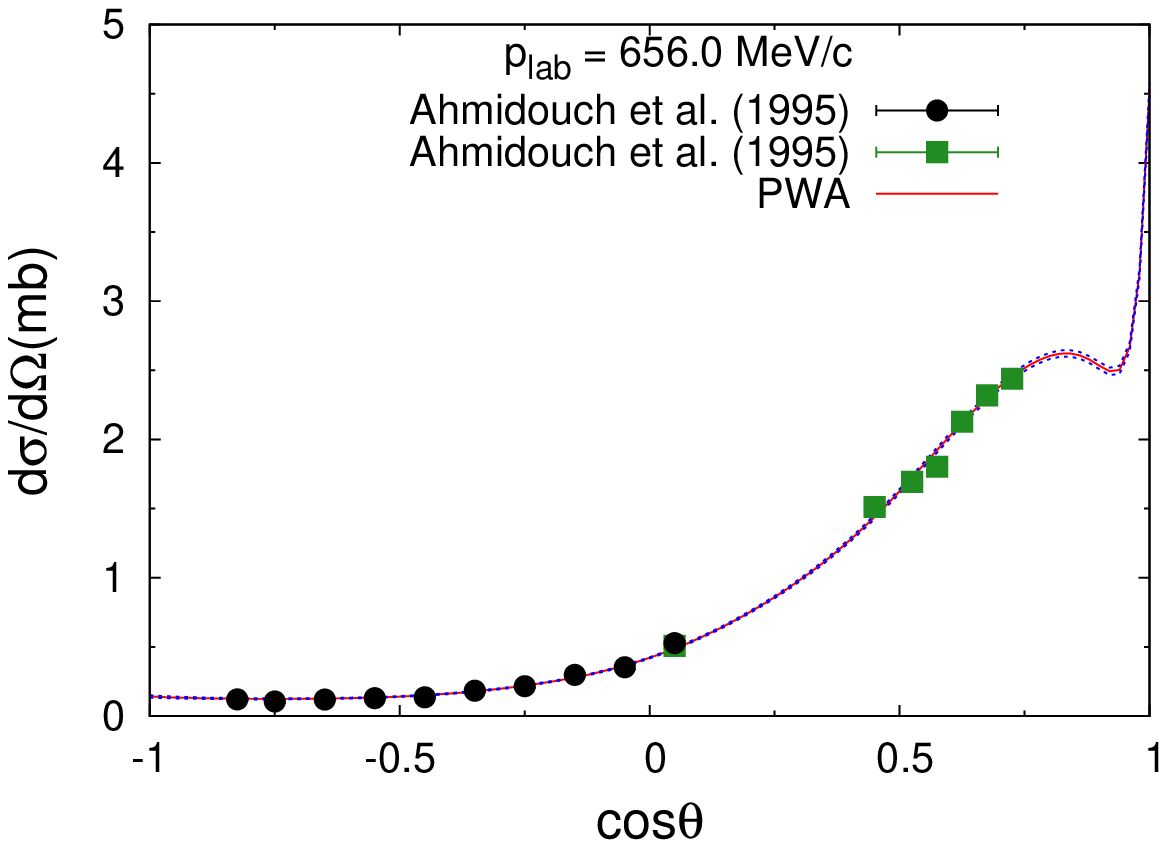}\hspace{1em}
   \includegraphics[width=0.48\textwidth]{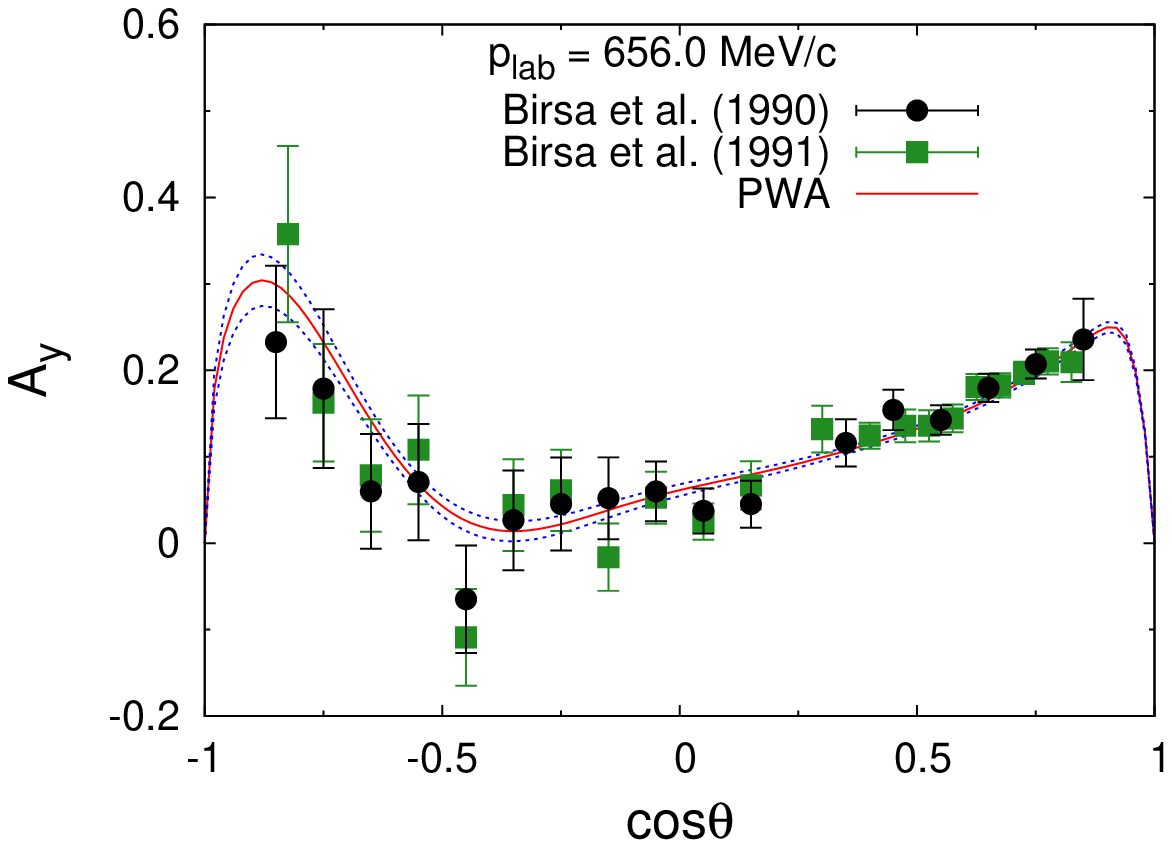}
\caption{\label{656_ce} (Color online)
Differential cross sections and analyzing powers for charge-exchange
scattering as function of angle in the center-of-mass system.
The PWA result is given by the drawn red line and the dotted blue lines
indicate the one-sigma uncertainty region. The fit has
for Ahmidouch \textit{et al}.~\cite{Ahm95} 
$\chi^{2}_{\textrm{min}}=12.9$ for 10 points $d\sigma/d\Omega$ at backward angles,
$\chi^{2}_{\textrm{min}}=14.6$ for 7 points $d\sigma/d\Omega$ at forward angles;
for Birsa \textit{et al}.~\cite{Bir90}
$\chi^{2}_{\textrm{min}}=11.2$ for 17 points $A_{y}$;
for Birsa \textit{et al}.~\cite{Bir91}
$\chi^{2}_{\textrm{min}}=23.5$ for 21 points $A_{y}$.}
\end{figure}

\begin{figure}
   \centering
   \includegraphics[width=0.48\textwidth]{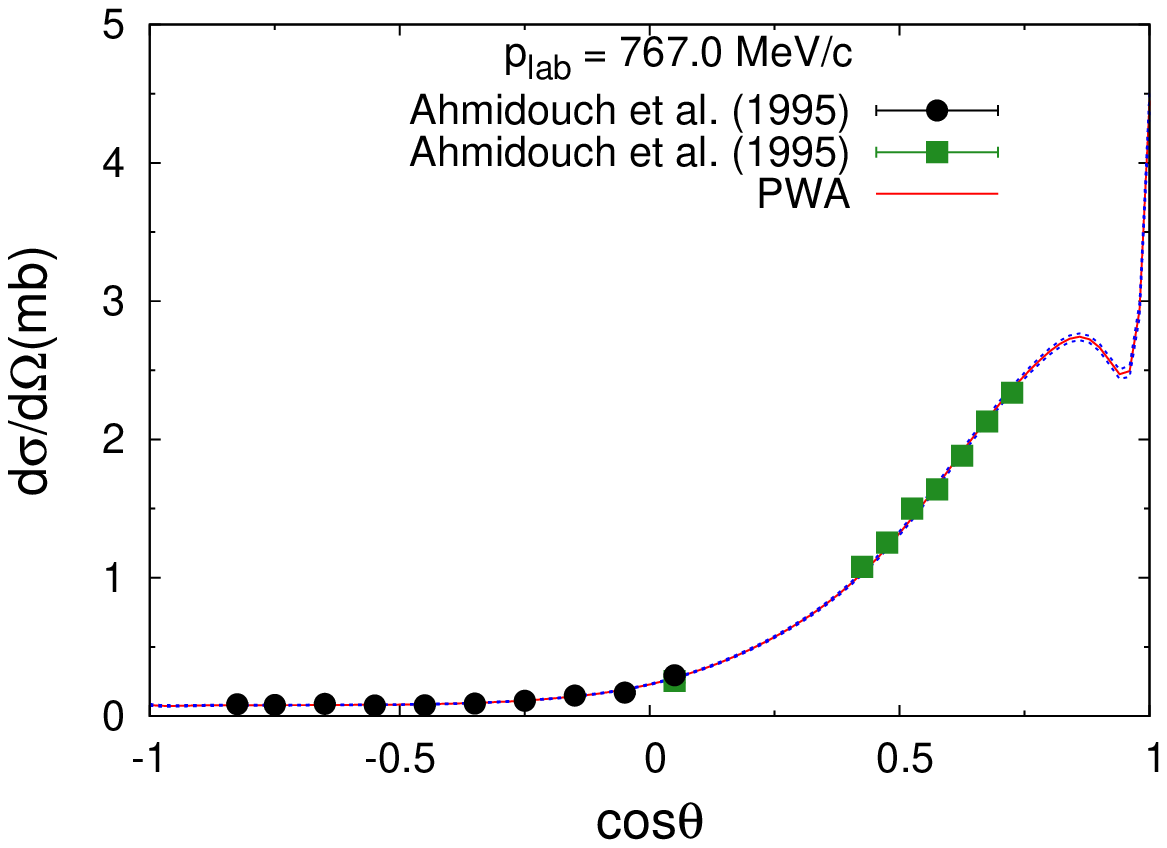}\hspace{1em}
   \includegraphics[width=0.48\textwidth]{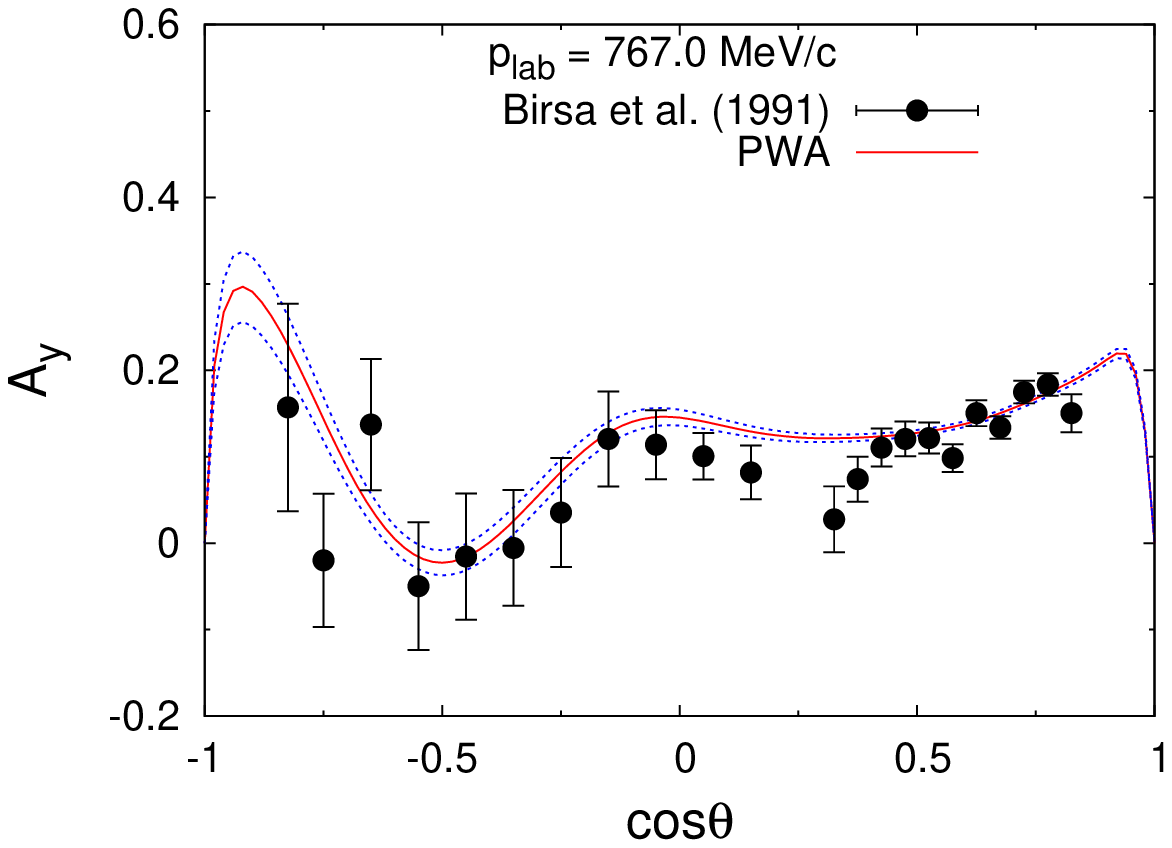}
\caption{\label{767_ce} (Color online)
Differential cross sections and analyzing powers for charge-exchange
scattering as function of angle in the center-of-mass system.
The PWA result is given by the drawn red line and the dotted blue lines
indicate the one-sigma uncertainty region. The fit has
for Ahmidouch \textit{et al}.~\cite{Ahm95}
$\chi^{2}_{\textrm{min}}=9.1$ for 10 points $d\sigma/d\Omega$ at backward angles,
$\chi^{2}_{\textrm{min}}=9.6$ for 8 points $d\sigma/d\Omega$ at forward angles;
for Birsa \textit{et al}.~\cite{Bir91}
$\chi^{2}_{\textrm{min}}=28.0$ for 22 points $A_{y}$.}
\end{figure}

\begin{figure}
   \centering
   \includegraphics[width=0.48\textwidth]{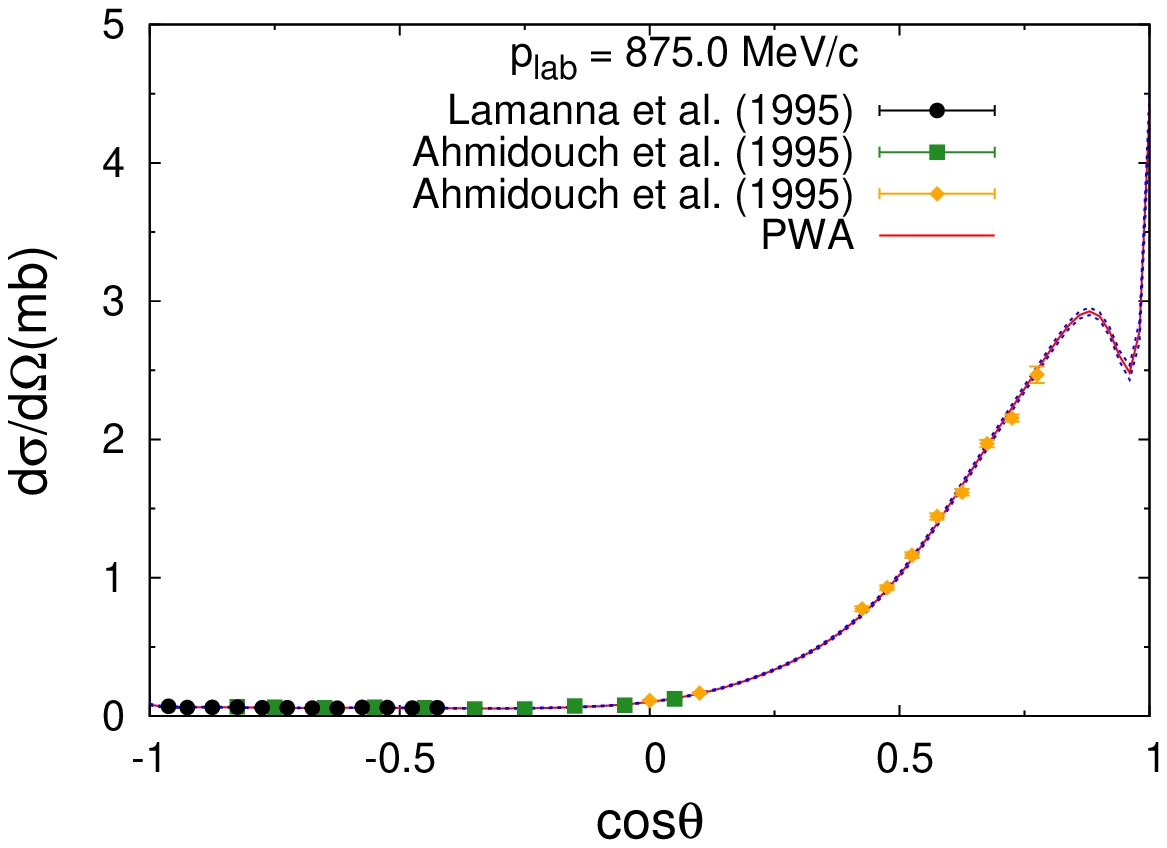}\hspace{1em}
   \includegraphics[width=0.48\textwidth]{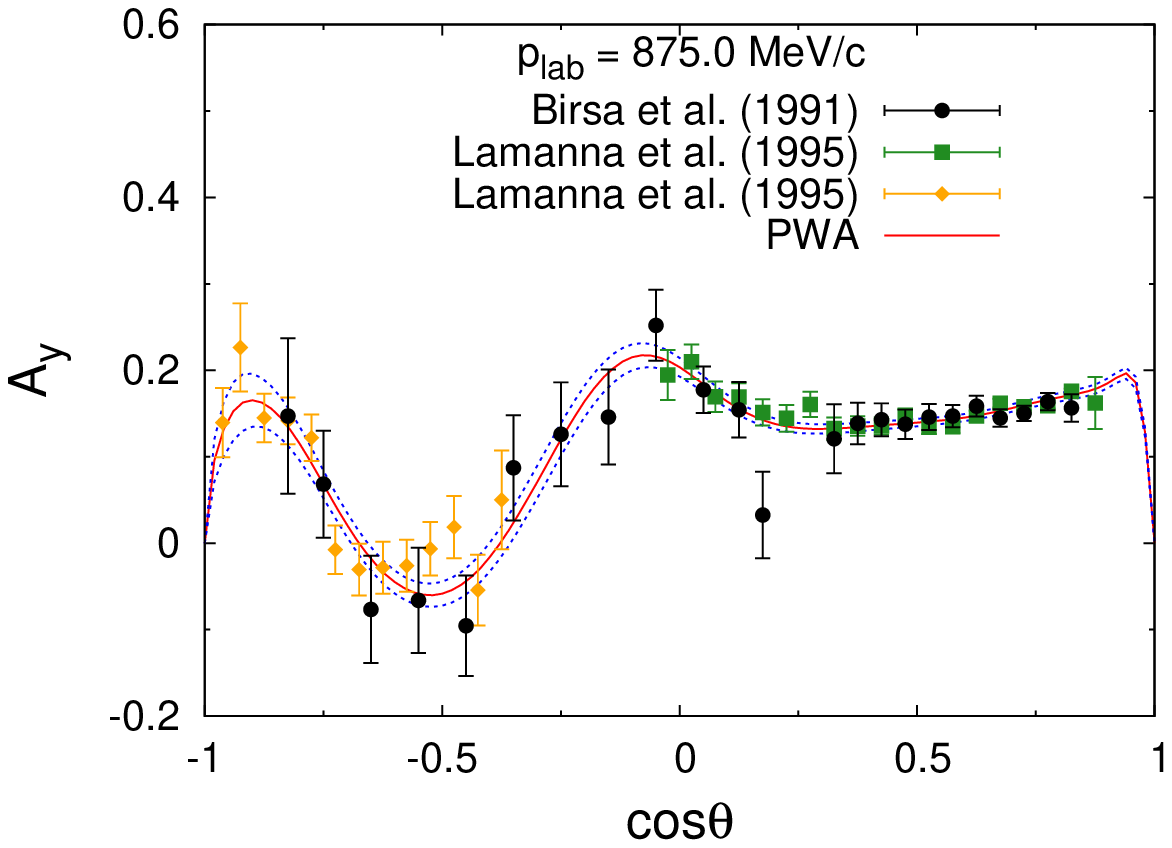}
\caption{\label{875_ce} (Color online)
Differential cross sections and analyzing powers for charge-exchange
scattering as function of angle in the center-of-mass system.
The PWA result is given by the drawn red line and the dotted blue lines
indicate the one-sigma uncertainty region. The fit has
for Lamanna \textit{et al}.~\cite{Lam95}
$\chi^{2}_{\textrm{min}}=8.4$ for 12 points $d\sigma/d\Omega$;
for Ahmidouch \textit{et al}.~\cite{Ahm95}
$\chi^{2}_{\textrm{min}}=8.1$ for 10 points $d\sigma/d\Omega$ at backward angles,
$\chi^{2}_{\textrm{min}}=20.9$ for 10 points $d\sigma/d\Omega$ at forward angles;
for Birsa \textit{et al}.~\cite{Bir91}
$\chi^{2}_{\textrm{min}}=12.1$ for 23 points $A_{y}$;
for Lamanna \textit{et al}.~\cite{Lam95}
$\chi^{2}_{\textrm{min}}=19.2$ for 19 points $A_{y}$ at forward angles,
$\chi^{2}_{\textrm{min}}=14.0$ for 13 points $A_{y}$ at backward angles.}
\end{figure}

\begin{figure}
   \centering
   \includegraphics[width=0.48\textwidth]{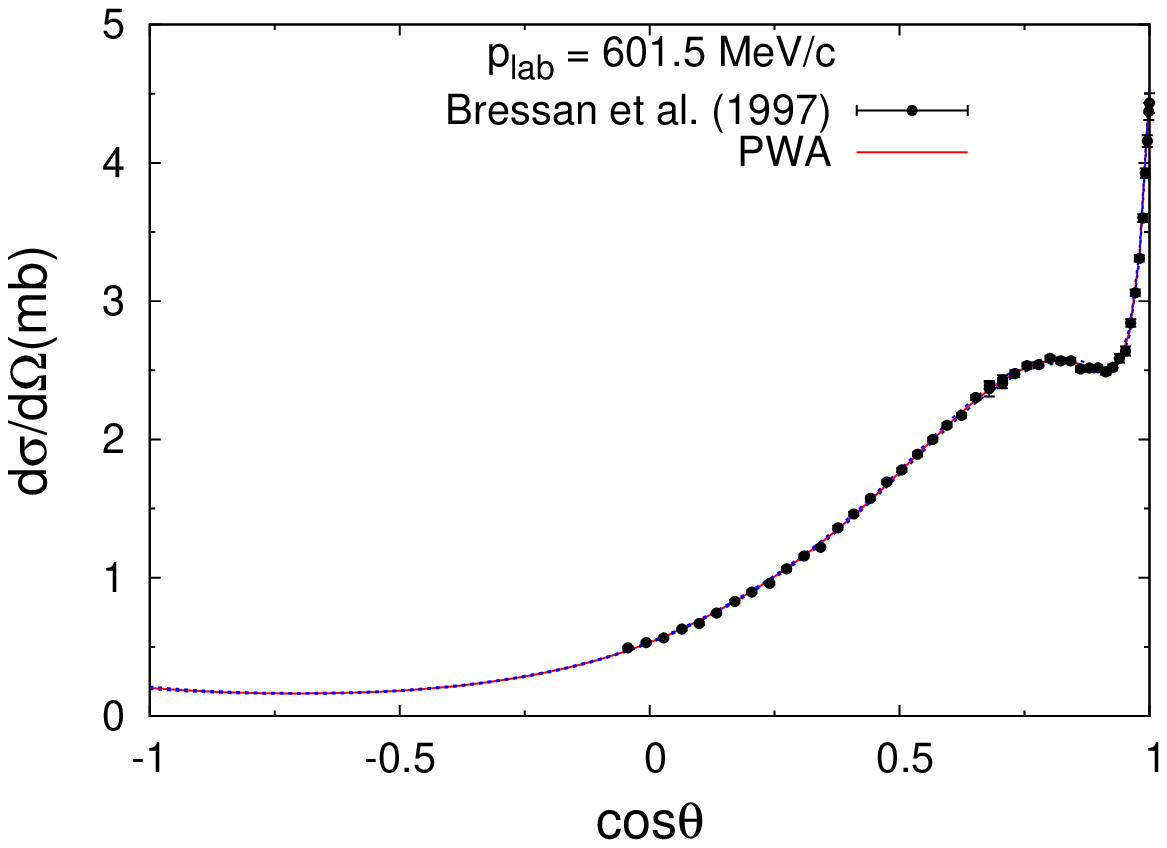}\hspace{1em}
   \includegraphics[width=0.48\textwidth]{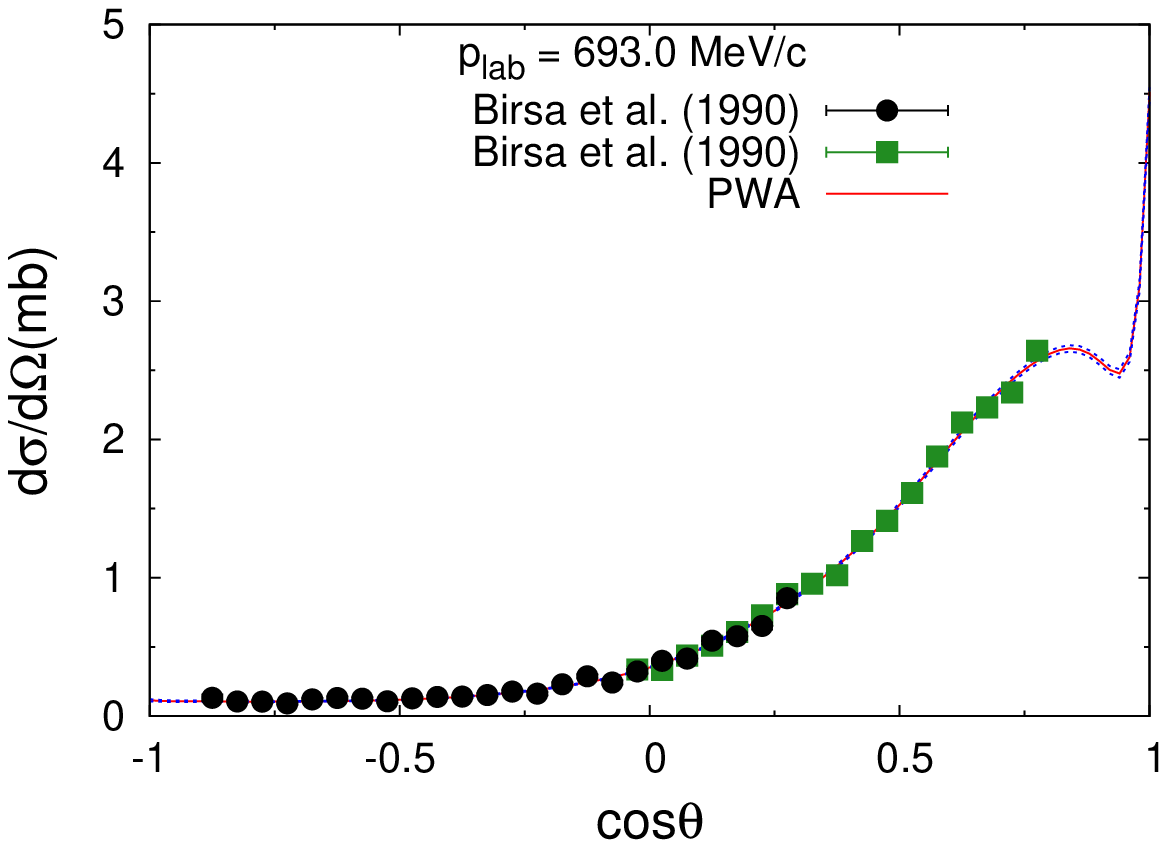}   
\caption{\label{601.0_ce} (Color online)
Differential cross sections $d\sigma/d\Omega$ for charge-exchange
scattering as function of angle in the center-of-mass system.
The PWA result is given by the drawn red line and the dotted blue lines
indicate the one-sigma uncertainty region. The fit has
for Bressan \textit{et al}.~\cite{Bre97}
$\chi^{2}_{\textrm{min}}=37.8$ for 47 points;
for Birsa \textit{et al}.~\cite{Bir90} 
$\chi^{2}_{\textrm{min}}=37.8$ for 24 points
at backward angles,
$\chi^{2}_{\textrm{min}}=20.4$ for 17 points 
at forward angles.}
\end{figure}

In Figs.~\ref{679_el}, \ref{790_el}, and \ref{860_el} the differential cross sections
$d\sigma/d\Omega$ and the analyzing powers $A_{y}$ are shown for elastic scattering
$\overline{p}p\rightarrow\overline{p}p$ at momenta near 690,
790, and 860 MeV/$c$, respectively. In general, the uncertainty on the PWA prediction
for the differential cross sections is determined by the accuracy of the data. For the analyzing
powers, on the other hand, the theoretical uncertainties are in general smaller than the
errors of the data points. The theoretical uncertainty is very small for forward angles.
For backward angles, where there are no data available, this
uncertainty increases. Fig.~\ref{Dyy_el} shows the very limited data available for
the depolarization $D_{yy}$ for elastic scattering at 679, 783, and 886 MeV/$c$.
There are only a few data points in the backward
hemisphere and the data points have large error bars. In this case, the theoretical
uncertainty for the PWA prediction is much smaller than these error bars, which
implies that there is little new information in these data and that the fit would not
change significantly if they were left out of the fit. The theoretical uncertainty is
again very small for forward angles.

Figs.~\ref{546_ce}, \ref{656_ce}, \ref{767_ce}, and \ref{875_ce} show the differential cross
sections $d\sigma/d\Omega$ and the analyzing powers $A_{y}$ for charge-exchange scattering
$\overline{p}p\rightarrow\overline{n}n$ at 546, 656, 767, and 875 MeV/$c$, respectively.
Like for the elastic case, one observes that, in general, the uncertainty on the PWA prediction
for the differential cross sections is determined by the accuracy of the data. For the analyzing
powers, on the other hand, the theoretical uncertainties are in general smaller than the
errors of the data points. For some of the differential cross-section measurements, we
introduced different normalization parameters for the data in the forward and in the
backward hemisphere, which were taken with different detectors. The charge-exchange
differential cross section is highly anisotropic, because of the contributions of many,
high-$\ell$ partial waves. It has a ``spike'' at the most forward angles and it is flat at
backward angles. It exhibits a very typical dip-bump structure at forward angles, which is
due to the interference of the OPE interaction with a background due to short-range
interactions~\cite{Lea76}. The precise form of this structure evolves rapidly as function
of energy, from a rather flat plateau structure at 546 MeV/$c$ to a pronounced
dip-bump structure at 875 MeV/$c$. The structure was measured accurately at 601 MeV/$c$
by the PS206 experiment at the end of the LEAR era~\cite{Bir94,Bre97}. The high-quality
charge-exchange differential cross sections from Ref.~\cite{Bre97} are shown in
Fig.~\ref{601.0_ce}. At the time of Ref.~\cite{Tim94}, only the data at 693 MeV/$c$
shown in Fig.~\ref{601.0_ce} were available~\cite{Bir90}, but these differential cross
sections did not pin down the dip-bump structure. The PWA of Ref.~\cite{Tim94}
predicted a more pronounced structure for this data set.

\begin{figure}
   \centering
   \includegraphics[width=0.50\textwidth]{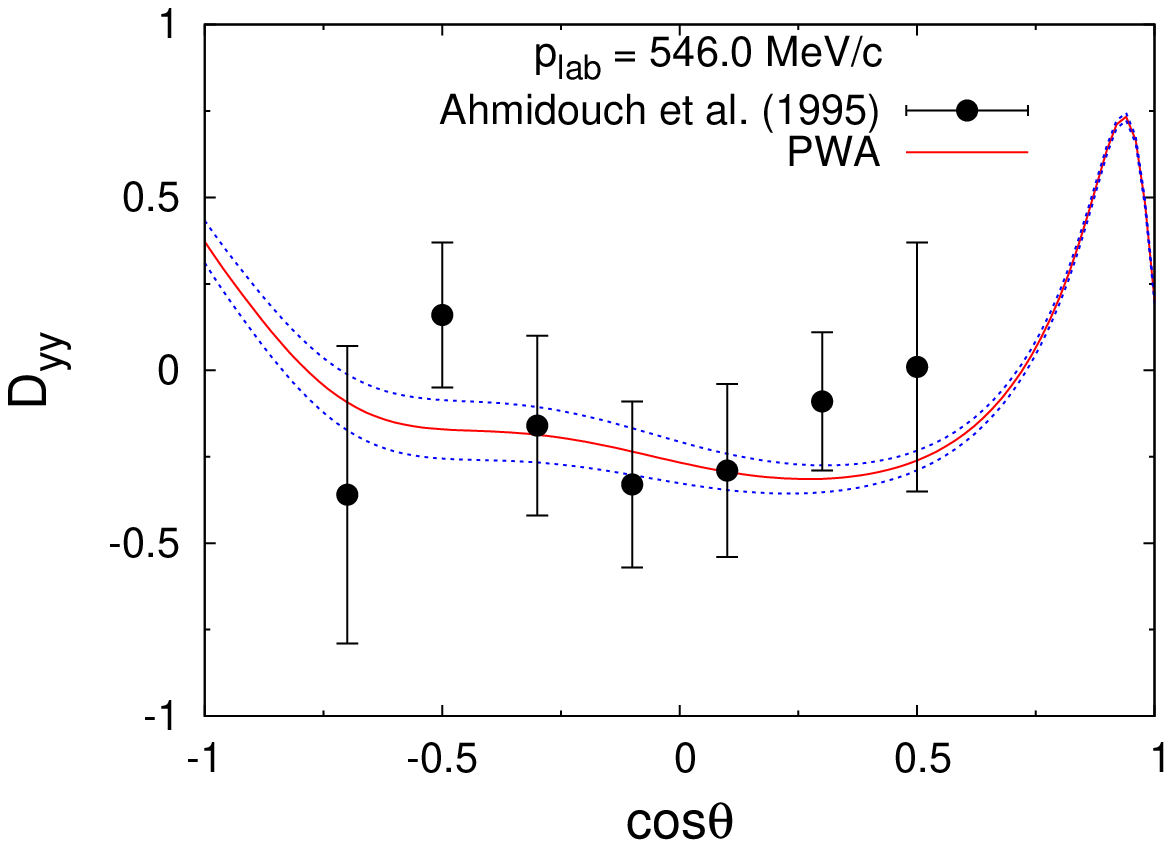}\\ \vspace{0.4em}
   \includegraphics[width=0.50\textwidth]{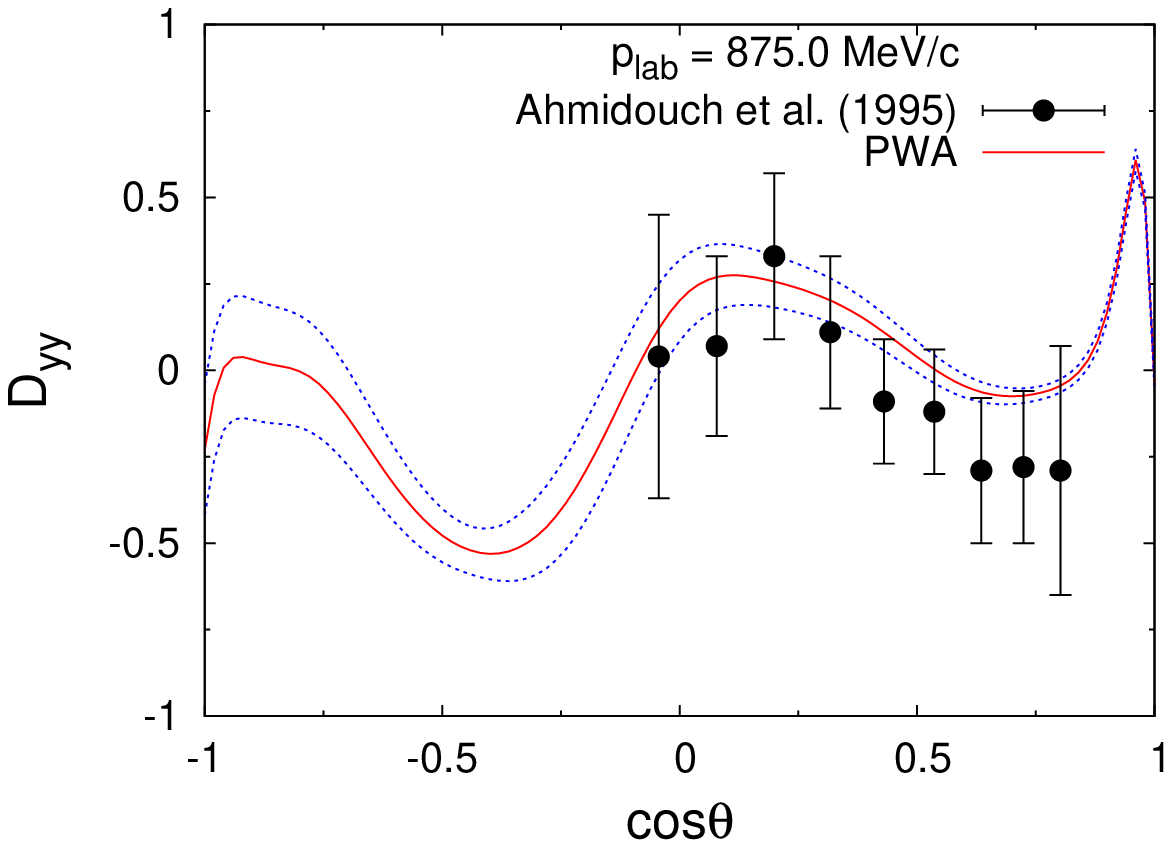}\\ \vspace{0.4em}
   \includegraphics[width=0.50\textwidth]{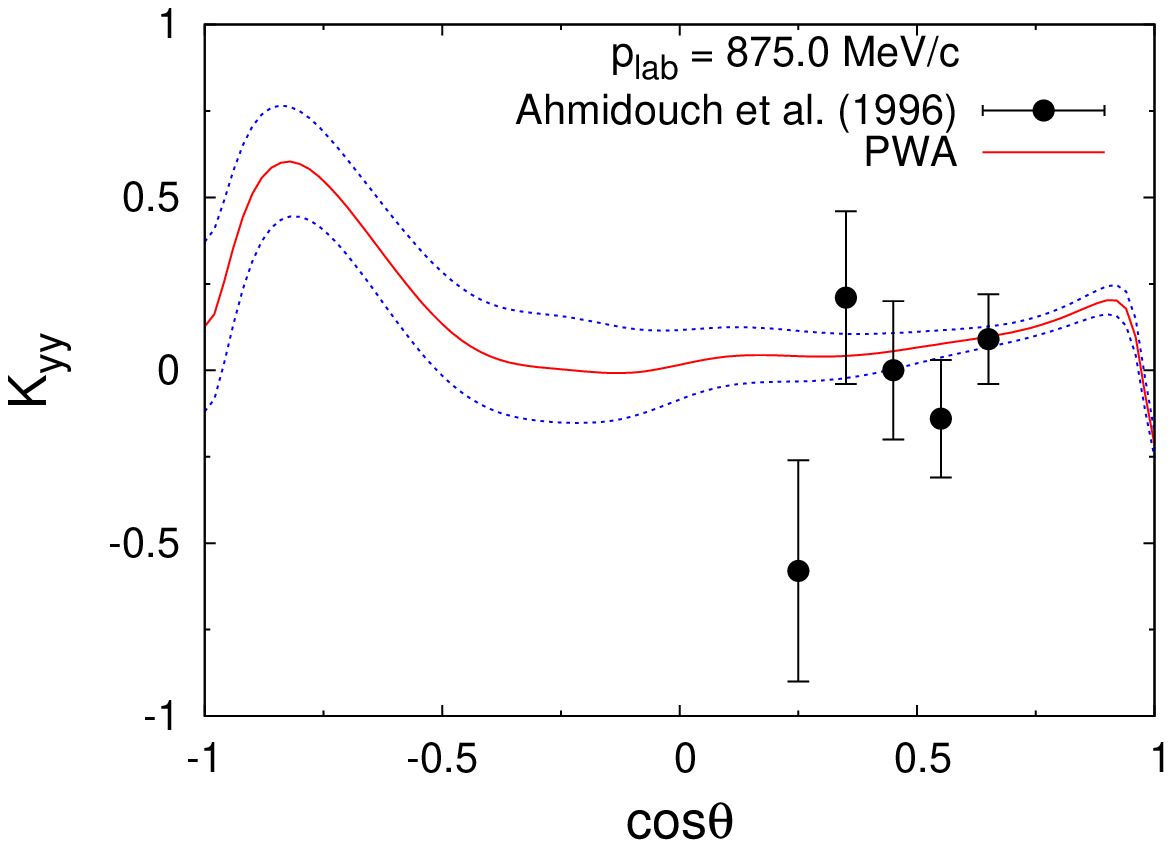}
\caption{\label{Dyy_ce} (Color online) Differential
depolarizations and spin transfers for charge-exchange
scattering as function of angle in the center-of-mass system.
The PWA result is given by the drawn red line and the dotted blue lines
indicate the one-sigma uncertainty region. The fit has for
Ahmidouch {\it et al.}~\cite{Ahm95d} at $p_{\textrm{lab}}=546.0$ MeV/$c$
$\chi^{2}_{\textrm{min}}=4.9$ for 7 points $D_{yy}$;
for Ahmidouch {\it et al.}~\cite{Ahm95d} (Birsa {\it et al.}~\cite{Bir93})
at $p_{\textrm{lab}}=875.0$ MeV/$c$ $\chi^{2}_{\textrm{min}}=5.1$ for 9 points $D_{yy}$;
for Ahmidouch {\it et al.}~\cite{Ahm96} at $p_{\textrm{lab}}=875.0$ MeV/$c$
$\chi^{2}_{\textrm{min}}=5.9$ for 5 points $K_{yy}$.}
\end{figure}

In Fig.~\ref{Dyy_ce} the few data sets available for the depolarization $D_{yy}$
at 546 and 875 MeV/$c$ and the spin transfer $K_{yy}$ at 875 MeV/$c$ in
charge-exchange scattering are
shown. The data points have large error bars, and also in this case the theoretical
uncertainty for the PWA prediction is much smaller than these error bars. This
demonstrates that spin observables are, of course, important, but they improve
a good energy-dependent PWA only if they are precise enough~\cite{Tim95}.
The theoretical uncertainty is again very small for forward angles. 

\newpage
\section{Phase-shift and inelasticity parameters} \label{sec:Phases}
In this section we present results for the $\overline{N}\!N$ $S$ matrix. The
$S$ matrix for the coupled $\overline{p}p$ and $\overline{n}n$ channels from 
our PWA suffices to construct the complete scattering amplitudes and hence
the observables. For the uncoupled partial waves with $\ell=J$, $s=0,1$ or
$\ell=1$, $J=0$, the $S$ matrix is $2\times2$, while for the coupled partial
waves with $\ell=J \pm 1$ ($J\ge1$), $s=1$, it is a $4\times4$ matrix. We
give numerical values at a number of momenta. Other results are available
upon request. The matrix elements of $S^C_{C+N}$ for different partial waves for the
elastic and charge-exchange reactions are given in Tables \ref{Tab:smtrx_uncoup},
\ref{Tab:smtrx_coup_el}, and \ref{Tab:smtrx_coup_ce} for $p_{\textrm {lab}} = 100$
to $1000$ MeV/$c$. The $S$ matrices are symmetric for the coupled partial waves
in the case of elastic $\overline{p}p$ and $\overline{n}n$ scattering, but they are not
symmetric in the case of charge-exchange scattering $\overline{p}p\leftrightarrow\overline{n}n$,
as one can see from Table \ref{Tab:smtrx_coup_el} for $\overline{p}p\rightarrow\overline{p}p$
and from Table \ref{Tab:smtrx_coup_ce} for $\overline{p}p\rightarrow\overline{n}n$.

For illustrative purposes we also present phase-shift and inelasticity parameters
assuming that isospin symmetry is exact (we take then the average nucleon and
pion mass and set the electromagnetic interaction to zero).
In that case, the parametrization of the $S$ matrix can be done in a transparent way,
similar to the procedures used for $N\!N$ scattering (above the pion-production
threshold).
 
For the uncoupled partial waves with $\ell=J$, $s=0,1$ or $\ell=1$, $J=0$,
the $S$ matrix, for isospin $I=0$ or $I=1$, is a $1 \times 1$ matrix that can
be written as
\begin{equation}
   S^J  = \eta \exp(2i\delta)~,
\label{Eq:uncoup}   
\end{equation}
where $\delta$ is the phase shift and $\eta$ ($0\le\eta\le1$) is the inelasticity due
to the annihilation into mesonic channels.  The $S$ matrix for the uncoupled waves
is thus given in terms of two parameters, which are functions of energy.
For high values of $\ell$, where there is almost no annihilation, $\eta\rightarrow1$.

For the partial waves with $\ell=J \pm 1$ ($J\ge1$), $s=1$, coupled by a tensor
force,  the $S$-matrix, for isospin $I=0$ or $I=1$, is a $2 \times 2$ matrix that can
be parametrized by the generalized ``bar-phase'' convention~\cite{Bry81} 
\begin{equation}
   S^J = \exp(i\bar{\delta}) \,
               \exp(i\bar{\varepsilon}_{J}\sigma_x) \,\,
       H^J  \, \exp(i\bar{\varepsilon}_{J}\sigma_x) \,
               \exp(i\bar{\delta})~,
\end{equation}
where $\bar{\delta}$ is a $2\times 2$ diagonal matrix with real entries
$\bar{\delta}_{J-1,J}$ and $\bar{\delta}_{J+1,J}$, and $\bar{\varepsilon}_{J}$
is the mixing angle for the coupled partial waves; $\sigma_{x}$ is the first Pauli matrix.
The matrix $H^J$ is used to parametrize the inelasticity. Different ways to write $H^J$
can be found in the literature. We will follow the parametrization of Ref.~\cite{Kla83},
in which one writes
\begin{equation}
   H^J = \exp(-i\omega_J\sigma_y) \,
               \left( \begin{array}{cc}
                      \eta_{J-1,J} &       0        \\
                          0        &   \eta_{J+1,J}
                      \end{array} \right) \,
               \exp(i\omega_J\sigma_y)~,
\label{Eq:Klarsfeld}               
\end{equation}
where $\eta_{J-1,J}$ and $\eta_{J+1,J}$ are real numbers with $0 \leq\eta_{J\mp1,J}\leq 1$,
and $\omega_J$ is the mixing angle for the inelasticity; $\sigma_y$ is the second Pauli matrix.
The $S$ matrix for these coupled waves is thus given in terms of six parameters. 

From the numerical results of the PWA, one has to extract for each energy the phase-shift
and inelasticity parameters from the numerical values of the $S$ matrix. For the uncoupled
partial waves this is easy. In order to obtain the phase-shift and inelasticity parameters for
the coupled partial waves,  the algorithm of Ref.~\cite{Bry81} is used.
One can write the $S$ matrix as
\begin{equation}
        S^J = 
   \left( \begin{array}{cc}
     R_{11}\exp(2i\delta_{11}) & iR_{12}\exp(2i\delta_{12}) \\
    iR_{12}\exp(2i\delta_{12}) &  R_{22}\exp(2i\delta_{22})
   \end{array} \right)~,
\end{equation}
where $R_{ij}$ and $\delta_{ij}$ are real numbers.
When one defines the auxiliary phases
\begin{eqnarray}
   \theta_{a} & \equiv & \delta_{11}-\bar{\delta}_{J-1,J}~, \nonumber \\
   \theta_{b} & \equiv & \delta_{22}-\bar{\delta}_{J+1,J}~, \\
   \delta'     & \equiv & \delta_{11}+\delta_{22}-2\delta_{12}~, \nonumber
\end{eqnarray}
it follows that
\begin{eqnarray}
   \tan2(\theta_a+\theta_b) & = &
   \frac{R^2_{12}\sin2\delta'}{R_{11}R_{22}+R^2_{12}\cos2\delta'}~, \nonumber \\
   \tan(\theta_a-\theta_b) & = &
   \frac{R_{22}-R_{11}}{R_{11}+R_{22}}\tan(\theta_a+\theta_b)~.
\end{eqnarray}
From this the phase-shift parameters $\bar{\delta}_{J-1,J}$ and $\bar{\delta}_{J+1,J}$
can be obtained. The mixing angle $\bar{\varepsilon}_{J}$ is given by
\begin{equation}
   \tan2\bar{\varepsilon}_{J} = 
   \frac{2R_{12}\cos(\theta_a+\theta_b-\delta')}
        {R_{11}\cos2\theta_a + R_{22}\cos2\theta_b}~.
\end{equation}
The elements of the matrix $H^J$ can then be related to the parameters obtained.
One finds
\begin{eqnarray}
   2H_{11}\cos2\bar{\varepsilon}_{J} & = &
      R_{11}(1+\cos2\bar{\varepsilon}_{J})\cos2\theta_a  +
      R_{22}(1-\cos2\bar{\varepsilon}_{J})\cos2\theta_b~, \nonumber \\
   2H_{22}\cos2\bar{\varepsilon}_{J} & = &
      R_{11}(1-\cos2\bar{\varepsilon}_{J})\cos2\theta_a +
      R_{22}(1+\cos2\bar{\varepsilon}_{J})\cos2\theta_b~, \\
    H_{12}\cos2\bar{\varepsilon}_{J} & = &
      R_{12}\sin(\delta'-\theta_a-\theta_b)~, \nonumber
\end{eqnarray}
from which one can determine the values of $H_{11}$, $H_{22}$, and $H_{12}$. 
By using Eq.~(\ref{Eq:Klarsfeld}), the remaining parameters
$\eta_{J-1,J}$, $\eta_{J+1,J}$, and $\omega_J$
can be obtained via
\begin{eqnarray}
    \eta_{J-1,J} + \eta_{J+1,J} & = & {\rm Tr}  \, H^J~, \nonumber \\
       \eta_{J-1,J}\,\eta_{J+1,J} & = & {\rm det} \, H^J~, \\
                  \tan2\,\omega_J & = & 2H_{12}/(H_{11}-H_{22})~. \nonumber
\end{eqnarray}

If one extracts the values of the parameters for one single energy, there can be
ambiguities~\cite{Tim95}. In order to ensure continuity as function of energy  one
can always change the values of these parameters in such a way that the corresponding
$S$-matrix elements are not changed. In the case of uncoupled partial waves, one
can change $\delta$ by 180$^\circ$ and keep $\eta$ unchanged, as can be
seen from Eq.~(\ref{Eq:uncoup}). In the case of the coupled partial waves, for
instance, one can change $\bar{\delta}_{J-1,J}$ or $\bar{\delta}_{J+1,J}$ by 180$^\circ$ 
and at the same time change the signs of $\bar{\varepsilon}_{J}$ and $\omega_{J}$, while
keeping $\eta_{J-1,J}$ and $\eta_{J+1,J}$ unchanged; one can also change both $\bar{\delta}_{J-1,J}$
and $\bar{\delta}_{J+1,J}$ by 180$^\circ$ at the same time and keep $\eta_{J-1,J}$, $\eta_{J+1,J}$,
$\bar{\varepsilon}_{J}$, and $\omega_{J}$ unchanged. 
In the limit where $\eta_{J\mp1,J}=1$, $\bar{\delta}_{J\mp1,J}=0$, and
$\bar{\varepsilon}_{J}=0$, one can choose $\omega_{J}=0$ in order to keep continuity,
although $\omega_{J}$ can take any value in this case, but the corresponding $S$-matrix
elements are unchanged. 

The results of the phase-shift and inelasticity parameters are given in 
Tables \ref{Tab:phas_uncoup}, \ref{Tab:phas_coup12}, and \ref{Tab:phas_coup34}
for $p_{\textrm {lab}} = 100$ to $1000$ MeV/$c$.
A convenient way to plot the $S$ matrix, or equivalently $T=(S-1)/i$, as function of energy
is to use Argand diagrams. In Fig.~\ref{Fig:argand_SPD} Argand diagrams are shown for the
uncoupled partial waves  and in Fig.~\ref{Fig:argand_mixSDPFDG} for the coupled ones
assuming isospin symmetry.

\begin{table}
\caption{$S$-matrix elements of the uncoupled partial waves for
$\overline{p}p\rightarrow\overline{p}p$ and $\overline{p}p\rightarrow\overline{n}n$.}
\tabcolsep=0.5em
\footnotesize
\renewcommand{\arraystretch}{0.69}
\begin{tabular}{cc|rrrrr}
\hline
\hline
\multicolumn{2}{c}{$p_{\textrm{lab}}$(MeV/$c$)} & $100\hspace{2.5em}$ & $200\hspace{2.5em}$ 
& $300\hspace{2.5em}$ & $400\hspace{2.5em}$ & $500\hspace{2.5em}$\\
\hline
\multirow{10}{*}{\rotatebox{90}{\mbox{\normalsize{$\overline{p}p\rightarrow\overline{p}p$}}}} 
&$^{1}S_{0}$ & $0.596-0.193i$ & $0.351-0.296i$ & $0.162-0.300i$ & $0.032-0.242i$ &  $-0.033-0.157i$ \\ 
&$^{3}P_{0}$ & $0.883-0.023i$ & $0.657-0.199i$ & $0.450-0.287i$ & $0.216-0.248i$ &  $-0.007-0.145i$ \\ 
&$^{1}P_{1}$ & $0.987+0.023i$ & $0.916+0.054i$ & $0.806+0.015i$ & $0.688-0.076i$ &  $0.563-0.181i$ \\ 
&$^{3}P_{1}$ & $0.989-0.030i$ & $0.898-0.093i$ & $0.745-0.173i$ & $0.618-0.238i$ &  $0.524-0.253i$ \\ 
&$^{1}D_{2}$ & $1.000+0.002i$ & $0.996+0.020i$ & $0.977+0.050i$ & $0.925+0.074i$ &  $0.835+0.072i$ \\ 
&$^{3}D_{2}$ & $1.000-0.003i$ & $0.998-0.025i$ & $0.981-0.050i$ & $0.938-0.064i$ &  $0.869-0.071i$ \\ 
&$^{1}F_{3}$ & $1.000+0.000i$ & $1.000+0.004i$ & $0.999+0.015i$ & $0.997+0.031i$ &  $0.989+0.051i$ \\
\hline
\hline
\end{tabular}
\label{Tab:smtrx_uncoup}
\end{table}

\begin{table}
\caption{$S$-matrix elements of the coupled partial waves for $\overline{p}p\rightarrow\overline{p}p$.}
\tabcolsep=0.5em
\footnotesize
\renewcommand{\arraystretch}{1.12}
\begin{tabular}{cc|rrrrr}
\hline
\hline
\multicolumn{2}{c}{$p_{\textrm{lab}}$(MeV/$c$)} & $100\hspace{2.5em}$ & $200\hspace{2.5em}$ 
& $300\hspace{2.5em}$ & $400\hspace{2.5em}$ & $500\hspace{2.5em}$\\
\hline
\multirow{3}{*}{\rotatebox{90}{\mbox{$^{3}S_{1}$\,--\,$^{3}D_{1}$}}} 
&$S_{11}$ &  $0.514-0.307i$   & $0.207-0.290i$   & $0.039-0.195i$   &  $-0.039-0.096i$   &  $-0.071-0.015i$ \\
&$S_{22}$ &  $0.998+0.003i$   & $0.974+0.015i$   & $0.926+0.025i$   &  $0.865+0.034i$   &  $0.798+0.044i$ \\ 
&$S_{12}$ &  $0.004-0.015i$   & $-0.010-0.058i$   & $-0.031-0.093i$   &  $-0.056-0.108i$   &  $-0.077-0.103i$ \\
\cline{1-2}
\multirow{3}{*}{\rotatebox{90}{\mbox{$^{3}P_{2}$\,--\,$^{3}F_{2}$}}} 
&$S_{11}$ &  $0.971+0.021i$   & $0.807+0.089i$   & $0.558+0.111i$   &  $0.355+0.101i$   &  $0.220+0.095i$ \\
&$S_{22}$ &  $1.000+0.000i$   & $0.998+0.003i$   & $0.991+0.008i$   &  $0.980+0.012i$   &  $0.964+0.018i$ \\
&$S_{12}$ &  $0.000-0.003i$   & $0.007-0.022i$   & $0.016-0.038i$   &  $0.014-0.047i$   &  $0.004-0.055i$ \\
\cline{1-2}
\multirow{3}{*}{\rotatebox{90}{\mbox{$^{3}D_{3}$\,--\,$^{3}G_{3}$}}} 
&$S_{11}$ &  $1.000+0.000i$   & $0.998+0.010i$   & $0.981+0.055i$   &  $0.919+0.146i$   &  $0.791+0.247i$ \\
\hline
\hline
\end{tabular}
\label{Tab:smtrx_coup_el}
\end{table}

\begin{table}
\caption{$S$-matrix elements of the coupled partial waves for $\overline{p}p\rightarrow\overline{n}n$.}
\tabcolsep=0.5em
\footnotesize
\renewcommand{\arraystretch}{0.85}
\begin{tabular}{cc|rrrrr}
\hline
\hline
\multicolumn{2}{c}{$p_{\textrm{lab}}$(MeV/$c$)} & $100\hspace{2.5em}$ & $200\hspace{2.5em}$ 
& $300\hspace{2.5em}$ & $400\hspace{2.5em}$ & $500\hspace{2.5em}$ \\
\hline
\multirow{4}{*}{\rotatebox{90}{\mbox{$^{3}S_{1}$\,--\,$^{3}D_{1}$}}} 
&$S_{11}$ &  $-0.021-0.071i$  &  $-0.154-0.091i$  &  $-0.191+0.010i$  &  $-0.172+0.102i$  &  $-0.123+0.165i$  \\
&$S_{22}$ &  $0.000+0.000i$  &  $-0.014+0.025i$  &  $-0.049+0.069i$  &  $-0.091+0.114i$  &  $-0.129+0.154i$  \\
&$S_{12}$ &  $0.008-0.021i$  &  $0.003-0.148i$  &  $-0.042-0.228i$  &  $-0.098-0.269i$  &  $-0.146-0.280i$  \\
&$S_{21}$ &  $0.000+0.000i$  &  $-0.012-0.093i$  &  $-0.055-0.179i$  &  $-0.108-0.229i$  &  $-0.154-0.248i$  \\
\cline{1-2}
\multirow{4}{*}{\rotatebox{90}{\mbox{$^{3}P_{2}$\,--\,$^{3}F_{2}$}}} 
&$S_{11}$ &  $-0.001+0.000i$  &  $-0.077+0.013i$  &  $-0.162-0.026i$  &  $-0.181-0.058i$  &  $-0.170-0.053i$  \\
&$S_{22}$ &  $0.000+0.000i$  &  $-0.001+0.003i$  &  $-0.005+0.015i$  &  $-0.011+0.032i$  &  $-0.016+0.050i$  \\ 
&$S_{12}$ &  $0.000+0.000i$  &  $0.008-0.045i$  &  $0.027-0.099i$  &  $0.035-0.135i$  &  $0.026-0.162i$  \\
\hline
\hline
\end{tabular}
\label{Tab:smtrx_coup_ce}
\end{table}

\begin{table}
\caption{Phase-shift and inelasticity parameters of the uncoupled partial waves assuming isospin symmetry.
$\delta$ is given in degrees.}
\tabcolsep=0.4em
\footnotesize
\renewcommand{\arraystretch}{0.689}
\begin{tabular}{cc|rrrrrrrrrr}
\hline
\hline
\multicolumn{2}{c}{$p_{\textrm{lab}}$(MeV/$c$)} & $100$ & $200$ & $300$ & $400$ & $500$ & $600$ & $700$ & $800$ & $900$ & $1000$ \\
\hline
\multirow{2}{*}{$^{11}S_{0}$} & $\eta$ & $0.61$ & $0.48$ & $0.39$ & $0.31$ & $0.24$ & $0.18$ & $0.13$ & $0.11$ & $0.13$ & $0.16$ \\
                            & $\delta$ & $179.28$ & $169.97$ & $157.96$ & $145.40$ & $132.28$ & $117.71$ & $99.83$ & $77.61$ & $56.50$ & $41.44$ \\
\multirow{2}{*}{$^{31}S_{0}$} & $\eta$ & $0.73$ & $0.51$ & $0.34$ & $0.20$ & $0.09$ & $0.02$ & $0.10$ & $0.18$ & $0.25$ & $0.31$ \\
                            & $\delta$ & $-14.44$ & $-26.94$ & $-38.41$ & $-49.03$ & $-57.58$ & $-4.58$ & $4.39$ & $-3.48$ & $-11.88$ & $-20.11$ \\
\multirow{2}{*}{$^{13}P_{0}$} & $\eta$ & $0.82$ & $0.50$ & $0.39$ & $0.36$ & $0.35$ & $0.35$ & $0.34$ & $0.33$ & $0.31$ & $0.29$ \\
                            & $\delta$ & $179.87$ & $168.27$ & $148.53$ & $139.70$ & $113.53$ & $99.67$ & $87.73$ & $77.52$ & $68.97$ & $62.10$ \\
\multirow{2}{*}{$^{33}P_{0}$} & $\eta$ & $0.99$ & $0.95$ & $0.81$ & $0.55$ & $0.25$ & $0.05$ & $0.21$ & $0.33$ & $0.43$ & $0.49$ \\
\hline
\hline
\end{tabular}
\label{Tab:phas_uncoup}
\end{table}

\begin{table}
\caption{Phase-shift and inelasticity parameters of the coupled partial waves with $J=1, 2$ assuming isospin symmetry. 
$\bar{\delta}$, $\bar{\varepsilon}_{J}$ and $\omega_{J}$ are given in degrees.}
\tabcolsep=0.45em
\footnotesize
\renewcommand{\arraystretch}{1.13}
\begin{tabular}{cc|rrrrrrrrrr}
\hline
\hline
\multicolumn{2}{c}{$p_{\textrm{lab}}$(MeV/$c$)} & $100$ & $200$ & $300$ & $400$ & $500$ & $600$ & $700$ & $800$ & $900$ & $1000$ \\
\hline
\multirow{2}{*}{\raisebox{-2.5em}{\mbox{$^{13}S_{1}$}}} & $\eta_{S}$ & 
$0.63$ & $0.37$ & $0.22$ & $0.15$ & $0.14$ & $0.14$ & $0.16$ & $0.17$ & $0.19$ & $0.22$ \\ 
                      & $\bar{\delta}_{S}$ & 
$160.16$ & $141.90$ & $125.45$ & $111.64$ & $99.16$ & $86.70$ & $76.21$ & $69.19$ & $65.54$ & $64.21$ \\ 
\multirow{2}{*}{--}           & $\omega_{1}$ & 
$3.71$ & $9.61$ & $15.24$ & $20.32$ & $24.93$ & $29.08$ & $31.90$ & $33.49$ & $34.89$ & $37.29$ \\ 
                      & $\bar{\varepsilon}_{1}$ & 
$1.62$ & $6.90$ & $12.60$ & $17.51$ & $20.74$ & $21.16$ & $19.42$ & $16.88 $ & $14.46$ & $12.61$ \\ 
\multirow{2}{*}{\raisebox{3em}{\mbox{$^{13}D_{1}$}}} & $\eta_{D}$ & 
$1.00$ & $1.00$ & $0.99$ & $0.98$ & $0.96$ & $0.92$ & $0.88$ & $0.84$ & $0.79$ & $0.74$ \\  
                            & $\bar{\delta}_{D}$ & 
$0.17$ & $2.01$ & $5.83$ & $11.08$ & $16.81$ & $21.13$ & $22.53$ & $21.15$ & $17.91$ & $13.54$ \\ 
\cline{1-2}
\multirow{2}{*}{\raisebox{-2.5em}{\mbox{$^{33}S_{1}$}}} & $\eta_{S}$ & 
$0.63$ & $0.43$ & $0.32$ & $0.25$ & $0.20$ & $0.17$ & $0.14$ & $0.12$ & $0.13$ & $0.18$ \\
\hline
\hline
\end{tabular}
\label{Tab:phas_coup12}
\end{table}

\begin{table}
\caption{Phase-shift and inelasticity parameters of the coupled partial waves with $J=3, 4$ assuming isospin symmetry. 
$\bar{\delta}$, $\bar{\varepsilon}_{J}$ and $\omega_{J}$ are given in degrees.}
\tabcolsep=0.6em
\footnotesize
\renewcommand{\arraystretch}{1.13}
\begin{tabular}{cc|rrrrrrrrrr}
\hline
\hline
\multicolumn{2}{c}{$p_{\textrm{lab}}$(MeV/$c$)} & $100$ & $200$ & $300$ & $400$ & $500$ & $600$ & $700$ & $800$ & $900$ & $1000$ \\
\hline
\multirow{2}{*}{\raisebox{-2.5em}{\mbox{$^{13}D_{3}$}}} & $\eta_{D}$ & 
$1.00$ & $1.00$ & $0.97$ & $0.89$ & $0.74$ & $0.58$ & $0.47$ & $0.42$ & $0.41$ & $0.41$ \\ 
                      & $\bar{\delta}_{D}$ & 
$0.02$ & $0.47$ & $2.52$ & $7.07$ & $13.52$ & $19.85$ & $24.11$ & $24.78$ & $21.37$ & $14.52$ \\ 
\multirow{2}{*}{--}           & $\omega_{3}$ & 
$-0.24$ & $-1.04$ & $-1.71$ & $-1.68$ & $-0.63$ & $1.37$ & $4.05$ & $7.18$ & $10.92$ & $15.92$ \\ 
                      & $\bar{\varepsilon}_{3}$ & 
$-0.01$ & $-0.41$ & $-1.59$ & $-3.29$ & $-5.03$ & $-6.37$ & $-7.10$ & $-7.34$ & $-7.44$ & $-7.86$ \\ 
\multirow{2}{*}{\raisebox{3em}{\mbox{$^{13}G_{3}$}}} & $\eta_{G}$ & 
$1.00$ & $1.00$ & $1.00$ & $1.00$ & $1.00$ & $1.00$ & $1.00$ & $0.99$ & $0.99$ & $0.96$ \\ 
                            & $\bar{\delta}_{G}$ & 
$0.00$ & $0.03$ & $0.23$ & $0.57$ & $0.94$ & $1.26$ & $1.61$ & $2.16$ & $3.07$ & $4.40$ \\ 
\cline{1-2}
\multirow{2}{*}{\raisebox{-2.5em}{\mbox{$^{33}D_{3}$}}} & $\eta_{D}$ & 
$1.00$ & $1.00$ & $1.00$ & $0.99$ & $0.96$ & $0.91$ & $0.84$ & $0.76$ & $0.68$ & $0.61$ \\
\hline
\hline
\end{tabular}
\label{Tab:phas_coup34}
\end{table}

\begin{figure}[t]
   \centering
   \includegraphics[width=0.45\textwidth]{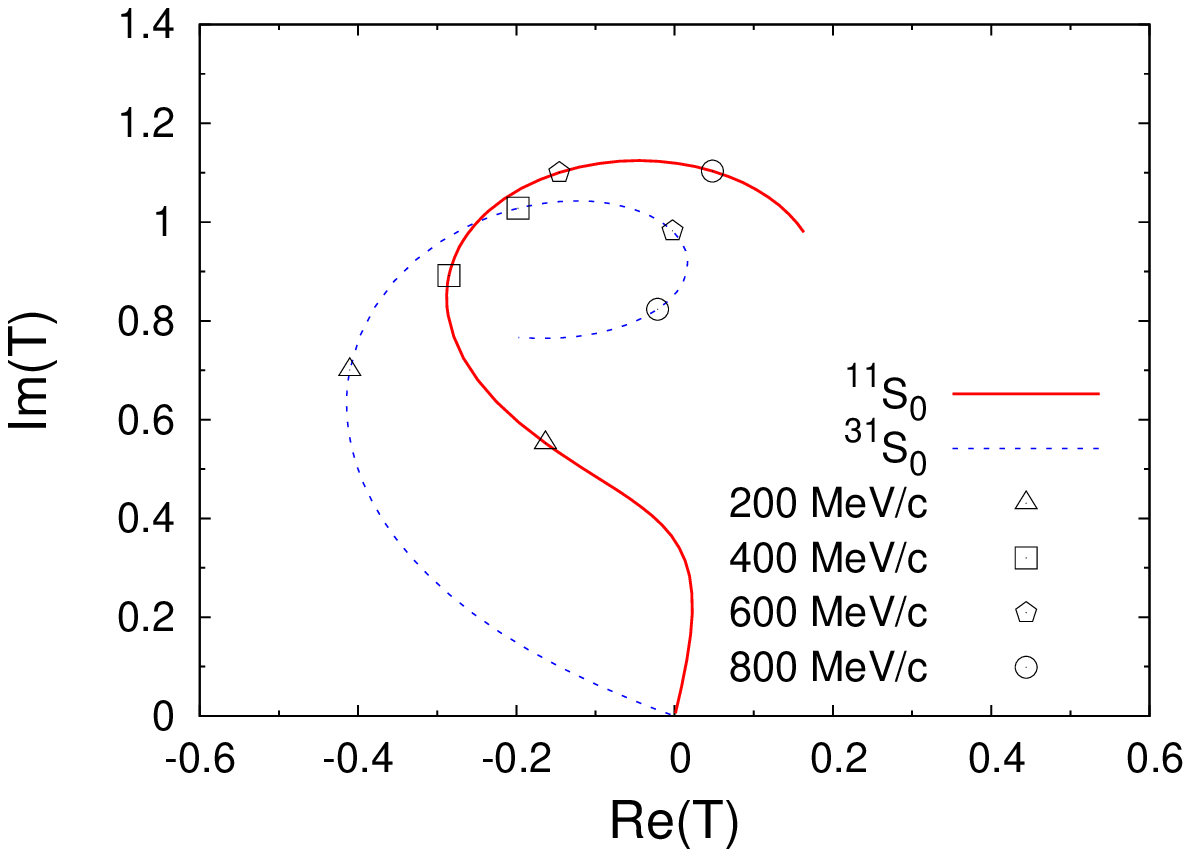} \hspace{5mm}
   \includegraphics[width=0.45\textwidth]{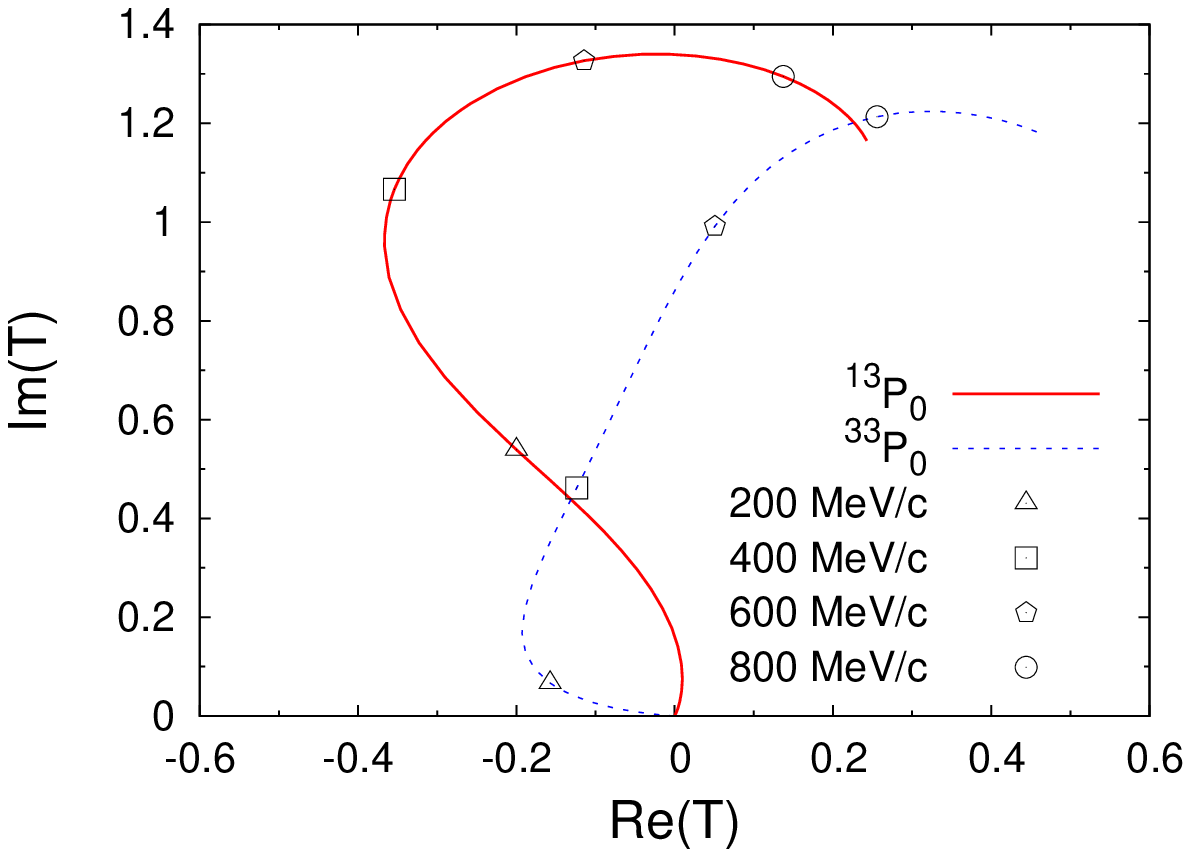}\\
   \includegraphics[width=0.45\textwidth]{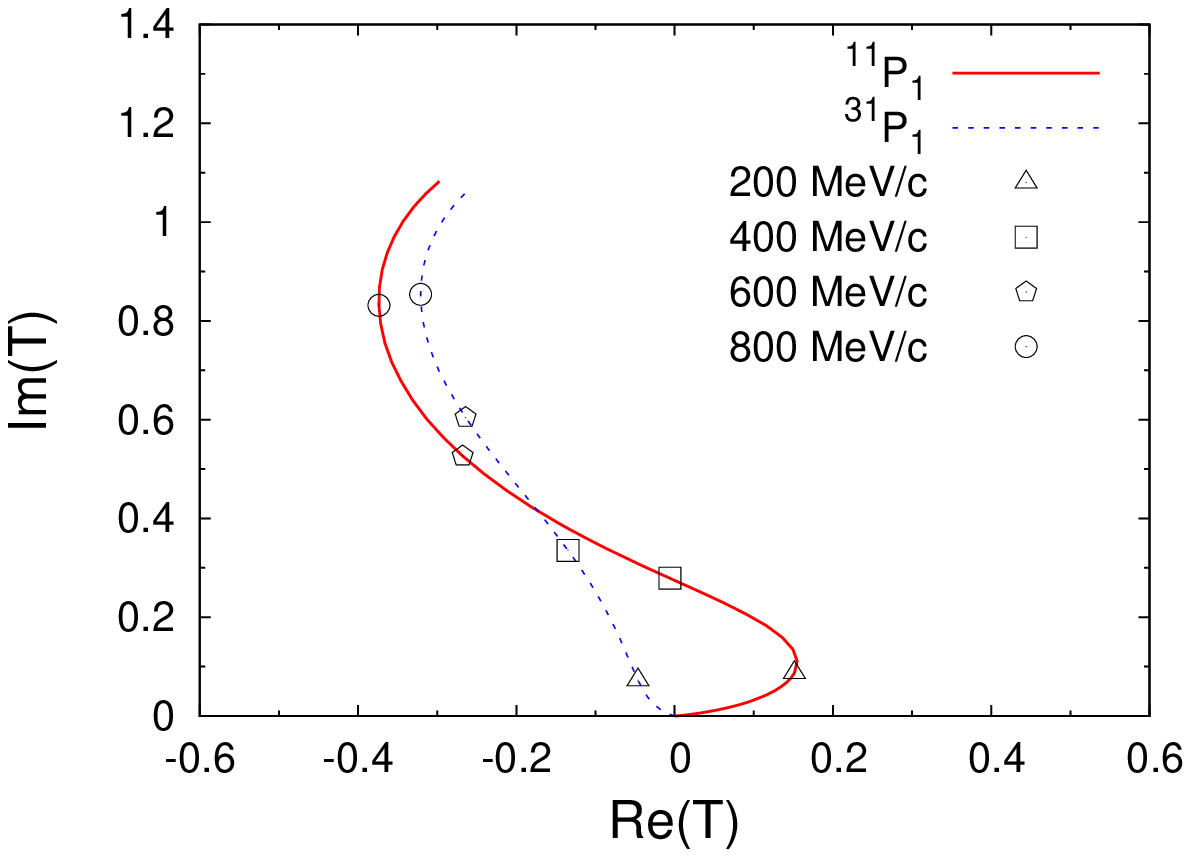} \hspace{5mm}
   \includegraphics[width=0.45\textwidth]{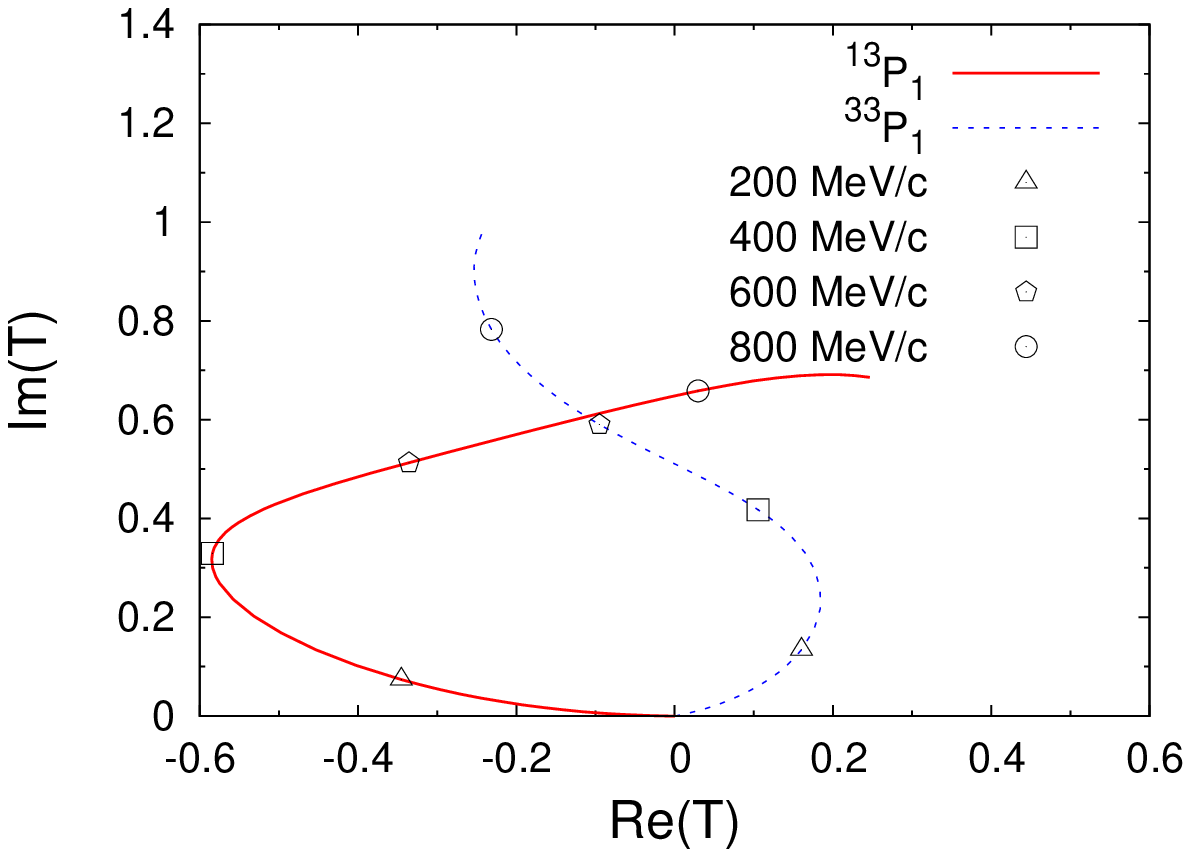}\\
   \includegraphics[width=0.45\textwidth]{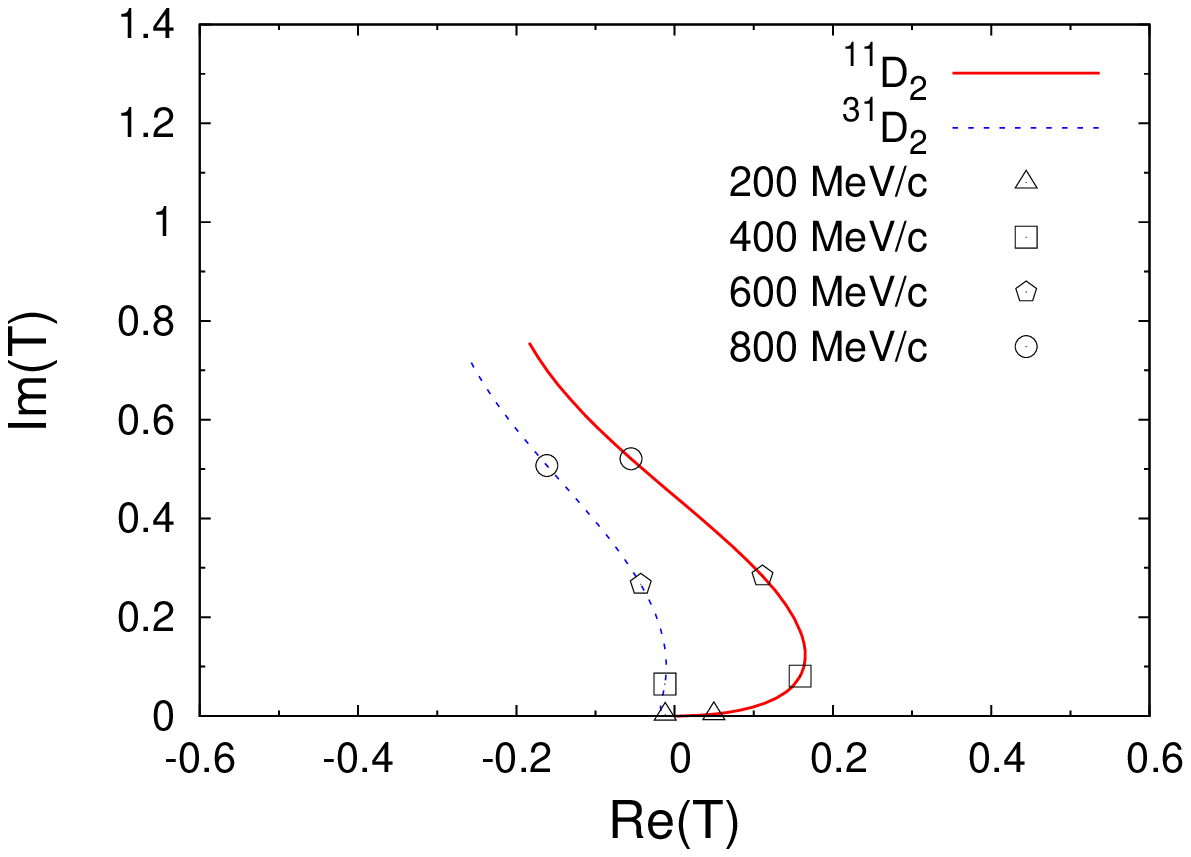} \hspace{5mm}
   \includegraphics[width=0.45\textwidth]{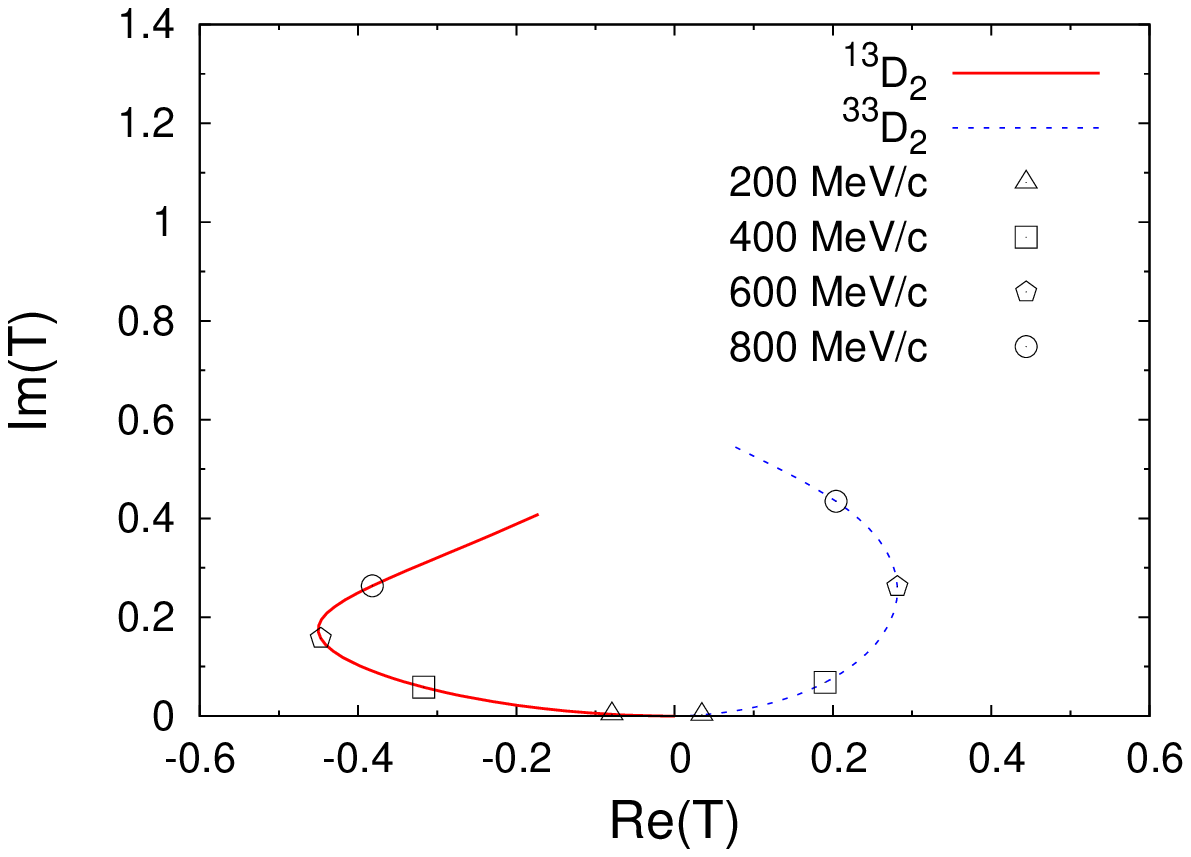}
\caption{\label{Fig:argand_SPD} (Color online)
The Argand diagrams for the uncoupled $S$, $P$, and $D$
waves, assuming isospin symmetry. The symbols on the lines denote
the values of the antiproton laboratory momenta.}
\end{figure}

\begin{figure}
   \centering
   \includegraphics[width=0.45\textwidth]{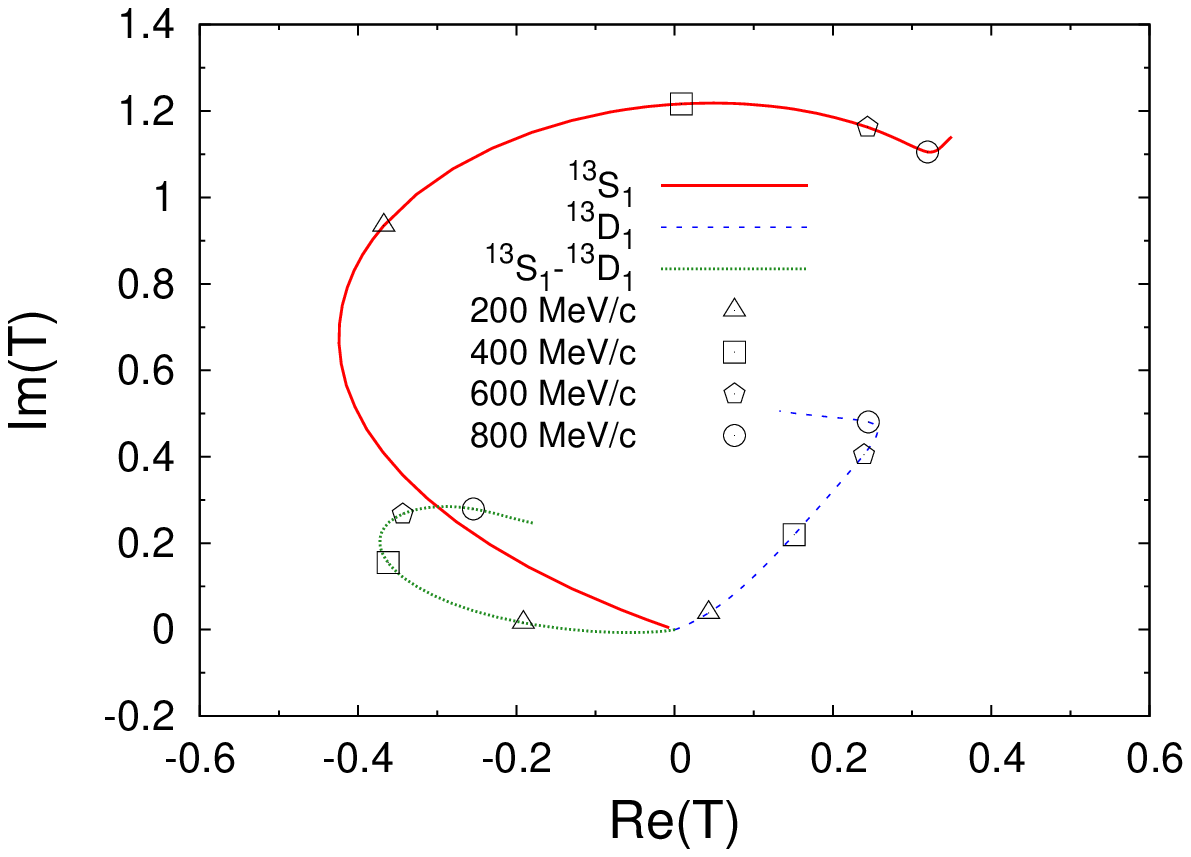} \hspace{5mm}
   \includegraphics[width=0.45\textwidth]{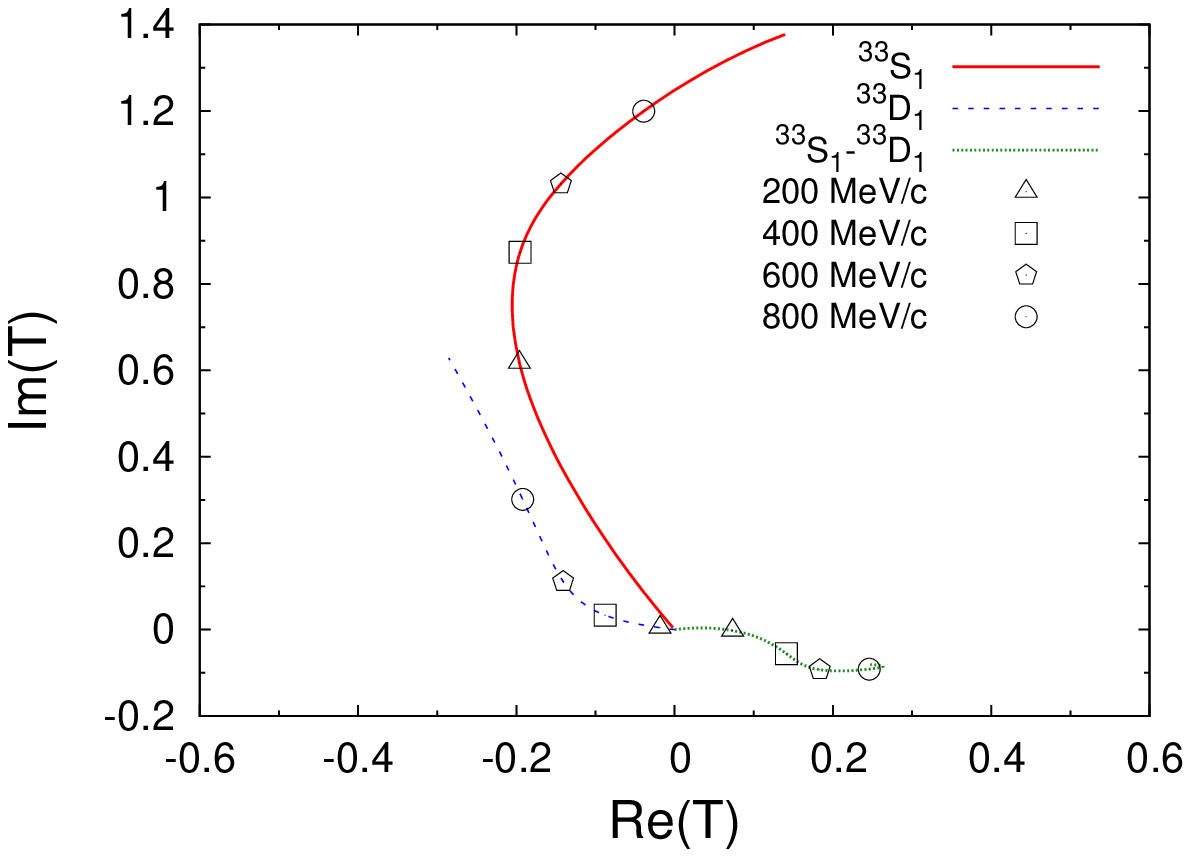}\\
   \includegraphics[width=0.45\textwidth]{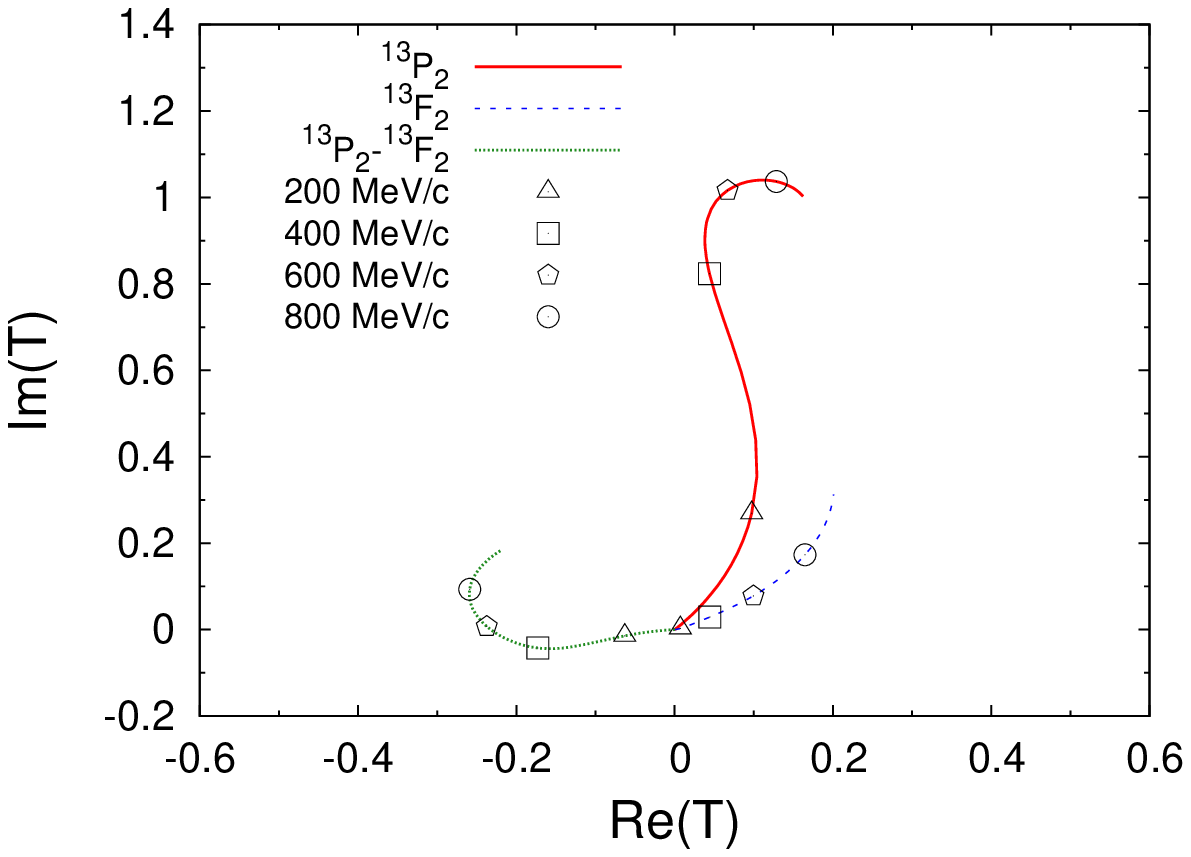} \hspace{5mm}
   \includegraphics[width=0.45\textwidth]{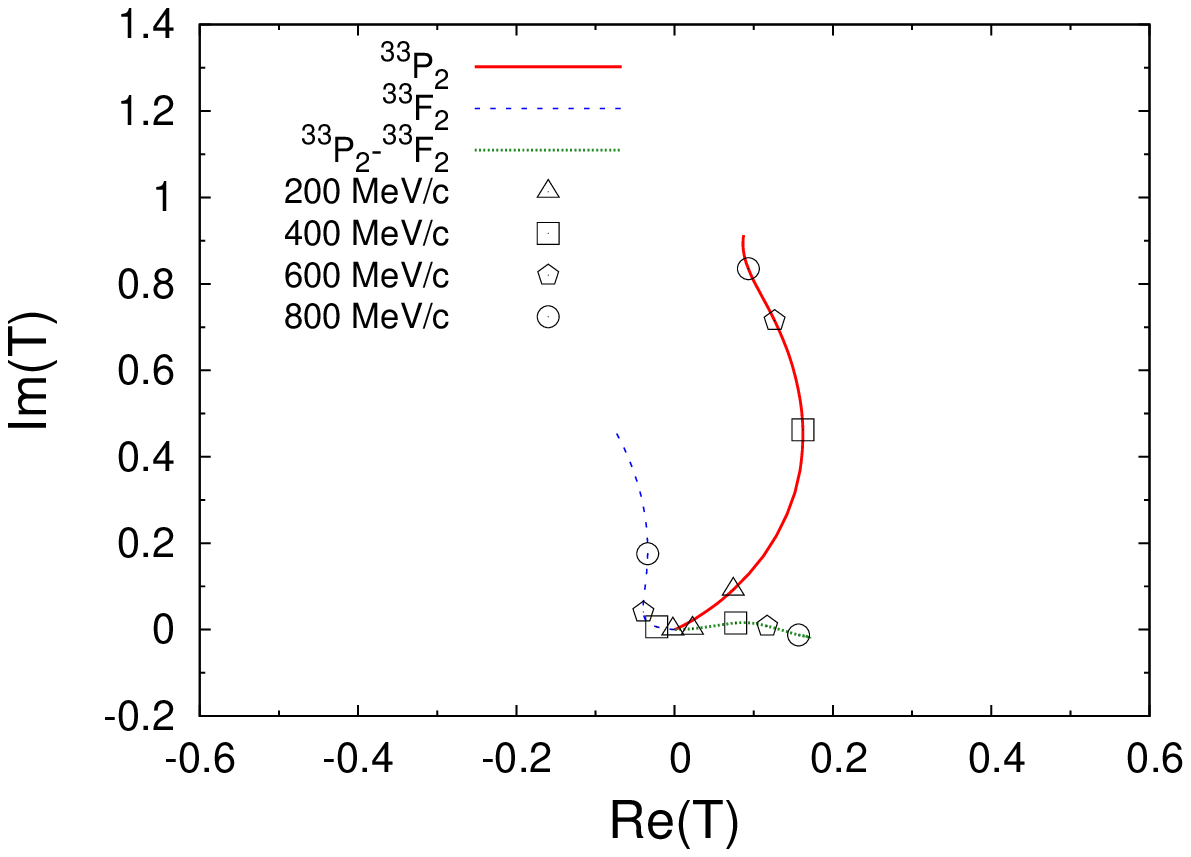}
   \includegraphics[width=0.45\textwidth]{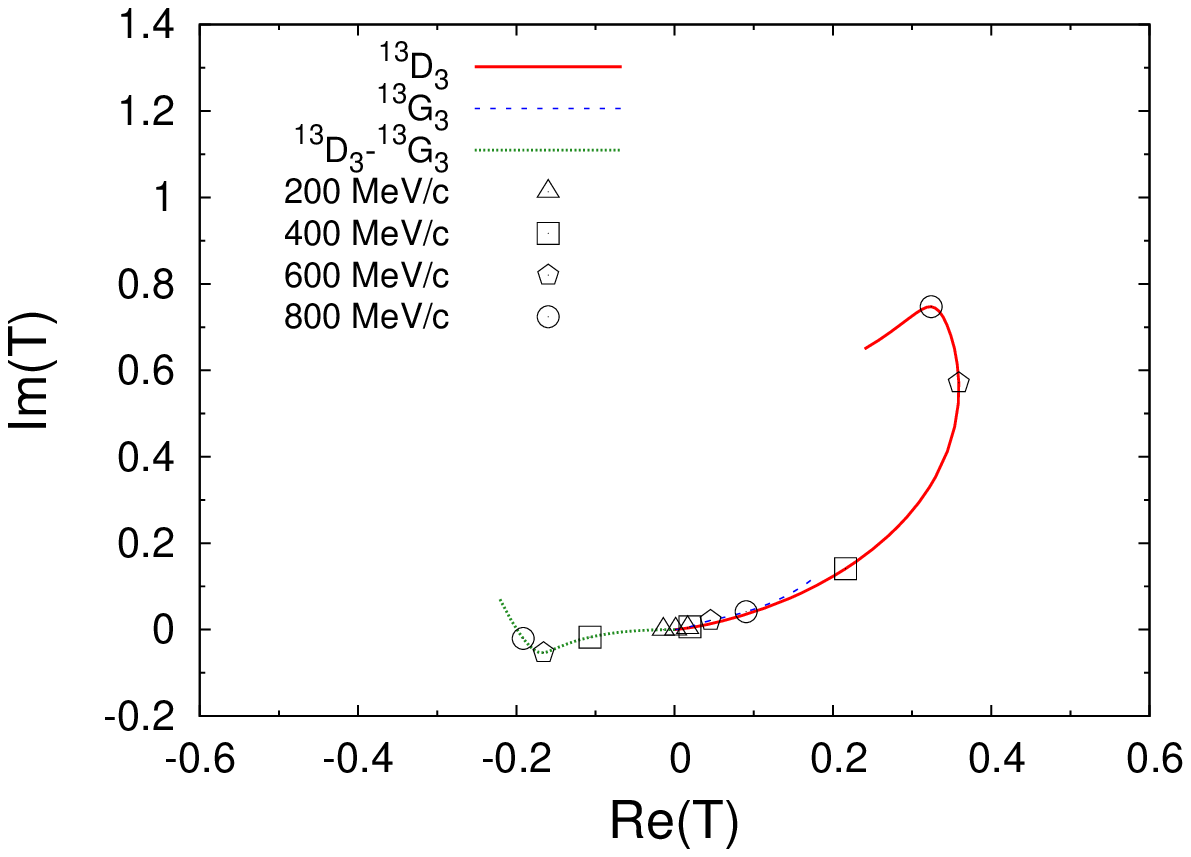} \hspace{5mm}
   \includegraphics[width=0.45\textwidth]{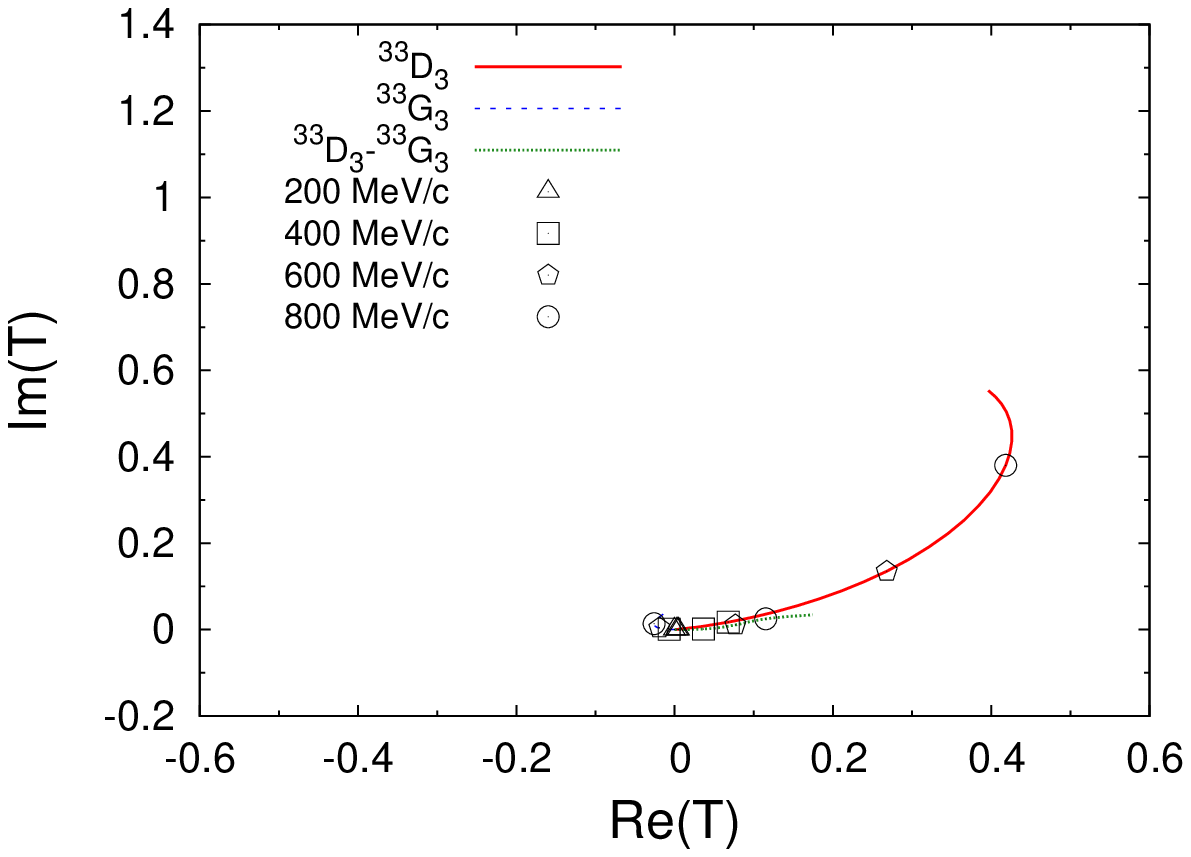}      
\caption{\label{Fig:argand_mixSDPFDG} (Color online)
The Argand diagrams for the coupled $S$-$D$, $P$-$F$,
and $D$-$G$ waves, assuming isospin symmetry.
The symbols on the lines denote
the values of the antiproton laboratory momenta.}
\end{figure}

\newpage
\section{Summary} \label{sec:Summary}
In summary, motivated by renewed experimental interest in low-energy antiproton-proton
scattering, we have presented a new energy-dependent PWA of all $\overline{p}p$ scattering
data below 925 MeV/$c$ antiproton laboratory momentum. We have improved the model
independence and quality of the PWA by using for the long-range interaction, next to the
electromagnetic potential, the charge-conjugated one- and two-pion exchange potential derived
from the effective chiral Lagrangian of QCD. We have updated the database and included the
high-quality differential cross sections and analyzing powers for charge-exchange scattering
$\overline{p}p\rightarrow\overline{n}n$ that were measured in the last years of operation
of LEAR. The final database contains 3749 scattering data, which are fitted with an
excellent $\chi^{2}_{\text{min}}/N_{\text{dat}}=1.000$ or $\chi^{2}_{\text{min}}/N_{\text{df}}=1.048$.
This implies that the long-range potential provides an excellent description of $\overline{p}p$
elastic and charge-exchange scattering, which we count as a success for chiral effective
field theory. Further improvement  of the PWA is certainly possible, but it will require additional
high-quality experimental data. Below 400 MeV/$c$, there are hardly scattering data available.
Spin observables will further constrain the PWA solution, provided they are precise enough.
The results presented in this paper will serve as the starting point for more specific investigations
of low-energy antiproton-proton scattering.

\section*{Acknowledgements}
We would like to thank our colleagues at KVI  for useful discussions. D. Zhou would like
to thank F. Jin for help with the figures and W. Kruithof for help with the tables.

\section*{Appendix}
In this Appendix, we study in more detail the statistical quality of the final
antiproton-proton database, by investigating the distribution of the contributions of the
$N_{\rm dat}=3749$ individual data points to the total $\chi^2$, $\chi_{\rm tot}^2$~\cite{Ber88}.
In the PWA, this distribution is given by
\begin{equation}
   P_{1,{\rm analysis}}(\chi^{2}) = 
       \frac{1}{N_{\rm dat}} \sum_{i=1}^{N_{\rm dat}} \delta(\chi^2-\chi_i^2) \ .
\label{P1a}
\end{equation}
In Fig.~\ref{fig:histogram} we plot this distribution as a histogram and compare
it to the theoretical $\chi^2$ distribution for 1 degree of freedom,
\begin{equation}
  P_1(\chi^2) = \frac{1}{\sqrt{2\pi}}\,t^{-1/2}e^{-t/2} \ .
\label{P1}
\end{equation}

\begin{figure}
\centering
\includegraphics[width=0.9\textwidth]{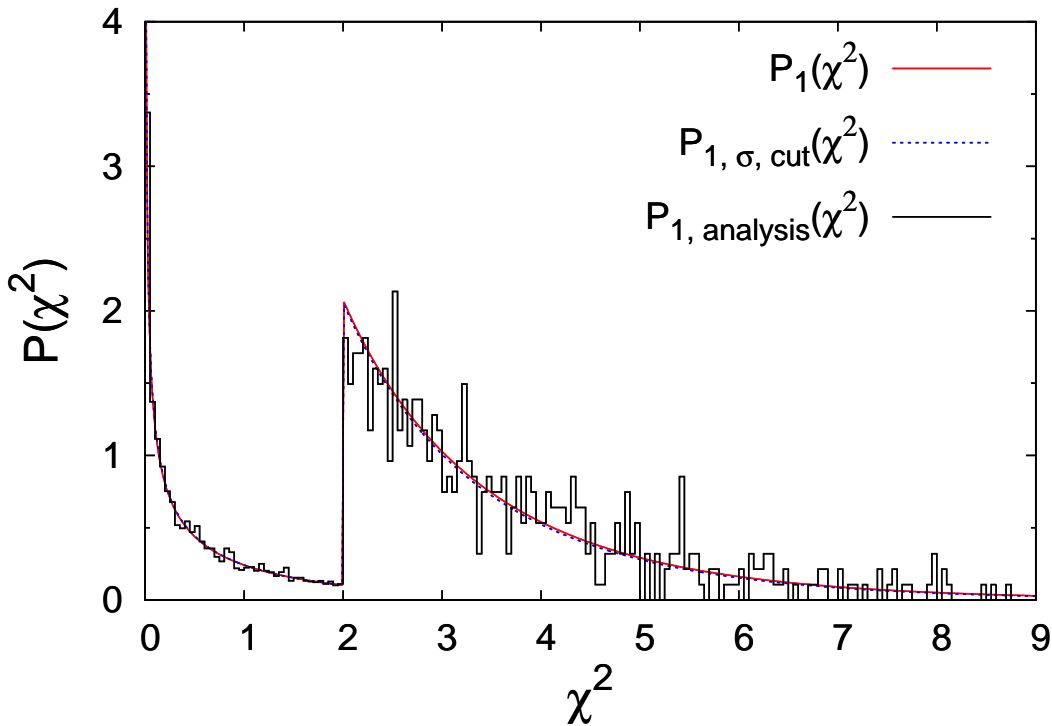}
\caption{ (Color online)
Probability distribution functions versus $\chi^{2}$. The tails, with
the values $\chi^{2}>2$, are enlarged by a factor of 20. The histogram contains
3749 data points in bins with $\Delta\chi^2=0.05$.}
\label{fig:histogram}
\end{figure}

In order to make this comparison quantitative, we give the moments, the central moments,
and the corresponding errors for the distributions. For a distribution $P(t)$, with $t\ge 0$,
we define the moments $\mu'_n$ and the central moments $\mu_n$ by
\begin{eqnarray}
   \mu'_n & = & \int_0^\infty dt P(t)\,t^n \ , \nonumber \\
   \mu_n  & = & \int_0^\infty dt P(t) (t-\mu'_1)^n \ ,
\end{eqnarray}
respectively. The errors on the moments are given by
\begin{equation}
  \sigma_{\mu'_n} = \left[ \frac{\mu'_{2n}-(\mu'_n)^2}{N_{\rm dat}} \right]^{1/2} \ ,
\end{equation}
and similarly for $\sigma_{\mu_n}$. The lowest moments and their errors are given
in Table \ref{tab:moments}. The agreement between the moments of $P_1(\chi^2)$
and $P_{1,{\rm analysis}}(\chi^{2})$ is reasonable, but not perfect.

In fact, for two reasons $P_1(\chi^2)$ is not the best distribution to compare to. First, while
the first moment of $P_1(\chi^2)$ is $\mu'_1=1$, that of $P_{1,{\rm analysis}}(\chi^{2})$
is $\mu'_1=\chi^2_{\rm tot}/N_{\rm dat}$. Since $\langle\chi^2_{\rm tot}\rangle=N_{\rm df}$,
we should compare to a narrower distribution
$P(\chi^2)=\beta^{-1}P_1(\beta^{-1}\chi^2)$ with $\beta=N_{\rm df}/N_{\rm dat}$.
Second, the data points with individual $\chi_i^2>9$ were rejected, which affects the
tail of the distribution and the higher moments. Therefore, it is better to compare
$P_{1,{\rm analysis}}(\chi^{2})$ to
\begin{equation}
  P_{1,\sigma,{\rm cut}}(\chi^{2}) = 
     \left[\sigma\sqrt{2}\gamma\left(\frac{1}{2},\frac{9}{2}\sigma^{-2}\right)\right]^{-1}\!
     (\chi^{2})^{-1/2} \,e^{-\chi^{2}/2\sigma^2} \theta(9-\chi^{2}) \ ,
\label{P1c}
\end{equation}
where $\gamma(s,z)=\int_0^z t^{s-1} e^{-t}dt$ is the lower incomplete gamma function and
$\sigma$ is a constant chosen to satisfy $\langle\chi^{2}\rangle=N_{\rm df}/N_{\rm dat}$;
in our case, $N_{\rm df}=3578$ and $N_{\rm dat}=3749$,  
therefore we have $\sigma=0.989$ and $\gamma(\frac{1}{2},\frac{9}{2\sigma^{2}})=1.768$.
The Heaviside step function $\theta(9-\chi^{2})$ removes the tail with
$\chi^{2}>9$. $P_{1,\sigma,{\rm cut}}(\chi^{2})$ is also plotted in Fig.~\ref{fig:histogram}
and its lowest moments with errors are given in
Table~\ref{tab:moments} as well. The agreement between the moments of
$P_{1,{\rm analysis}}(\chi^{2})$ and $P_{1,\sigma,{\rm cut}}(\chi^{2})$ is good,
which implies that the $\chi^2$ distribution of the PWA is close to what is expected
for statistical data.

\begin{table}
\caption{Moments $\mu'_n$ and central moments $\mu_n$ of the database of the
PWA and of the two theoretical probability distribution functions. The errors are given
for $N_{\rm dat}=3749$, where the contributions of the normalization data are included.}
\tabcolsep=2.5em
\renewcommand{\arraystretch}{0.85}
\begin{tabular}{c|ccc}
\hline\hline
&$P_{1}(\chi^{2})$&$P_{1,\sigma,{\rm cut}}(\chi^{2})$&$P_{1,{\rm analysis}}(\chi^{2})$ \\
\hline
$\mu'_{1}$ & 1.00 $\pm$ 0.02  & 0.95   $\pm$ 0.02  & 1.00  $\pm$ 0.02 \\
$\mu'_{2}$ & 3.00 $\pm$ 0.16  & 2.59   $\pm$ 0.11  & 2.80  $\pm$ 0.12 \\
$\mu'_{3}$ & 15.0 $\pm$ 1.6    & 10.7  $\pm$ 0.7      & 11.8 $\pm$ 0.8 \\
$\mu'_{4}$ & 105 $\pm$ 23      & 56  $\pm$ 5            & 62 $\pm$ 6 \\
$\mu_{2}$  & 2.00 $\pm$ 0.12  & 1.67   $\pm$ 0.08  & 1.80  $\pm$ 0.08 \\
$\mu_{3}$  & 8.0   $\pm$ 1.3    & 5.1   $\pm$ 0.5       & 5.4  $\pm$ 0.5 \\
$\mu_{4}$  & 60  $\pm$ 18       & 26.8  $\pm$ 3.1      & 28.5 $\pm$ 3.1 \\
\hline\hline
\end{tabular}
\label{tab:moments}
\end{table}

\end{document}